\documentclass[article]{revtex4-2} 
\usepackage{blindtext}
\usepackage{titlesec}

\usepackage{graphicx} 
\usepackage{natbib} 
\usepackage{hyperref}
\usepackage[usenames,dvipsnames]{color} 
\usepackage{amsmath} 
\usepackage{amssymb} 
\usepackage{ifthen} 
\usepackage{tikz}
\usetikzlibrary{decorations.pathmorphing,decorations.markings}
\usepackage{makeidx}

\makeindex

\def\be#1\ee{\begin{equation}#1\end{equation}}
\def\ba#1\ea{\begin{align}#1\end{align}}
\def\bg#1\eg{\begin{gather}#1\end{gather}}

\newcommand{\vect}[1]{\boldsymbol{#1}}

\def\shownote{1} 
\newcommand{\note}[1]{\ifthenelse{\shownote=1}{\textcolor{Red}{[[#1]]}}{}}
\def\showaddmat{1} 
\newcommand{\addmat}[1]{\ifthenelse{\showaddmat=1}{\textcolor{Gray}{[[#1]]}}{}}

\begin{document}

\title{A review on Aharonov-Bohm quantum machines: Thermoelectric heat engines and diodes}

\author{Salil Bedkihal $^1$, Jayasmita Behera $^2$, and Malay Bandyopadhyay $^2$}
\affiliation{$^{1}$ Dartmouth Engineering Thayer School, 15 Thayer Drive, Hanover, NH 03755, USA.\\
$^{2}$School of Basic Sciences, Indian Institute of Technology Bhubaneswar, Odisha, India 752050.}



\date{ \today}

\begin{abstract}
The study of heat-to-work conversion has gained considerable attention in recent years, highlighting the potential of nanoscale systems to achieve energy conversion in steady-state devices without the involvement of macroscopic moving parts. The operation of these devices is predicated on the steady-state flows of quantum particles, including electrons, photons, and phonons. This review examines the theoretical frameworks governing these steady-state flows within various mesoscopic or nanoscale devices, such as thermoelectric heat engines, particularly in the context of quantum dot Aharonov-Bohm interferometric configurations. Naturally, quantum interference effects hold great promise for enhancing the thermoelectric transport properties of these quantum devices by allowing more precise control over energy levels and transport pathways, thus improving heat-to-work conversion. Driven quantum dot Aharonov-Bohm networks offer an ideal platform for studying these engines, thanks to their ability to maintain quantum coherence and provide precise experimental control. Unlike bulk systems, nanoscale systems such as quantum dots reveal distinct quantum interference phenomena, including sharp features in transmission spectra and Fano resonances. This review highlights the distinction between optimization methods that produce boxcar functions and coherent control methods that result in complex interference patterns. This review reveals that the effective design of thermoelectric heat engines requires careful tailoring of quantum interference and the magnetic field-induced effects to enhance performance. In addition, We focus on the fundamental questions about the bounds of these thermoelectric machines. Particular emphasis is given to how magnetic fields can change the bounds of power or efficiency and the relationship between quantum theories of transport and the laws of thermodynamics. These machines with broken time-reversal symmetry provide insights into directional dependencies and asymmetries in quantum transport. We offer a thorough overview of past and current research on quantum thermoelectric heat engines using the Aharonov-Bohm effect and present a detailed review of three-terminal Aharonov-Bohm heat engines, where broken time-reversal symmetry can induce a coherent diode effect. Our review also covers bounds on power and efficiency in systems with broken time-reversal symmetry. We close the review by presenting open questions, summaries, and conclusions.
\end{abstract}

\newpage 
\maketitle
\tableofcontents
\section{Introduction}
Optimal conversion of heat to work is in the limelight for its immense applications in efficient energy-harvesting technologies. In this perspective, the role of quantum
coherence in energy harvesting in an efficient way is the central dogma. Recent progress in nanotechnology
has allowed us to investigate quantum-dot (QD) based thermoelectric heat machines where quantum interference effects can play a crucial role. Unlike the bulk systems, these nanoscale devices do not have any macroscopic moving parts (i.e., no turbines, pistons, etc.). Rather, they work on the principles of the steady-state currents of microscopic particles (electrons, phonons, etc.). Typically,
one considers a nanostructure made of a few QDs or a single-molecule junction between two thermal reservoirs maintained
at different temperatures and electrochemical potentials for such nanoscale machines. In such nanoscale devices, the striking role of symmetries, phase coherence, the effect of an external magnetic field, and quantum interference effects are investigated extensively \cite{study1, study2, study3, study4,study5,study6,study7}. Several studies of heat engines using the quantum system as a working fluid have been proposed and even experimentally realized \cite{expt1, expt2, expt3}.
Particle-exchange (PE) heat engines represent a promising advancement over traditional cyclical heat engines, especially for applications requiring miniaturization and low-power operation. Unlike cyclical engines, which rely on moving parts and are thus impractical for miniaturization, PE engines utilize energy filtering to control thermally driven particle flows between heat reservoirs, making them ideal for integration into solid-state materials. Theoretically, PE engines are predicted to reach thermodynamically ideal efficiency limits like cyclical engines.
Recent experimental work on a PE heat engine based on a quantum dot embedded in a semiconductor nanowire \cite{expt2} showed that under maximum power conditions, the efficiency $\eta$ aligned with the Curzon-Ahlborn efficiency, and the maximum $\eta$ exceeded 70\% of the Carnot efficiency while maintaining a finite power output. These results suggest that PE engines can approach thermodynamic limits for power conversion, with significant implications for applications such as hot-carrier photovoltaics, on-chip cooling, and energy harvesting in quantum technologies. Moreover, the inclusion of Aharonov-Bohm flux in quantum-dot networks enables the controlled manipulation of quantum interference effects, thereby allowing the exploration of quantum coherence in thermoelectric operation.\\
\indent
The celebrated Aharonov-Bohm (AB) interferometer can be considered a demonstrative example of a phase-tunable quantum device. The AB rings offer a tunable system to study the role of quantum effects in heat and charge transport \cite{AB1, AB2, AB3}. In this review, we present a comprehensive review of quantum coherent transport in driven AB interferometers with particular emphasis on (i) quantum interference enhanced thermoelectric behavior (two terminal systems) (ii) diode behavior resulting from many-body effects (Symmetries of transport in linear and non-linear regime) and
(iii) enhanced thermoelectric effects by broken time-reversal symmetry (three terminal system set-up. Linear and non-linear regime).\\
\subsection{Focus of this Review}
In the context of sustainable energy harvesting and green energy supply, it is always useful to investigate the fundamental dynamic mechanisms that control heat and particle transport. The main goal of the present review is to understand basic mechanisms and fundamental results on the steady-state conversion of heat to work in the framework of the Aharonov-Bohm (AB) interferometer.\\
\indent We start this manuscript with a short overview of the Landauer-Buttiker scattering theory for non-interacting electron systems. In Sec. III, we discuss coherent control of thermoelectric transport in a two-terminal AB heat engine. We investigate the essential role of various quantum interference mechanisms to obtain optimal output power and efficiency. Further, we explore the possible avenues to engineer transmission probability to design efficient thermoelectric heat engines. In Sec. IV, we elaborately demonstrate different probe techniques to mimic various elastic and inelastic effects. We provide a systematic analysis of the magnetic field and gate voltage symmetries of charge current and heat current in an AB interferometer, potentially far from equilibrium. Finally, we outline that a double-dot AB interferometer can act as a diode when two conditions are met simultaneously: (i) many-body effects are incorporated in the form of inelastic scattering, and (ii) broken-time reversal symmetry. In the subsequent section, we discuss the bounds on output power and efficiency in two-terminal systems with time-reversal symmetry and demonstrate the extensive changes observed in the bounds on power and efficiency when time-reversal symmetry is broken. We then review the bounds on power and efficiency for multi-terminal setup, particular emphasis is given to three-terminal setup. Finally, we also present bounds on power and efficiency in minimally nonlinear heat engines. In conclusion, we summarize our findings and present a few relevant open questions in the field of nanoscale thermoelectric heat engines.\\

\subsection{Literature Review}
To the best of our knowledge, we cite several relevant works and apologize for those that we overlook. The AB effect was observed in metallic loops and later on in the semiconductor heterostructures as a periodic modulation of current with the magnetic flux, with a periodicity of $\Phi_{0} = h/ne$ where n is an integer \cite{washburn1986, yacoby1995, holleitner2001}. Specifically this effect has been demonstrated in mesoscopic rings, with a single quantum dot structure integrated into one of the arms in a ring, and in double-dot structures. These experiments, and others, demonstrated that charge transport in these mesoscopic systems is phase coherent. Therefore, AB devices offer tunable systems and a natural laboratory to study the interplay of quantum interference and many-body effects in solid-state environments.\\
\indent
In the context of the AB interferometer, we emphasize the fundamental questions: Is electron transfer through quantum dot structures phase coherent, or incoherent? How do electron-electron and electron-phonon interactions influence phase-coherent transport? Conversely, what role do interference phenomena play in many-body effects, such as the formation of the Kondo resonance? These inquiries have been explored in numerous experimental and theoretical studies, which have detected the presence of quantum coherence in mesoscale and nanoscale objects using AB interferometry, as seen in Refs. \cite{yacoby1995, holleitner2001,verduijn2013, hackenbroich2001, schuster1997,konig2001, li2009, tokura2007, liu2007, boese2002, rai2012, sigrist2006, golosov2007, hofstetter2001, sun2002, malecki2010, hod2006}.
In particular, oscillations in the conductance resonances of an AB interferometer, with either one or two quantum dots embedded in its arms have been experimentally demonstrated in Refs. \cite{yacoby1995, holleitner2001}, indicating the presence of quantum coherence. Interestingly, AB oscillations were also observed in the co-tunneling regime, suggesting the involvement of phase coherence in such processes \cite{sigrist2006}. Experimental study of quantum transport through a parallel configuration of two coherently coupled silicon dopants forming an AB interferometer \cite{verduijn2013}, has demonstrated that the Kondo effect can be modulated coherently by changing the magnetic flux. This device was also shown to exhibit phase-coherent transport in the sequential tunneling regime.\\
\indent
The steady-state properties of the quantum dot AB interferometer have been extensively investigated \cite{hackenbroich2001, imry2002}, aiming to explore coherence effects in electron transmission within mesoscopic and nanoscale structures \cite{yacoby1995, schuster1997}. Particularly, the role of electron-electron (e-e) interactions in AB interferometry has been considered in Refs. \cite{konig2001, li2009, tokura2007, liu2007, boese2002, li2009a, kashcheyevs2006} revealing asymmetric interference patterns \cite{entinwohlman2012} and the enhancement \cite{boese2002} or elimination \cite{aharony2005} of Kondo physics. Recent works have further explored the possibility of magnetic field control in molecular transport junctions \cite{rai2011, hod2006, entinwohlman2012, rai2012}.\\
\indent
A systematic theoretical analysis conducted in Ref. \cite{konig2001} argued that e-e repulsion effects, leading to spin-flipping channels for transferred electrons, induce dephasing. In other studies, e-e repulsion effects were ignored \cite{kubala2002}, incorporated using a mean-field scheme \cite{golosov2007}, or treated perturbatively using Green’s function formalism \cite{sztenkiel2007, liu22007}. Most of the above studies typically focused on the steady-state limit, analyzing the conductance, a linear response quantity, or the current behavior, often in the case of infinite bias \cite{li2009, tokura2007}.\\
\indent
The double-quantum dot AB interferometer provides an important realization of a qubit, where magnetic flux can control interference effects. Entangled states of electrons are also of interest in solid-state quantum computing. Investigating the entanglement of electrons in a double-dot AB interferometer via transport noise was suggested by Loss {\it et al.} \cite{loss2000}.\\
\indent
Nonlinear transport measurements have been performed recently on AB rings connected to two leads by Leturcq \textit{et al.} \cite{leturcq2006phys, leturcq2006physica}, reporting that the even (odd) conductance terms are asymmetric (symmetric) in a magnetic field. It was also argued that these observations were insensitive to geometric asymmetries in the ring.\\
\indent
Magnetic field asymmetries of transport in mesoscopic conductors coupled to an environment have been theoretically studied by Kang \textit{et al.} \cite{sanchez2008phys}. The model system used in this work is a quantum dot conductor coupled to another conductor (treated as an environment) via a Coulomb interaction. This allows energy exchange between the conductor and the environment, without particle exchange. The environment was then driven out of equilibrium by applying a voltage bias. It was found that the interaction between the conductor and the environment causes magneto-asymmetry even in the linear regime if the environment is maintained out-of-equilibrium.\\
\indent
Magneto-asymmetries of transport were also explored in a double-dot interferometer coupled capacitively to an external fermionic environment. While in early studies, the capacitive interaction was either treated at the mean-field level or perturbatively \cite{sanchez2008phys}, in subsequent studies numerically exact results were obtained. Our work verified symmetry relations by calculating the current in a double-dot setup, using a numerically exact influence functional path integral method. We also derived nonlinear transport symmetries analytically using Buttiker probes.  Necessary and sufficient conditions for diode behavior in driven AB quantum-dot interferometers were obtained \cite{Bedkihal2013, PhysRevB.90.235411_Bedkihal2014}.\\
\indent
The AB interferometer network can also be used as an energy harvester or heat engine. The second law of thermodynamics ensures that the efficiency $\eta$ of such a heat engine is bounded by the Carnot value $\eta_C =1-\frac{T_c}{T_h}$ with $T_c$ and $T_h$ are the temperatures of the cold and hot reservoirs, respectively. Only when the device can act as a perfect energy filter this bound can be reached on the price of vanishing power \cite{Mahan1996,ref4, ref5}. In the past two decades, researchers have been investigating the relationship between thermoelectric power and efficiency beyond this limiting case \cite{engine4,engine5,engine6,engine7}. In this context, one may mention some landmark works by Whitney \cite{PhysRevLett.112.130601,engine9}, who recognized the transmission function that leads to optimal efficiency at a given power. As a result, the upper bounds on the output power of quantum coherent devices can be calculated. However, the efficiency of these devices is strictly smaller than $\eta_C$ at finite power in all the cases. Based on a reciprocating quantum heat engine by Allahverdyan {\it et al.} results confirm the general expectation that a conventional cyclic heat engine can approach Carnot efficiency only in the quasistatic limit with its output power necessarily to be zero \cite{engine10}.\\
\indent
On the other hand, Benenti {\it et al.} demonstrated that if one breaks the symmetry between the off-diagonal Onsager coefficients by applying an external constant magnetic field, it might enhance the efficiency of a thermoelectric heat engine substantially such that $\eta_C$ seems to be achievable even at finite power \cite{ref6}. An encouraging platform for investigating such aspects is provided by the multiterminal setup and a stronger bound on efficiency compared to the Benenti {\it et al.} \cite{ref6} was found by Brandner {et al.} \cite{ref18}. Further, they have moved forward one step by investigating that not only efficiency but also power can be bounded \cite{engine13}. They provide strong numerical evidence that power P indeed has a new bound. Their analysis demonstrates that the output power of the device vanishes whenever its efficiency reaches the upper limit corresponding to the respective number of terminals. Finally, they extrapolate their results for the arbitrary number of terminals (n), and arrive at the conjecture for a universal bound on P, which rules out any option of finite power at Carnot efficiency even in the limit $n\rightarrow\infty$ \cite{ref19}.\\



\subsection{Quantum Coherence: Importance and applications}
In traditional thermoelectrics, the relaxation length at room temperature is usually of the order of the mean free path of the carriers, a scale way smaller than the thermoelectric structure. Thus, the transport phenomenon can be well described by the Boltzmann-transport theory which assumes a local equilibrium at each point in thermoelectrics.\\
\indent
On the other hand, the structure is much smaller than this relaxation length in the nanoscale thermoelectrics. This originates new kind of physics, like quantum interference effects or strong correlation effects. The transport properties of the system are non-local in every sense and can not be described by Boltzmann's transport theory.\\
\indent
Since the thermalization is weak in these nanoscale devices, the particles keep their energy and the coherence of their wavefunction as they pass through the device. As a result, one may expect interference between the different paths that a particle can travel and the quantum correlation can build up which can give rise to different new effects in nanoscale thermoelectrics that can not be found in classical ones. Thus, we can at least demand that coherence related to the quantum mechanical phase can be manipulated by experimentalists to improve the performance of a heat engine. In addition, the AB phase supplies another parameter to the experimentalist to tune to optimize the machine's performance. In this context, the effect of Breit-Wigner resonance and Fano resonance through AB interferometer is needed to be examined in details. \\
\indent
The coherence of electron transport processes through an AB interferometer has been typically identified and characterized via conductance oscillations in magnetic fields. However, in a double-dot AB structure, a device including two dots, both connected to biased metal leads, it is imperative that the relative phase between the two dot states (or charge states) should similarly convey information on electron coherence-decoherence dynamics. The coherence dynamics of the double-dot AB interferometer have been studied using the non-perturbative influence functional path integral method. The effects of e-e interactions on the relative phase between charge states of double-dot were quantified. It has been found that in the non-interacting limit, the relative phase between dot states is localized to $\pm{\pi/2}$ for magnetic flux $\phi\neq 2\pi$, the real part of coherence between dot-states was found to have a periodicity of $4\pi$. It has been shown that the phase localization behavior is destroyed in the limit of large inter-dot Coulomb interaction where double occupancy is forbidden. This coherent control of double-dot states using magnetic flux and voltage bias can be useful in quantum information processing. In other words, non-equilibrium effects can be useful to generate inter-dot coherence that can be exploited to design charge and spin qubits.\\
\indent
The quantum coherence of electron transport through nanoscale semiconductor quantum dots (QDs) represents a critical aspect, thoroughly explored in numerous experiments. A single quantum dot, often referred to as an ``artificial atom," when coupled with ideal leads (which can act as a heat bath), functions similarly to an impurity atom in bulk metal. Therefore, mesoscopic systems containing QDs offer an excellent platform for studying quantum interference in a controlled manner, which is not easily accessible in bulk solids. Since such a system is a typical representation of a quantum working fluid coupled to a hot and cold bath, it allows the probing of the role of quantum interference effects in heat-to-work conversion in a controllable fashion.\\
\indent
When an electron transmits through two coupling paths—discrete energy levels and a continuous band—the interference between these two channels results in a distinctive asymmetric line shape in the electron conductance, known as the Fano effect. Such line shapes have been observed in linear transport through quantum dot interferometers \cite{XIONG2005216}.\\
\indent
Recent measurements of the Fano resonance have been conducted using an alternative setup, specifically an AB interferometer with a quantum dot embedded in one of its arms. In this configuration, the quantum dot is in the Coulomb blockade regime, facilitating electron resonant tunneling, while the other arm serves as the continuous path. This setup realizes a well-defined Fano system with the added advantage of allowing independent control of both interfering paths \cite{XIONG2005216}.\\
\indent
The tunable Fano system exhibits unique properties. In addition to the complex expression of the Fano parameter, the phase evolution demonstrates behavior contrary to previous observations. The Landauer conductance through the AB interferometer includes a term dependent on the magnetic flux through the ring. In this ``closed" two-terminal geometry, unitarity and time-reversal symmetry impose the Onsager relation. However, in the mesoscopic structures, this relation requires specific phase jumps across each quantum dot resonance, which is often observed by opening the interferometer by attaching ``lossy" leads.\\
\indent
Despite extensive theoretical investigation, the unique behavior of the measured phase in this system remains puzzling. Using a non-interacting model, it is possible to derive an analytical expression for the conductance, incorporating the factor of dephasing, which extends the Fano form. In this closed system, the unitarity of the coherent part of the tunnel current is broken due to dephasing, although the total current remains unitary. Consequently, when coherent conductance is measured, it does not obey the Onsager relation for a two-terminal system and exhibits a non-zero phase shift, leading to continuous phase variation. It is this broken symmetry of the transmission that can be exploited to design diodes using quantum-dot networks as discussed in subsequent sections.\\
\subsection{Quantum heat engines: Efficiency and Bounds}
The AB quantum dot networks have also been investigated as thermoelectric quantum heat engines. The canonical set-up consists of a network of interconnected quantum dots with AB phases appearing as complex tunneling amplitudes between the sites. The magnetic flux-dependent thermoelectric transport in two-terminal and three-terminal setups has been intensively investigated in the linear and non-linear regimes. The theory of linear irreversible thermodynamics provides relations between thermodynamic fluxes and thermodynamic forces,
\begin{equation}
    \mathbf{J} = \mathbf{LX}. \label{eq:1.5}
\end{equation}
Here $\mathbf{J}$ is a column vector denoting the heat and particle current fluxes, $\mathbf{X}$ denotes a column vector of thermodynamic forces related to the temperature and voltage bias, and $\mathbf{L}$ is the Onsager matrix. Its diagonal elements are conductances, and the off-diagonal elements are related to the Seebeck and Peltier coefficients \cite{saito2011, balachandran2013,onsager1931a, onsager1931b, casimir1945}. The Onsager-Casimir symmetry dictates that time-reversal symmetry results in reciprocal relations between linear response coefficients \cite{onsager1931a, onsager1931b, casimir1945}, $L_{i,j} = L_{j, i}$. In the presence of a magnetic field $\mathbf{B}$, the reciprocity relation becomes
\begin{equation}
    L_{i,j}(\mathbf{B}) = L_{j,i}(-\mathbf{B}). \label{eq:1.6}
\end{equation}
From the above equation, we can immediately see that the conductances (diagonal matrix elements) are even functions of the magnetic field. This symmetry is known as the “phase rigidity” of linear conductance in a two-terminal AB interferometer.

In the non-linear regime, Onsager-Casimir symmetries need not hold. A prominent example of this breakdown is the asymmetry of the differential conductance out-of-equilibrium. This effect has been attributed to e-e interactions in the system, resulting in an asymmetric charge response under the reversal of a magnetic field, leading to a magneto-asymmetric differential conductance.

The nanoscale systems offer a promising platform for efficient heat-to-work conversion. The strong demand for cost-effective and environmentally friendly energy sources drives research in this field. Enhancing the efficiency of thermoelectric materials is a central focus of current research. A typical thermoelectric setup comprises a system in contact with two reservoirs—left (L) and right (R)—at different temperatures and chemical potentials. In the linear response regime, the performance of a bulk thermoelectric device is characterized by a dimensionless parameter known as the figure of merit, \( ZT \). This parameter is a combination of transport coefficients: electrical conductivity (\( \sigma \)), thermal conductivity (\( \kappa \)), thermopower (\( S \)), and temperature (\( T \)). The figure of merit is defined as \( ZT = \left(\frac{\sigma S^2}{\kappa}\right)T \). The efficiency can be expressed as:
\begin{equation}
\eta = \eta_c \frac{\sqrt{ZT + 1} - 1}{\sqrt{ZT + 1} + 1},
\end{equation}
where \( \eta_c = 1 - \frac{T_c}{T_H} \) is the Carnot efficiency, achieved in the limit as \( ZT \to \infty \). The linear response approximation is often valid for bulk systems due to the possibility of large temperature differences across the sample with minimal temperature gradients. In contrast, nanoscale systems experience temperature and electrical potential gradients at the nanometer scale, which can lead to nonlinear effects. Nevertheless, the performance of nanoscale devices can still be analyzed using an expression analogous to the efficiency equation above \cite{balachandran2013}.
Thermodynamics does not set an upper limit on \( ZT \), but the interdependence between electric and thermal transport properties makes it challenging to increase \( ZT \) beyond 1. It has been suggested that breaking time-reversal symmetry could enhance thermoelectric performance because, in such systems, the efficiency depends on (i) the magnetic field asymmetry of the thermopower and (ii) the figure of merit. Therefore, understanding how many-body interactions and phase-breaking processes affect transport in nanoscale systems in the nonlinear regime is crucial for enhancing efficiency.

With this outset the rest of the paper is organized as follows. We begin by introducing the theoretical modeling approach based on Landauer-B\"{u}ttiker scattering theory. Next, we provide a comprehensive review of coherent control in thermoelectric transport, discussing various quantum interference mechanisms and their role in engineering transmission probability—key factors in designing efficient thermoelectric engines. In the subsequent section, we outline the sufficient conditions for diode behavior in systems with broken time-reversal symmetry within the Büttiker probe formalism, accompanied by a brief review of probe techniques. We then review the bounds on power and efficiency in two-terminal systems with time-reversal symmetry and in multi-terminal systems where time-reversal symmetry is broken. Bounds on power and efficiency in minimally nonlinear heat engines are presented. Finally, we summarize our findings and present open questions in the field that merit exploration to achieve optimal power and efficiency configurations

\section{Method}
The steady-state and transient behavior of driven quantum-dot AB interferometers has been studied using various methods such as Markovian master equations, perturbative Green's function-based methods, non-perturbative deterministic path integrals based on influence functional, and hierarchical equations of motions \cite{Keldysh1965, Schwinger1961, Kadanoff1962, SBPath}. The complexity of this problem varies depending on the many-body interactions, such as electron-electron and electron-phonon interactions, or both. The non-equilibrium many-body effects in these systems exhibit rich behavior that is beyond the scope of this article. However, the non-interacting limit also shows non-trivial behavior, such as phase localization of relative phases between charge states,
$4\pi$ periodicity of the real part of electronic coherence, and the breaking of phase localization in the infinite electron-electron repulsion limit where double occupancy is forbidden.

The non-equilibrium Green's function (NEGF) technique was rigorously developed by Schwinger in a classic mathematical paper \cite{Schwinger1961}, treating the Brownian motion of a quantum oscillator. The next significant development in this field was by Kadanoff and Baym, who derived quantum kinetic equations \cite{Kadanoff1962}. A diagrammatic expansion in powers of the coupling to the environment was developed by Keldysh \cite{Keldysh1965}, with the key idea of contour ordering. These initial developments occurred in the early 1960s. Another important advancement in the context of quantum transport was an explicit derivation of a formula for the transmission function in terms of Green's function by Caroli {\it et al.} \cite{Caroli1971}.

Most of the foundational work in quantum thermoelectrics and finite-time thermodynamics reviewed in this article is based on Landauer-Büttiker scattering theory \cite{Landauer1957, buttiker1984, Buttiker1986, Buttiker1986b}, which is equivalent to the NEGF technique for the non-interacting case. The effects of many-body interactions can be incorporated into the scattering theory description using Büttiker probes. Most of the foundational results and efficiency bounds on thermoelectric heat engines with broken time-reversal symmetry can be understood using this framework.

In the scattering formalism of Landauer and Büttiker, interactions between particles are neglected. Considering a multi-terminal setup, one can express the charge current from $\nu$ to $\nu'$ terminal in terms of the transmission probability $T_{\nu,\nu'}(\omega)$, a function which depends on the energy of the incident electron,
\begin{align}
I_\nu(\phi) &= \int_{-\infty}^{\infty} d\omega \left[
    \sum_{\nu' \ne \nu} T_{\nu,\nu'}(\omega, \phi) f_\nu(\omega)
    - \sum_{\nu' \ne \nu} T_{\nu',\nu}(\omega, \phi) f_{\nu'}(\omega)
\right].
\end{align}
The Fermi-Dirac distribution function $f_\nu(\omega) = \left[ e^{\beta_\nu(\omega - \mu_\nu)} + 1 \right]^{-1}$ is defined in terms of the chemical potential $\mu_\nu$ and the inverse temperature $\beta_\nu$. The magnetic field is introduced via an Aharonov-Bohm flux $\Phi$ applied perpendicular to the conductor, with the magnetic phase $\phi = 2\pi\Phi/\Phi_0$. The transmission function $T_{\nu,\nu'}$ can be expressed as,
\begin{align}
T_{\nu,\nu'} &= \text{Tr} \left[ \Gamma_\nu G^+ \Gamma_{\nu'} G^- \right].
\end{align}
Here the matrices \( G^+ \) and \( G^- = (G^+)^\dagger \) are the retarded and advanced Green’s functions, respectively and it is defined as
\begin{equation}
G^+(\omega) =\Big[\omega - H_S - \sum_{\nu}\Sigma_{\nu}^+(\omega)\Big]^{-1},
\end{equation}
where $H_S$ is the Hamiltonian of the subsystem. Here, \( \Sigma_{\nu}^+(\omega) \) are the self-energies and \( \Gamma_{\nu}(\omega) \) are the hybridization matrices defined as follows:
\begin{equation}
    \Sigma_{\nu}^+(\omega)=V^{\nu}g_{\nu}^+(\omega)V^{\nu\dagger},\,\,\
    \Gamma_{\nu}(\omega)=\frac{1}{2\pi i}(\Sigma_{\nu}^{-}-\Sigma_{\nu}^{+}),
\end{equation}
where, $\Sigma_{\nu}^{-}=(\Sigma_{\nu}^{+})^{\dagger}$, the subsystem-reservoir coupling matrices are denoted by $V^{\nu}$, and $g_{\nu}^+$ are the isolated reservoir Green’s functions.

With these tools, one can perform an analysis of Aharonov-Bohm heat engines and gain significant insights into (i) coherent control of thermoelectric effects, (ii) the trade-off between power and efficiency, (iii) bounds on efficiency in linear and non-linear response regimes, and (iv) the necessary and sufficient conditions for diode behavior under broken time-reversal symmetry. In the subsequent sections, we present a comprehensive review of the results in this field.

\section{Coherent control of thermoelectric transport in two-terminal Aharonov-Bohm heat engines}
\begin{figure}[h] 
  \centering
  \includegraphics[width=1.0\textwidth]{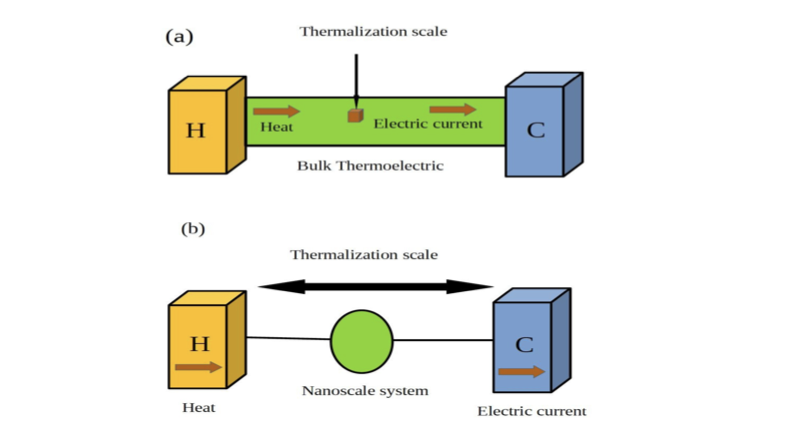}
  \caption{(a) In conventional thermoelectrics, electrons quickly reach local equilibrium, allowing the temperature to vary smoothly and be described by Boltzmann equations.
  (b) The nanoscale structure is of similar size
or smaller than the length scale on which electrons relax to a local equilibrium. The system's physics becomes more complex, showing significant non-equilibrium effects. Without local thermalization, quantum interference effects emerge that would normally be suppressed by decoherence. The steady-state quantum coherence in such a system can be exploited for energy harvesting.}
  \label{fig:bulk}
\end{figure}
For over 40 years, bulk semiconductor thermoelectrics have been used to convert heat into electrical power, prominently featured in space missions such as the Curiosity rover on Mars, which relies on a thermoelectric generator powered by plutonium-238 \cite{ambrosi2019rtgs}. These generators are well-suited for space applications due to their durability and lack of moving parts, as demonstrated by Voyager 1’s generator, which continues to function beyond the solar system. Thermoelectrics also offer refrigeration capabilities through Peltier cooling. However, despite their utility, the efficiency of these devices remains relatively low. For instance, the Curiosity rover’s thermoelectric generator operates at about 5 percent efficiency, significantly below the theoretical maximum efficiency predicted by the Carnot cycle for the temperature differences involved. Although the theoretical maximum efficiency corresponds to zero power output, the practical question is how close we can approach the Carnot limit and whether efficiency can be improved using quantum interference effects.

In conventional thermoelectrics, electrons achieve local equilibrium rapidly, allowing the temperature to exhibit a smooth variation described by Boltzmann equations. On the other hand, when the nanoscale structure is comparable to or smaller than the length scale over which electrons relax to local equilibrium, the system's physics becomes more intricate, manifesting significant non-equilibrium effects. Figure \ref{fig:bulk} shows a schematic of the thermalization scale in bulk and nanoscale thermoelectrics. In the absence of local thermalization, quantum interference effects arise, which would typically be suppressed by decoherence. Exploiting steady-state quantum coherence in such systems offers potential for energy harvesting.

Optimal conversion of heat to work is crucial for efficient energy-harvesting technologies. The influence of quantum coherence in enhancing energy harvesting is a vibrant area of research. Recent advancements in nanotechnology have enabled the study of quantum-dot (QD) thermoelectric heat machines, where quantum interference effects can play a significant role. Unlike traditional machines, these thermoelectric-based devices lack macroscopic moving parts, such as turbines or pistons.

Before moving to specifics of coherent control of AB quantum heat engines, we discuss the operation of quantum-dot heat engines briefly. When an electron with energy \(E\) transfers from reservoir 1 to reservoir 2 through a nanostructure, its energy is conserved. The work performed in this process is given by \(\mu_2 - \mu_1\), where \(\mu_1\) and \(\mu_2\) represent the electrochemical potentials of the two reservoirs. Concurrently, heat is extracted from reservoir 1, equal to \(E - \mu_1\), and added to reservoir 2, amounting to \(E - \mu_2\). This results in a net sum of zero for both heat and work, consistent with the first law of thermodynamics.

According to Clausius, the entropy change in a reservoir is determined by the change in heat multiplied by the reservoir’s inverse temperature \(\beta_j\). In Landauer scattering theory, the electron flow from reservoir 1 to reservoir 2 at energy \(E\) is described by \(T(E) \left(f(x_1) - f(x_2)\right)\), where \(T(E)\) is the transmission probability and \(f(x_j)\) denotes the Fermi distribution function with \(x_j = \beta_j (E - \mu_j)\). The overall entropy change due to the electron flow is proportional to \((x_2 - x_1)\left(f(x_1) - f(x_2)\right)\). Since \(f(x)\) decreases monotonically, the entropy change cannot be negative, thereby satisfying the second law of thermodynamics.

For thermodynamic reversibility, electron transmission should occur only at the energy where the Fermi distribution functions in both reservoirs are equal, given by \(E^* = \mu_1 + \frac{\beta_2 (\mu_2 - \mu_1)}{\beta_2 - \beta_1}\). At this energy \(E^*\), there is no net flow of heat or electric current. Introducing electrons within a very narrow energy range just above \(E^*\) allows for minimal entropy production and approaches Carnot efficiency. This ideal scenario can be realized using a quantum dot or molecule precisely tuned to \(E^*\). Nonetheless, practical challenges, such as heat flow through phonons and photons, limit efficiency. To improve bulk thermoelectrics, techniques like incorporating nanostructures are investigated to optimize their energy spectra and enhance overall performance. While achieving reversibility is challenging, it is desirable from a practical standpoint to design an artificial quantum thermoelectric system that exhibits sharp transmission features due to quantum interference effects. In this context, systems composed of multiple coupled quantum-dot nanostructures present interesting possibilities. The key question is whether we can design a transmission with sharp features that allow us to enhance efficiency and power output. The alternative ways of increasing efficiency and power include three terminal systems with broken time-reversal symmetry which is discussed in subsequent sections.

 The classical problem of best thermoelectrics, originally believed to have been solved by Mahan and Sofo \cite{Mahan1996}, was revisited and discussed in the quantum limit. The thermoelectric figure of merit ($ZT$) was expressed as a function of electronic transmission probability using the Landauer–Büttiker formalism, which was able to address thermoelectric transport ranging from ballistic to diffusive regimes. The study proposed applying the calculus of variations to search for the optimal transmission probability that maximizes $ZT$. For the sake of completeness, we outline an argument based on variational calculus as follows: In the study of two-terminal quantum thermoelectric heat engines, the transmission function plays a crucial role in determining the device's efficiency and power output. An ideal energy filter, represented by a boxcar function, can significantly enhance performance by selectively allowing electrons within a specific energy range to contribute to transport processes.

A quantum thermoelectric heat engine consists of two reservoirs at different temperatures (\(T_H\) and \(T_C\)) and chemical potentials (\(\mu_H\) and \(\mu_C\)) connected by a quantum conductor. The goal is to convert thermal energy into electrical energy efficiently. The performance of such devices is characterized by the electrical conductance (\(G\)), the Seebeck coefficient (\(S\)), and the thermal conductance (\(\kappa\)).

To enhance the efficiency of a thermoelectric device, it is beneficial to employ an energy filter that selectively allows electrons within a specific energy window to pass through the conductor. This energy filtering can be mathematically represented by a boxcar function in the transmission probability \(T(E)\).

The boxcar function, \(T(E)\), is defined as:
\begin{equation}\label{eq:Tbox}
T(E) =
\begin{cases}
1 & \text{if } E_1 \leq E \leq E_2, \\
0 & \text{otherwise},
\end{cases}
\end{equation}
where \(E_1\) and \(E_2\) define the energy window through which electrons can pass.

To find the optimal transmission function that maximizes the efficiency of the thermoelectric device, we employ variational calculus. The efficiency \(\eta\) of a thermoelectric engine is given by:
\begin{equation}
\eta = \frac{P}{J_Q},
\end{equation}
where \(P\) is the power output and \(J_Q\) is the heat current absorbed by the quantum system from the hot reservoir.
The power output \(P\) and the heat current \(J_Q\) are expressed as:
\begin{equation}
P = e \int_{-\infty}^{\infty} (E - \mu) T(E) [f_H(E) - f_C(E)] dE,
\end{equation}
\begin{equation}
J_Q = \int_{-\infty}^{\infty} (E - \mu_H) T(E) [f_H(E) - f_C(E)] dE,
\end{equation}
where \(e\) is the electronic charge, \(\mu\) is the chemical potential, and \(f_{H/C}(E)\) are the Fermi-Dirac distribution functions for the hot and cold reservoirs.

The variational problem is to find the transmission function \(T(E)\) that maximizes \(\eta\) subject to physical constraints. Using the method of Lagrange multipliers, we introduce a functional:
\begin{equation}
\mathcal{L}[T(E)] = \eta[T(E)] - \lambda \left( \int_{-\infty}^{\infty} T(E) dE - N \right),
\end{equation}
where \(\lambda\) is the Lagrange multiplier and \(N\) is a normalization constant.

Taking the functional derivative of \(\mathcal{L}\) with respect to \(T(E)\) and setting it to zero gives:
\begin{equation}
\frac{\delta \mathcal{L}}{\delta T(E)} = \frac{\partial \eta}{\partial T(E)} - \lambda = 0.
\end{equation}

Solving this variational equation yields the optimal transmission function. For an ideal energy filter, the solution is a boxcar function as defined in Eq. (\ref{eq:Tbox}). It was revealed that the optimal transmission probability is a boxcar function instead of a delta function as proposed by Mahan and Sofo, leading to $ZT$ values exceeding the well-known Mahan–Sofo limit \cite{Mahan1996}. Furthermore, the realization of the optimal transmission probability in topological material systems was suggested. This work defined the theoretical upper limit for quantum thermoelectrics, which is of fundamental significance to the development of thermoelectrics. The sharper transmission probability leads to Carnot efficiency with vanishing power output and the broader box car shape function leads to reduced efficiency but with increased power output \cite{PhysRevLett.112.130601}. So it is interesting to explore the transmission engineering of nanostructure using coherent control such that a boxcar-like shape may be realized using quantum interference effects.

The impact of dimensionality on the electronic performance of thermoelectric devices was elucidated using the Landauer formalism. This formalism established the relationship between thermoelectric coefficients, the transmission function \( T(E) \), and the energy distribution of conducting channels, \( M(E) \) \cite{kim2009influence}. Applying the Landauer formalism enabled a comprehensive comparison of device performance across different dimensional systems, ranging from ballistic to diffusive transport limits. This method also provided a concrete interpretation of the ``transport distribution,'' a concept typically analyzed through the Boltzmann transport equation approach.

Quantitative comparisons of thermoelectric coefficients in one, two, and three dimensions indicated that conducting channels could be more effectively utilized in lower-dimensional systems. However, to fully exploit the advantages of reduced dimensionality, a high packing density was essential, requiring the thicknesses of quantum wells or wires to be minimized. Investigations into engineering \( M(E) \) into a delta-function form revealed the potential for approximately 50\% improvement in performance compared to bulk semiconductors. Although the shape of \( M(E) \) improved with decreasing dimensionality, enhanced performance was not solely determined by lower dimensionality. Instead, it was contingent upon both the shape and magnitude of \( M(E) \).
The benefits of optimizing the shape of \( M(E) \) appeared relatively modest, strategies to increase the magnitude of \( M(E) \) offered significant potential for enhancing the performance of thermoelectric devices.
Since quantum interference can be tuned using the Aharonov-Bohm phase, multiply connected quantum-dot networks are suitable for transmission engineering.
In the coherent control mechanism, quantum coherence and interference effects within the quantum conductor are leveraged to influence the transport properties. The design of the transmission function \(T(E)\) in this case would involve quantum interference effects, potentially leading to more complex shapes than a simple boxcar function.

In a quantum dot system, for example, coherence and interference can be controlled through external fields or by engineering the dot's coupling to the leads. The transmission probability \(T(E)\) can exhibit resonant peaks and dips due to these coherent effects, which may be tailored to approximate the desired energy filtering behavior.
While coherent control might not produce a perfect boxcar function, it can achieve a similar effect by designing interference patterns that selectively enhance or suppress transmission at specific energies. This approach leverages the wave nature of electrons and the resulting interference patterns to control transport properties.

To summarise, the boxcar function arises as an optimal transmission function in the theory of two-terminal quantum thermoelectric heat engines. It represents an ideal energy filter that maximizes efficiency by allowing only electrons within a specific energy range to contribute to transport. This approach is rooted in optimal control theory and variational calculus, where the transmission function is optimized to enhance the performance of the thermoelectric device. Additionally, coherent control provides an alternative method to achieve selective energy filtering by leveraging quantum interference effects, although it may result in more complex transmission functions.

Recent advancements in thermoelectric materials research have significantly focused on enhancing the thermoelectric figure of merit (\( ZT \)) of materials, particularly Bi\(_2\)Te\(_3\) alloys, through the application of quantum-well superlattice structures \cite{PARK2023107723}. Since the 1960s, Bi\(_2\)Te\(_3\) alloys have been the benchmark for thermoelectric applications due to their relatively high \( ZT \). However, the progress in increasing \( ZT \) has been gradual and incremental. A promising approach to improve the thermoelectric performance of these materials involves the preparation of quantum-well superlattice structures. These structures consist of alternating thin layers of different materials, creating quantum wells that significantly alter the electronic and thermal properties of the material.

Extensive research includes comprehensive calculations to examine the effect of such layering on \( ZT \), particularly for highly anisotropic materials like Bi\(_2\)Te\(_3\). Anisotropy, in this context, refers to the directional dependence of a material's properties, which is a critical factor in thermoelectric performance. The calculations indicate that the orientation of the superlattice multilayers is crucial; for highly anisotropic materials, aligning the layers in specific orientations can substantially enhance the figure of merit. These findings suggest that quantum-well superlattice structures have the potential to significantly increase the thermoelectric figure of merit, thereby improving their efficiency as thermoelectric coolers.

The magnetic-flux-dependent dispersions of sub-bands in topologically protected surface states of a topological insulator nanowire manifest as Aharonov-Bohm oscillations (ABOs) observed in conductance measurements, reflecting the Berry phase of \(\pi\) due to the spin-helical surface states \cite{KWON2023105691}. Thermoelectric measurements have been utilized to probe variations in the density of states at the Fermi level of the surface state of a topological insulator nanowire (Sb-doped Bi\(_2\)Se\(_3\)) under external magnetic fields and an applied gate voltage. The ABOs observed in the magneto-thermovoltage exhibited $180^\circ$ out-of-phase oscillations depending on the gate voltage values, which can be used to tune the Fermi wave number and the density of states at the Fermi level. The temperature dependence of the ABO amplitudes indicated that phase coherence was maintained up to \(T = 15 \, \textrm{K}\). These findings suggest that thermoelectric measurements could be a valuable tool for investigating the electronic structure at the Fermi level in various quantum materials.

The body of work in this area opens new avenues for enhancing the performance of thermoelectric materials through the use of quantum-well superlattice structures. By precisely controlling the orientation and layering within these structures, it may be possible to achieve significant improvements in the thermoelectric figure of merit, paving the way for the next generation of high-efficiency thermoelectric devices. The introduction of the AB flux into quantum-well superlattice structures can add an intriguing dimension to the study of thermoelectric materials. By threading a magnetic flux through the superlattice, the electronic phase coherence can be manipulated, potentially leading to novel quantum interference effects. These effects could further optimize the electronic and thermal transport properties, thereby enhancing the thermoelectric figure of merit (\( ZT \)) beyond what is achievable with conventional methods. This approach opens up new pathways for exploiting quantum phenomena in the design of high-efficiency thermoelectric devices.

Low-dimensional and nanoscale materials are promising candidates for thermoelectric heat engines capable of converting heat into electrical work. The discrete energy levels in these materials create sharp transmission resonances, enabling natural energy filtering effects. This characteristic allows the system to function as a heat engine and refrigerator by exchanging particles with external reservoirs. In the linear response regime, the performance of a thermoelectric heat engine is characterized by the figure of merit, $ZT=GS^{2}T/\kappa$, where $G$ is the electrical conductance, $T$ is the temperature and $\kappa$ is the thermal conductance. The higher the value of $ZT$, the higher the thermal efficiency.
Recently, it has been observed that materials with a density of states characterized by sharp peaks exhibit a high value of $ZT$ \cite{lambert2016mesoscopic}. The presence of sharp peaks in transmission and density of states can be attributed to quantum interference effects. Therefore, exploiting and controlling quantum coherence is crucial for enhancing thermal efficiency. The AB flux naturally enables the tunability of quantum coherence, facilitating optimal efficiency. This leads to a richer field of coherent control in thermoelectric effects.

Quantum interference effects in coupled and driven nanostructures can stem from two main processes: (i) interference between continuum and discrete degrees of freedom known as Fano resonance, and (ii) interference arising solely from the internal coherent dynamics of the nanoscale structure.
The Fano resonance arises from the interference among the electrons circulating through the system's available channels: the discrete levels and the continuous conduction bands. Generally, by detuning the system from the bound state in the continuum (BIC) conditions, such as changing the hybridization between the leads and the dots, the BICs transform into Fano resonances (quasi-BICs). These resonances exhibit dips in the transmission spectra and sharp peaks in the density of states (DOS). In the case of the AB interferometer, the Fano resonance can be tuned by varying the magnetic flux. One can then optimize the system configuration to enhance thermal efficiency.

The previous section shows that charge and heat currents can be expressed in terms of the energy-dependent transmission function. Therefore, engineering the transmission function is crucial for achieving optimal thermal efficiency. The quantum interference effects can be broadly classified into two resonances: (i) Breit-Wigner resonance and (ii) Fano resonance. The transmission function typically consists of the sum of these two contributions.
The Breit-Wigner resonance describes the transmission probability through a quantum dot with a discrete energy level interacting weakly with the conduction electrons in the leads. The transmission probability \( T_{\text{BW}} \) is given by:
\begin{equation}
T_{\text{BW}}(E) = \frac{\Gamma^2}{(E - E_0)^2 + \Gamma^2},
\end{equation}
where \( E \) is the energy of the incoming electron,
   \( E_0 \) is the resonance energy (the energy level of the quantum dot), and \( \Gamma \) is the resonance width (related to the coupling strength between the dot and the leads).

The Fano resonance arises due to the interference between a discrete energy state and a continuum of states. In a triple dot system, this can occur when one dot's discrete energy level interferes with the continuum states provided by the other dots and leads. The transmission probability \( T_{\text{Fano}} \) is given by:
\begin{equation}
T_{\text{Fano}}(E) = \frac{(q + \epsilon)^2}{1 + \epsilon^2},
\end{equation}
where \( \epsilon = \frac{E - E_0}{\Gamma} \) is the reduced energy, \( q \) is the Fano parameter describing the asymmetry of the resonance, and \( E_0 \) and
\( \Gamma \) are the same as in the Breit-Wigner formula.

\begin{figure}[h] 
  \centering
  \includegraphics[width=0.5\textwidth]{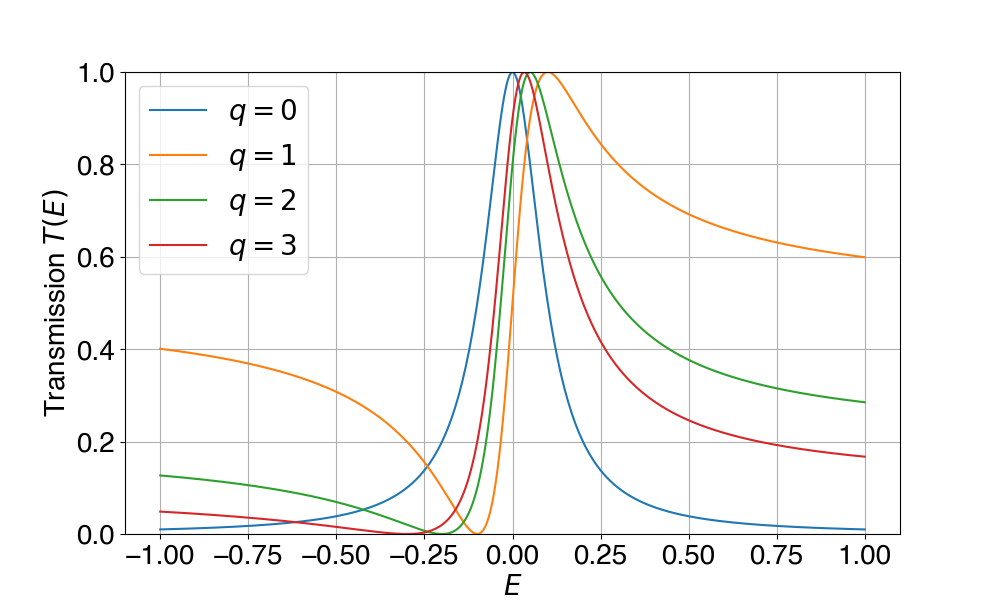}
  \caption{The transmission function, which describes the probability of electron transport through a system, typically includes contributions from both Breit-Wigner and Fano resonances. Breit-Wigner Resonance: This type of resonance occurs due to the discrete energy levels of a quantum system, leading to a Lorentzian-shaped peak in the transmission function ($q=0$). Fano Resonance: This arises from the interference between discrete quantum states and a continuum of states, resulting in an asymmetric line shape in the transmission function.}
  \label{fig:transmission}
\end{figure}

In Fig. \ref{fig:transmission} we show the Breit-Wigner and Fano resonance contributions to the transmission probability. Depending upon the geometrical arrangement of the quantum-dots network, a richer structure of resonances can be obtained depending upon the symmetries of the system as discussed in subsequent sections.

An example of a coupled quantum-dot AB network is a triangular triple-dot AB interferometer.
In this setup, each quantum dot acts as a localized state where an incoming electron from the source terminal (S) can tunnel between the dots, potentially circulating the loop before exiting to the drain terminal (D).
Given the Hamiltonian for a triangular triple quantum dot system:
\begin{equation}
H_{\text{TQD}} =
\begin{pmatrix}
\varepsilon_d & t e^{i\phi/3} & t e^{-i\phi/3} \\
t e^{-i\phi/3} & \varepsilon_d & t e^{i\phi/3} \\
t e^{i\phi/3} & t e^{-i\phi/3} & \varepsilon_d
\end{pmatrix},
\end{equation}
where \(\varepsilon_d\) is the on-site energy of each dot, \(t\) is the tunneling amplitude, and \(\phi\) is the AB phase. Dot 1 is coupled to the source, Dot 3 to the drain, and Dot 2 is not directly coupled to the source or drain.\\

\textbf{Breit-Wigner Resonance:}
\begin{equation}
\begin{split}
\mathcal{M}_{\text{BW}}^{(1 \to 2 \to 3)} &= (-i\sqrt{\Gamma_1}) \cdot \frac{1}{E - \varepsilon_d + i\Gamma_1/2} \cdot (-i t e^{i\phi/3}) \cdot \frac{1}{E - \varepsilon_d} \\
& \quad \cdot (-i t e^{-i\phi/3}) \cdot \frac{1}{E - \varepsilon_d + i\Gamma_3/2} \cdot (-i\sqrt{\Gamma_3}).
\end{split}
\end{equation}

\textbf{Fano Resonance:}
\begin{equation}
\mathcal{M}_{\text{Direct}} = (-i\sqrt{\Gamma_1}) \cdot \frac{1}{E - \varepsilon_d + i\Gamma_1/2} \cdot (-i\sqrt{\Gamma_3}) \cdot \frac{1}{E - \varepsilon_d + i\Gamma_3/2},
\end{equation}
\begin{equation}
\begin{split}
\mathcal{M}_{\text{Indirect}} &= (-i\sqrt{\Gamma_1}) \cdot \frac{1}{E - \varepsilon_d + i\Gamma_1/2} \cdot (-i t e^{i\phi/3}) \cdot \frac{1}{E - \varepsilon_d} \\
& \quad \cdot (-i t e^{-i\phi/3}) \cdot \frac{1}{E - \varepsilon_d + i\Gamma_3/2} \cdot (-i\sqrt{\Gamma_3}).
\end{split}
\end{equation}

\textbf{Total Transmission Amplitude:}
\begin{equation}
\mathcal{M}_{\text{total}} = \mathcal{M}_{\text{Direct}} + \mathcal{M}_{\text{Indirect}} + \mathcal{M}_{\text{BW}}^{(1 \to 2 \to 3)}.
\end{equation}

\textbf{Transmission Probability:}
\begin{equation}
T(E) = |\mathcal{M}_{\text{total}}|^2.
\end{equation}
\indent
In networks of tunnel-coupled quantum dots, two main time scales govern the system dynamics: (i) internal coherent dynamics determined by inter-dot tunneling $t$ and (ii) tunneling rate from dots to source/drain electrodes $\Gamma$ related to the hybridization of dot states with the source and drain regions. In our recent analysis, we found that by equating these two time scales, we achieve a compact transmission spectrum where each state equally contributes to the transmission. This results in maximally constructive interference, enhancing thermal efficiency \cite{PhysRevB.108.165419}. Please note that, as opposed to the box car function, this transmission function has three peaks with unit transmission and two dips. Here, the separation between these peaks is crucial to achieve large efficiency and power as shown in \cite{PhysRevB.108.165419}.

In the linear response regime, the electrical and the thermal conductance can be written as:
\begin{equation}
G = \frac{2e^2}{h} \int_{-\infty}^{\infty} \mathcal{T}(E,\phi) \left( -\frac{\partial f(E, \mu, T)}{\partial E} \right) dE,
\end{equation}
where \( \mathcal{T}(E,\phi) \) is the transmission function, \( f(E, \mu, T) \) is the Fermi-Dirac distribution function:
  \begin{equation}
  f(E, \mu, T) = \frac{1}{e^{(E - \mu)/k_B T} + 1},
  \end{equation}
 \( \mu \) is the chemical potential,
 \( e \) is the electron charge,
 \( h \) is Planck's constant, and
 \( k_B \) is Boltzmann's constant.
The thermal conductance \( \kappa \) can be expressed as:
\begin{equation}
\kappa = \frac{2}{h T} \int_{-\infty}^{\infty} \mathcal{T}(E,\phi) (E - \mu)^2 \left( -\frac{\partial f(E, \mu, T)}{\partial E} \right) dE.
\end{equation}
These expressions for conductance are derived from the Landauer-Büttiker formalism. The transmission function \( \mathcal{T}(E,\phi) \) represents the probability that an electron at energy \( E \) will transmit through the system. The Fermi-Dirac distribution function \( f(E, \mu, T) \) describes the occupancy of electronic states at energy \( E \), chemical potential \( \mu \), and temperature \( T \).

The electric conductance \( G \) is proportional to the integral of the transmission function weighted by the derivative of the Fermi-Dirac distribution, which essentially counts the number of electronic states available for conduction near the Fermi level.
The thermal conductance \( \kappa \) is proportional to the integral of the transmission function weighted by the energy difference from the chemical potential squared, reflecting the contribution of each energy state to the heat transport. Hence the transmission probability is crucial to optimise the thermoelectric figure of merit.

In the linear response regime, thermopower can be expressed in terms of the derivative of the transmission function around Fermi energy.
The thermopower (Seebeck coefficient) \( S \) can be expressed in terms of the derivative of the transmission function \( T(E,\phi) \) as follows \cite{lambert2016mesoscopic}:
\begin{equation}
S = -\frac{\pi^2 k_B^2 T}{3e} \left. \frac{d \ln T(E,\phi)}{dE} \right|_{E = E_F}.
\end{equation}

The energy variable is represented by \( E \), and \( E_F \) indicates the Fermi energy. The term \( \frac{d \ln T(E,\phi)}{dE} \) is the derivative of the natural logarithm of the transmission function concerning energy, evaluated at the Fermi energy \( E_F \).
Sharper derivatives of the transmission function around the Fermi energy result in larger thermopower. The presence of AB flux enables tuning of the resonance position and sharpness, thus allowing for coherent control of thermoelectric effects.
Coupled double quantum dots (c-2QD) connected to leads have been widely adopted as prototype model systems to verify interference effects on quantum transport at the nanoscale. An analytic study of the thermoelectric properties of c-2QD systems pierced by a uniform magnetic field has been performed in a linear response regime \cite{Menichetti_2018}. Fully analytic and easy-to-use expressions were derived for all the kinetic functionals of interest. Within Green’s function formalism, the results allowed a simple and inexpensive procedure for the theoretical description of thermoelectric phenomena for different chemical potentials and temperatures of the reservoirs, different threading magnetic fluxes, dot energies, and interdot interactions. Moreover, they provide an intuitive guide to parametrize the system Hamiltonian for the design of best-performing realistic devices. It is found that the thermopower \(S\) can be enhanced by more than ten times and the figure of merit \(ZT\) by more than a hundred times by the presence of a threading magnetic field. Most importantly, it is shown that the magnetic flux also increases the performance of the device under maximum power output conditions.
The kinetic functionals in the linear response regime can be written in the following form
\begin{equation}\label{eq:Kn}
    K_{n}=\int_{-\infty}^{\infty}d\omega T(\omega,\phi)\frac{(\omega-\mu)^{n}}{{(K_{B}T)}^{n}}\left(-\frac{\partial f}{\partial\omega}\right),
\end{equation}
where $n=0,1,2$.
For a tunnel-coupled parallel double-dot AB interferometer, the transmission function is Lorentzian, centered at the bonding state without a magnetic field and for an even number of flux quanta. For an odd number of flux quanta, the Lorentzian is centered at the anti-bonding state.
Applying a magnetic field can transform a simple, unstructured Lorentzian function into a sharply structured function with alternating peaks and valleys (including zero minima), which greatly improves the associated thermoelectric properties. Typically, the transmission function can be described qualitatively as a combination of a Lorentzian-like curve around the bonding state and a Fano-like curve around the anti-bonding state (or vice versa, depending on the magnetic field), with the separation determined by the coupling energy \( \left| t_d \right| \). From Eq. (\ref{eq:Kn}), we can see that the sharper features in the transmission probability are also reflected in the kinetic functional \cite{Menichetti_2018}.

Thermoelectric effects were studied in an AB interferometer with an embedded quantum dot in the Kondo regime \cite{PhysRevB.67.165313}. The AB flux-dependent transmission probability exhibits an asymmetrical shape due to Fano interference between the direct tunneling path and the Kondo-resonant tunneling path through the quantum dot. The sign and magnitude of the thermopower were shown to be modulated by the AB flux and the direct tunneling amplitude. Additionally, the thermopower is anomalously enhanced by the Kondo correlation in the quantum dot near the Kondo temperature (\(T_K\)). The Kondo correlation in the quantum dot also induces crossover behavior in diagonal transport coefficients as a function of temperature. The amplitude of an AB oscillation in electric and thermal conductances is small at temperatures far above \(T_K\) but becomes enhanced as the system is cooled below \(T_K\). The AB oscillation is pronounced in the thermopower and Lorenz number within the crossover region near the Kondo temperature.

The thermoelectric effects of an AB interferometer with a quantum dot (QD) embedded in each of its arms are investigated, taking into account the intra-dot Coulomb interaction between the electrons in each QD \cite{liu2011role}. Using Green's function methods and the equation of motion (EOM) technique, it is found that the Seebeck coefficient and Lorenz number can be strongly enhanced when the chemical potential sweeps the molecular states associated with the Fano line-shapes in the transmission spectra, due to quantum interference effects between the bonding and antibonding molecular states.
Enhancement of the thermoelectric effects occurs between the two groups of conductance peaks in the presence of strong intra-dot Coulomb interaction, as a transmission node is developed in the Coulomb blockade regime. In this case, the maximum value of the Lorenz number approaches
$L = \frac{10 \pi^2 k_B^2}{3e^2}$.
The thermoelectric conversion efficiency in the absence of phonon thermal conductance, described by the figure of merit $ZT$, approaches 2 at room temperature.

The role of quantum coherence in thermoelectric transport properties of a rectangular AB ring at low temperature were investigated using a theoretical approach based on Green's functions \cite{Pye2016}. The oscillations in the transmission coefficient as the magnetic field was varied were used to tune the thermoelectric response of the ring. Large thermopowers were achievable, which, in conjunction with low conductance, resulted in a high thermoelectric figure of merit. The effects of single-site impurities and more general Anderson disorder were explicitly considered to evaluate their impact on the Fano-type resonances in the transmission coefficient. Importantly, it was shown that even with moderate levels of disorder, the thermoelectric figure of merit remained significant, enhancing the potential of such structures for thermoelectric applications.

A quantum heat engine based on an AB interferometer in a two-terminal geometry is proposed, and its thermoelectric performance is investigated in the linear response regime \cite{study1}. Significant thermopower (up to approximately 0.3 mV/K) as well as $ZT$ values greatly exceeding unity can be achieved by adjusting the parameters of the setup and the temperature bias across the interferometer. This results in thermal efficiency at maximum power approaching 30\% of the Carnot limit, which is near the optimal efficiency at maximum power for a two-terminal heat engine.
The operation of the quantum heat engine can be finely tuned by altering the magnetic flux, the asymmetry of the structure, a side-gate bias voltage through a capacitively coupled electrode, and the transmission of the T junctions connecting the AB ring to the contacts. Exploration of the parameter space shows that the high performance of the AB two-terminal device as a quantum heat engine is stable over a wide range of temperatures and length imbalances, indicating strong potential for experimental realization.

A theoretical study was conducted to investigate how nonlinear thermoelectricity could be controlled and enhanced by quantum coherent control in nanostructures such as a quantum dot system or a single-molecule junction \cite{study4}. This work developed an analytical approach to analyze the optimal thermoelectric working conditions. In nanostructures, the typical temperature scale is much smaller than the resonance width, which largely suppresses thermoelectric effects. However, it was demonstrated that reasonably good thermoelectric performance could be achieved by regulating quantum coherence. By employing a quantum-dot interferometer (a quantum dot embedded in the ring geometry) as a heat engine, the study explored the idea of thermoelectric enhancement induced by the Fano resonance.
 It was shown that even when the temperature is much smaller than the resonant width, which is typically unfavorable for good thermoelectric performance, reasonably good thermoelectric performance could still be achieved by regulating quantum coherence via the Fano effect. This thermoelectric enhancement is effective in fully nonlinear regimes with appropriate parameters. Efficiency improvements up to ten times and power nearly five times have been reported theoretically. The optimal gate voltage that maximized nonlinear efficiency or power was also estimated. The significance of the linear thermopower at the operating temperature was also assessed alongside nonlinear efficiency. Quantum control by the Fano effect was posited as a promising and universal approach that could enhance thermoelectric performance.

\begin{figure}[h] 
  \centering
  \includegraphics[width=0.7\textwidth]{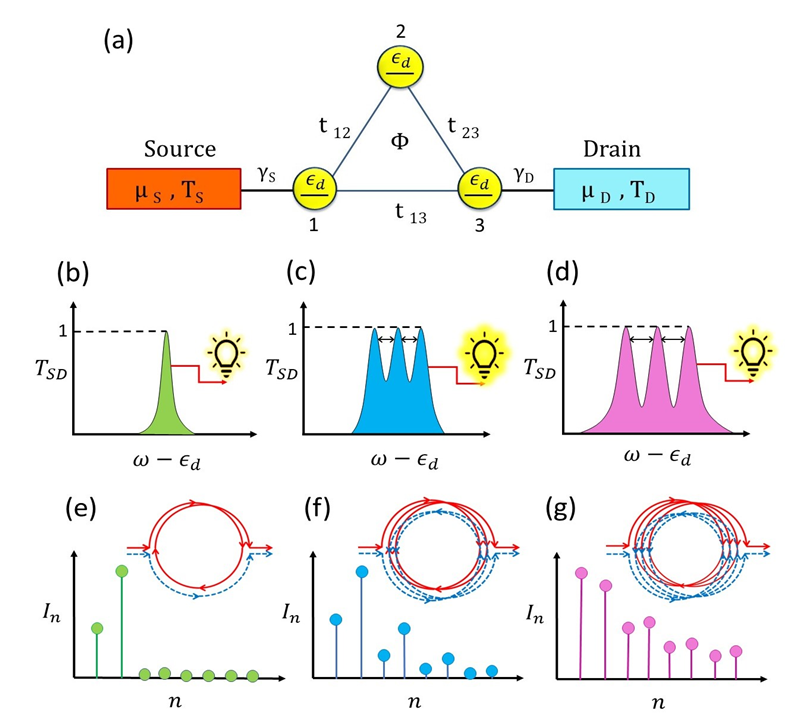}
  \caption{Reproduced from \cite{PhysRevB.108.165419}. (a) A triple quantum dot Aharonov-Bohm interferometer with dots 1 and 3 connected to the source and drain, respectively. A magnetic flux $\Phi$ pierces the triangular AB ring perpendicularly.
        (b)–(d) show transmission patterns with respect to energy for the regimes $t < \gamma$, $t = \gamma$, and $t > \gamma$, respectively.
        (b) A single symmetric peak around resonance is more efficient but generates less power ($t < \gamma$).
        (c) Three peaks equally spaced around the resonance are needed to achieve optimal power and efficiency, as shown in the figure with the brightest bulb ($t = \gamma$).
        (d) When the separation between the peaks increases, antiresonance becomes more pronounced, causing a decrease in power and efficiency ($t > \gamma$).
        The bottom set of figures shows harmonic patterns of charge current obtained by Fourier decomposition concerning $\phi$ ($n = 1, 2, 3, \ldots$). The different trajectories of electron circulation around the AB ring are represented by blue and red orbits.
        The presence of higher harmonic modes is necessary but not sufficient to achieve optimal power and efficiency.
        (e) shows only two dominant modes in the $t < \gamma$ regime, responsible for the least output power but the highest efficiency.
        (f) depicts optimal power efficiency in the $t = \gamma$ regime with a few higher harmonics.
        (g) shows that the dominance of higher harmonic modes is responsible for reducing both power output and efficiency in the $t > \gamma$ regime.}
  \label{fig:ring}
\end{figure}

The role of quantum coherence and higher harmonics resulting from multiple-path interference in nonlinear thermoelectricity in a two-terminal triangular triple-dot AB interferometer has been investigated recently \cite{PhysRevB.108.165419} as shown in Fig. \ref{fig:ring}.

\begin{figure}[t]
    \centering
    \includegraphics[width=7cm]{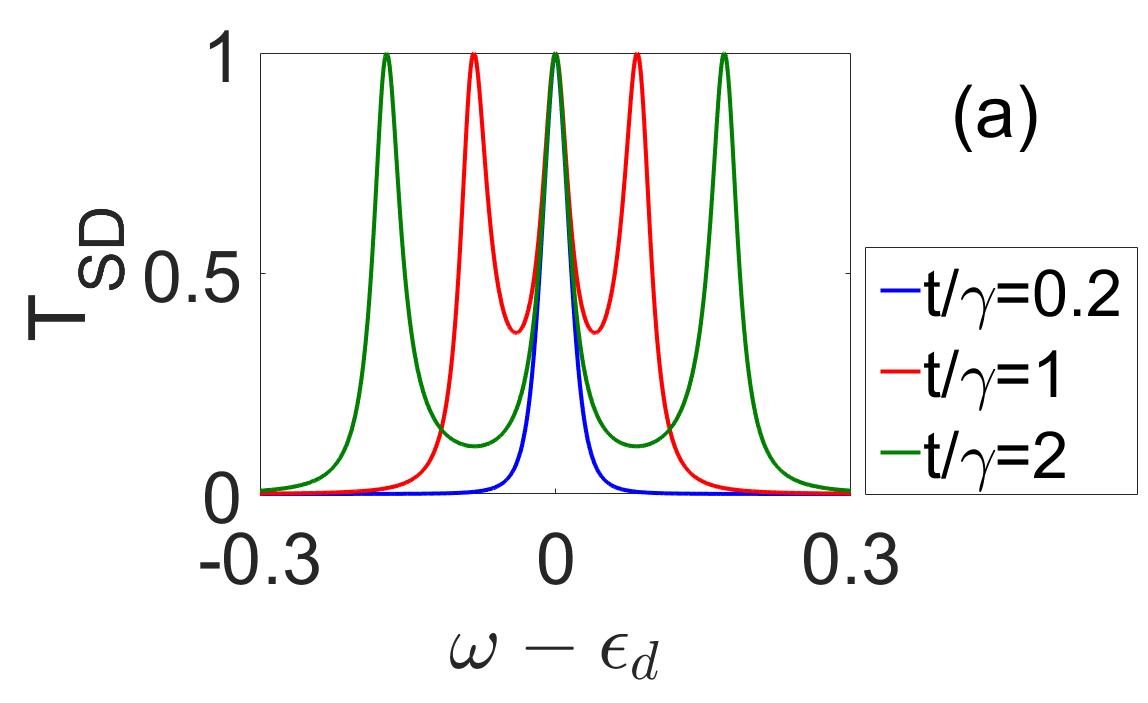}
    \includegraphics[width=7cm]{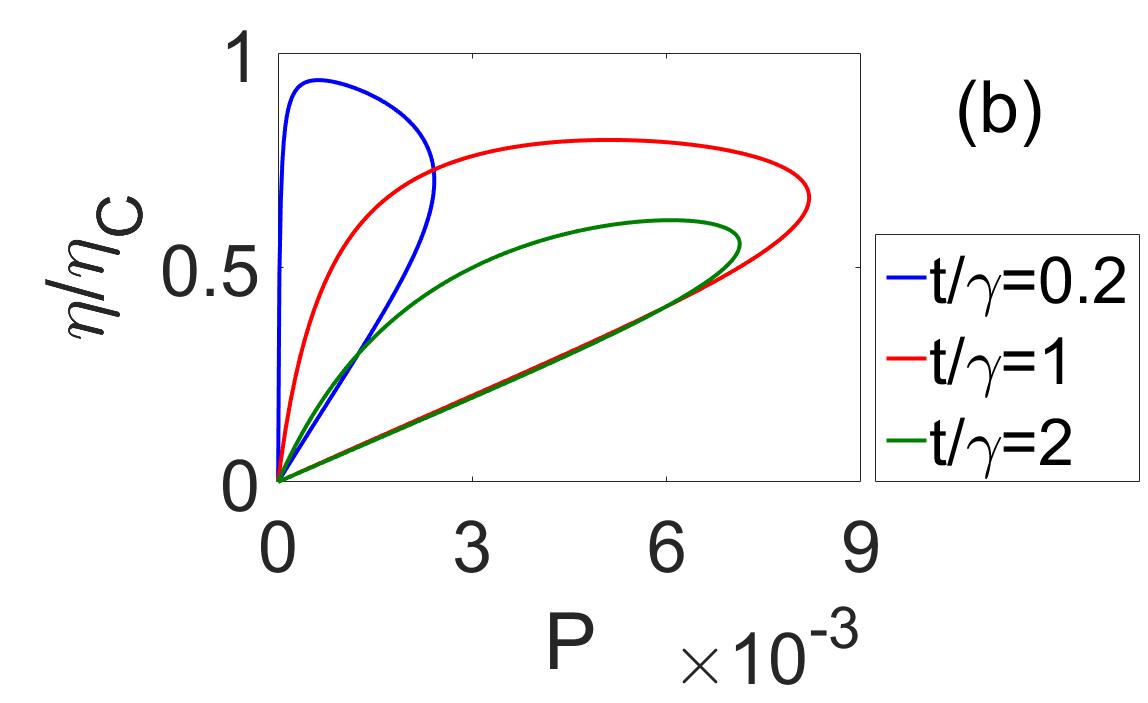}
    \caption{Reproduced from \cite{PhysRevB.108.165419}. (a) Transmission function and (b) Power-efficiency diagram for different values of tunneling strength $t$. Parameters used are $\gamma=0.05$, $\epsilon_d=8\gamma$, $\mu_S=-\mu_D$, $T_S=12\gamma$, $T_D=2\gamma$, $\phi=\pi/2$.}
    \label{power_effc_t}
\end{figure}

The trade-off between efficiency and power in the nonlinear regime was quantified for a setup of three non-interacting quantum dots arranged at the vertices of an equilateral triangle, with two dots connected to biased metallic reservoirs and a perpendicular magnetic flux. For a spatially symmetric configuration, optimal efficiency and power output were achieved when the inter-dot tunneling strength matched the dot-lead coupling, with an AB phase of $\phi = \pi/2$ as shown in Fig. \ref{power_effc_t}. For this particular configuration, maximal constructive quantum interference manifested as three closely spaced peaks as given by
\begin{equation}\label{trans}
    T_{SD}(\omega, \phi)=\frac{\gamma^2\big[t^4+2t^3\cos{\phi}(\omega-\epsilon_d)+t^2(\omega-\epsilon_d)^2\big]}{\Delta(\omega,\phi)},
\end{equation}
where
\begin{equation}
        \Delta(\omega,\phi)=\Big[(\omega-\epsilon_d)\Big((\omega-\epsilon_d)^2-3t^2-\frac{\gamma^2}{4}\Big)-2t^3\cos{\phi}\Big]^2
        +\gamma^2\Big[(\omega-\epsilon_d)^2-t^2\Big]^2.
\end{equation}
For $\phi=\pi/2$, the transmission peaks [$T_{SD}(\omega,\phi)=1$] are at positions
\begin{equation}\label{peaks}
    \omega=\epsilon_d, \hspace{0.5cm} \textrm{and} \quad \quad  \omega=\epsilon_d\pm\frac{1}{2}\sqrt{12t^2-\gamma^2}.
\end{equation}
For the above case, the optimal power-efficiency configuration was found to be at $t=\gamma$, $\phi=\pi/2$ for the mirror-symmetric set-up where the perfect constructive quantum interference occurs. In this regime, the time scale of internal coherent dynamics (as given by $1/t$) is equal to the tunneling rate from the triple-dot to the drain terminal.
The findings indicated that while higher harmonics are necessary, they are not sufficient alone to reach optimal power output. Notably, maximum constructive interference, indicated by three closely packed resonance peaks of unit transmission, boosted power output ($P_{\text{max}} \sim 2.35$ fW) nearly 3.5 times compared to scenarios where only a single channel contributed to transport (exhibiting Breit-Wigner peak only), with an efficiency of about $0.80\eta_c$, where $\eta_c$ is the Carnot efficiency. The effects of geometric asymmetries on efficiency and power output were also examined. An asymmetric setup, defined by the coupling ratio to the source and drain terminals ($x$), further increased the maximum power output to $P_{\text{max}} \sim 3.85$ fW for $x = 1.5$, maintaining the same efficiency as the symmetric case. The analysis concluded that output power and efficiency were optimal in the wide-band limit, with significant power reduction observed in the narrow-band case. Additionally, disorder effects drastically diminished the heat engine's performance.
We believe that making the gate voltage time-dependent could be an effective strategy to overcome the effects of the disorder. This problem can then be reduced to optimal quantum control of thermoelectric effects. This is a significant challenge because of the multidimensional nature of the problem. At this point, we would like to pose this as an open problem that requires urgent consideration.

\begin{figure}[t!]
\begin{center}
\begin{tikzpicture}

\draw[thick,->] (0,1) to[out=60, in=120] node[midway,above] {$t(E)$} (4,1);
\draw[thick,->] (0,1) to[out=-60, in=-120] node[midway,below] {$t(E)$} (4,1);

\draw[fill=red!30] (-1,0.5) rectangle (0,1.5) node[midway,rotate=90] {Hot};
\draw[fill=blue!30] (4,0.5) rectangle (5,1.5) node[midway,rotate=90] {Cold};

\node at (2,0.75) {$\phi$}; 

\end{tikzpicture}
\end{center}
\caption{Schematic of two path interferometer}
\label{fig:MZI}
\end{figure}
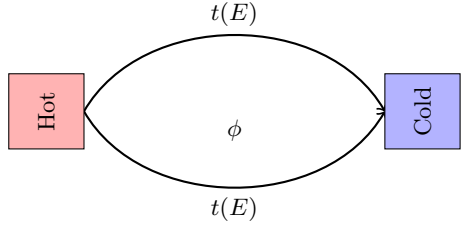

\begin{figure}[b!] 
  \centering
  \includegraphics[width=0.45\textwidth]{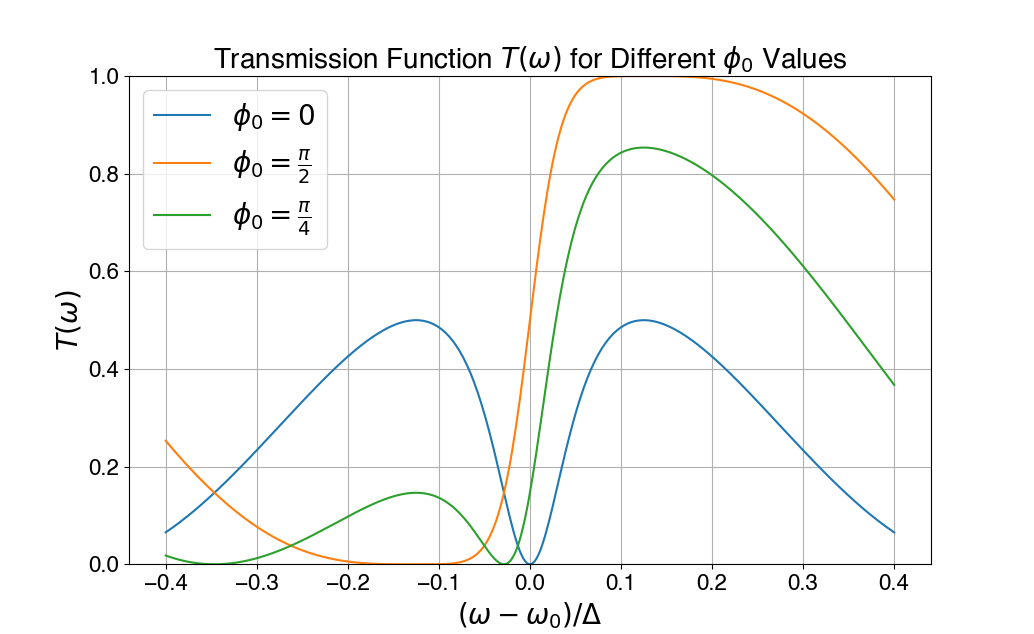}
  \caption{Reproduced from \cite{study3}. The transmission probability for two-path chiral edge thermoelectric interferometer.
  }
  \label{fig:transmission_2}
\end{figure}

A near-optimal heat engine implementation is proposed using a chiral edge state within an electronic Mach-Zehnder interferometer, featuring a mesoscopic capacitor coupled to one arm \cite{study3} as shown in Fig. \ref{fig:MZI}. Results show that the maximum power and corresponding efficiency can reach 90 percent and 83 percent, respectively, of the theoretical maximum. It was shown that the proposed heat engine is feasible with existing experimental techniques, and maintains robust performance against moderate dephasing. Quantum coherence is expected to play a fundamental role in the efficient heat-to-work conversion in nanoscale systems. An exemplary case is electronic interferometers leveraging chiral transport within quantum Hall edge channels, which can exhibit coherent oscillations with visibilities reaching up to 90 percent. A schematic of the two-terminal interferometer is depicted in Fig. \ref{fig:MZI}. An electron emitted from the hot reservoir (H) can travel via two distinct paths, $i = 1$ or $2$, to reach the cold reservoir (C). Each path includes a phase-coherent scatterer with transmission amplitude $t_i(\omega)$. The total amplitude to propagate from H to C is then $A(\omega) \propto t_1(\omega) + e^{i\phi_0} t_2(\omega)$, where the energy-independent path phase difference $\phi_0$ accounts for, for instance, an enclosed AB flux.
According to the Feynman path interference rule, the total transmission probability is given by  squared amplitude $|A(\omega)|^2$ is proportional to
\begin{equation}\label{eq:Aw2}
|A(\omega)|^2 \propto T_1 + T_2 + 2 \sqrt{T_1 T_2} \cos\left(\frac{\alpha_1(\omega) - \alpha_2(\omega) - \phi_0}{2}\right),
\end{equation}
where $T_1$ and $T_2$ are transmission coefficients, $\alpha_1(\omega)$ and $\alpha_2(\omega)$ are phases associated with paths 1 and 2, and $\phi_0$ represents an additional phase factor (e.g., AB flux). Given Eq. (\ref{eq:Aw2}), what is the best thermoelectric performance? This question has been recently addressed. It has been found that the optimal thermoelectric system is characterized by a step-like transmission probability in energy, where there is a rapid transition from zero to maximum transmission over a narrow energy range compared to the background temperature. This condition is satisfied for the probability $|A(\omega)|^2$ in Eq. (\ref{eq:Aw2}) if and only if:
\begin{enumerate}
    \item The total phase $\alpha_1(\omega) - \alpha_2(\omega) - \phi_0$ changes abruptly, as a function of energy, from 0 to $\pi$ (or $\pi$ to 0).
    \item The transmission probabilities of the two scattering channels are equal, $T_1 = T_2$.
\end{enumerate}
Here, $|A(\omega)|^2$ denotes the squared amplitude of the total amplitude $A(\omega)$, and $\alpha_1(\omega)$, $\alpha_2(\omega)$ are phases associated with paths 1 and 2, respectively. The phase $\phi_0$ represents an additional phase factor, such as Aharonov-Bohm flux.
To address how such an optimal interferometer could be experimentally realized,  a two-terminal electronic Mach-Zehnder interferometer implemented with edge states in a conductor within the integer quantum Hall regime was considered. The interferometer arms 1 and 2 have lengths \( L_1 \) and \( L_2 \), respectively. Arm 2 incorporated a mesoscopic capacitor, a small loop of length \( L \), connected to the edge via a quantum point contact with transparency \( 1 - \tau \). This capacitor effectively functions as a quantum dot, with a level spacing \( \Delta = \frac{2\pi\hbar v_D}{L} \), where \( v_D \) is the drift velocity of the edge state, coupled to the interferometer arm. Two additional quantum point contacts serve as beam splitters with transparencies \( \tau_A \) and \( \tau_B \), respectively. These transparencies \( \tau, \tau_A, \) and \( \tau_B \) are energy-independent and can be adjusted electrostatically between 0 and 1. For simplicity, it was assumed that the total path length difference is negligible, i.e., \( \frac{1}{2} \left( L_1 - (L_2 + L) \right) \frac{k_B T}{\hbar v_D} \ll 1 \), where \( T \) denotes the background temperature.
For this case, the transmission function is given by \cite{study3}
\begin{equation}
 T(\omega) = \frac{\left[ \sin\left( \frac{\phi_0}{2} \right) - \sqrt{\tau} \sin\left( \frac{\phi_0}{2} - \frac{2\pi (\omega - \omega_0)}{\Delta} \right) \right]^2}{1 - 2 \sqrt{\tau} \cos\left( \frac{2\pi (\omega - \omega_0)}{\Delta} \right) + \tau}.
\end{equation}
For $\phi_0=0$, the transmission probability is symmetric around the resonance, and owing to the particle-hole symmetry, the thermoelectric response vanishes. For $\phi_{0}=\pi/2$, the transmission probability is antisymmetric around the resonance, and then $\tau>0.5$, in the broad energy range, the transmission is approximately equal to $\Delta/2$ so that the system behaves as a low pass filter as shown in Fig. \ref{fig:transmission_2}.
The maximum power of the two-path thermoelectric interferometer was obtained using numerical optimization over the ranges \(0 \leq \tau \leq 1\), \(0 \leq \phi_0 \leq \pi\), \(\omega_0\), and \(\Delta\) for a given background temperature \(T\). Through this optimization, it was found that the maximum power is achieved for \(\tau = 0.61\) and \(\phi_0 = 0.52\pi\), along with \(\omega_0 = 1.17k_BT\) and \(\Delta = 24.2k_BT\). Under these specific parameters, we achieve \(P_{\text{max}} = 0.285 \frac{(k_B \Delta T)^2}{h}\), which corresponds to 90\% of the optimal value. The efficiency at maximum power is \(\eta_{\text{max}P} = 0.29 \eta_C\), which is 83\% of its upper bound.

Two terminal quantum spin Hall-based AB interferometers with Majorana bound states have been proposed as quantum thermoelectric heat engines and refrigerators \cite{mishra2023majorana}.
In the context of a two-terminal quantum spin Hall heat engine and refrigerator, Majorana bound states (MBSs), which arise from interacting Majorana fermions, can enhance thermoelectric performance.
 It has been theoretically shown that the coupling between individual MBS, as well as the coupling between MBSs and
the interferometer arms, can significantly affect performance. Tuning the AB flux can greatly improve performance and even outperform other contemporary quantum heat engines and refrigerators.
The above-mentioned works establish that designing a system with sharper transmissions with asymmetries or transmissions exhibiting maximal constructive quantum interference is crucial to obtaining optimal power-efficiency configurations of thermoelectric AB heat engines.

Recent studies have examined the interplay of AB flux and MBSs \cite{Mishra2023}. Dynamic properties of MBSs coupled with a double-quantum-dot (DQD) interferometer threaded with an AC magnetic flux have been investigated, leading to the derivation of time-averaged thermal current formulas. Photon-assisted local and nonlocal Andreev reflections contribute significantly to charge and heat transport. Modifications of source-drain electric conductance ($G$), electric-thermal conductance ($\xi$), thermal conductance ($\kappa_e$), Seebeck coefficient ($S_c$), and thermoelectric figure of merit ($ZT$) as functions of the AB phase have been numerically calculated. These coefficients exhibit a distinct shift in the oscillation period from $2\pi$ to $4\pi$ due to the attachment of MBSs. The applied AC flux enhances the magnitudes of $G$, $\xi$, and $\kappa_e$, with detailed enhancement behaviors depending on the energy levels of the DQD. While the coupling of MBSs leads to enhancements in $S_c$ and $ZT$, the application of AC flux suppresses the resonant oscillations. This investigation provides a clue for detecting MBSs by measuring the photon-assisted $S_c$ and $ZT$ oscillations versus the AB phase.

A model for a heat and temperature modulator using a voltage-biased mesoscopic AB interferometer has been proposed \cite{PhysRevResearch.6.013215}. In this model, the direction and amplitude of heat flow can be controlled via the AB or Rashba phases using electrical bias. This enables temperature variations of about 80 mK and heat flow of around 100 fW, with relative temperature changes up to 10 percent. The proposed setup allows for easy manipulation of heat flow and temperature using these phases and can produce spin-polarized heat flows without the need for ferromagnetic leads, making it useful for information processing in caloritronic systems.

The enhancement of linear and nonlinear thermoelectric performance in a nanoscale heat engine through structural modifications of a graphene rhombus dot has been analyzed \cite{Briones-Torres2021}. By evaluating phonon and electron transport, a nanostructure was identified that achieves high thermoelectric efficiency. Adjusting the junction bending angle, which can act as an AB phase, reduces phonon transport, while Fano-like resonances increase efficiency. A tunable local gate voltage on a double-bent graphene rhombus dot enhances efficiency and output power, especially at low temperatures (4 K). It was demonstrated that normalized linear-response plotting can predict nonlinear thermoelectric performance, and controlling quantum coherence is crucial for developing better nanoscale thermoelectric materials. An efficient thermoelectric device using a chiral organic molecule has been proposed recently \cite{PhysRevB.108.075407}. This design utilized the chiral-induced spin selectivity of these systems, particularly their strong spin-dependent transport properties. The analysis of the figure of merit and generated power for chiral-molecule-based heat engines showed that both could be significant, with the potential to exceed the performance of existing designs

In quantum-dot Aharonov-Bohm networks, the adjacency matrix \( A \) and the Hamiltonian \( H \) are key to understanding the system's properties \cite{tsuji2018quantum}. The adjacency matrix \( A \) is an \( N \times N \) matrix that represents the connectivity of \( N \) quantum dots, where each entry \( A_{ij} \) is 1 if there is a coupling between quantum dots \( i \) and \( j \), and 0 otherwise. This matrix captures the structural information of the network but does not account for coupling strengths or on-site energies.

The Hamiltonian \( H \) includes the adjacency matrix \( A \) and additional terms for coupling strengths and on-site energies. It is given by:
\begin{equation}
H = t A + V,
\end{equation}
where \( t \) is a scalar representing the coupling strength between adjacent quantum dots, and \( V \) is a diagonal matrix representing the on-site energies. The term \( t A \) accounts for the kinetic energy due to hopping between quantum dots, while \( V \) represents the local potential energy of each dot.

The Green's function \( G(\lambda) \) is defined as:
\begin{equation}
G(\lambda) = (\lambda I - H)^{-1}.
\end{equation}
Substituting \( H = t A + V \) into this expression, we obtain:
\begin{equation}
G(\lambda) = (\lambda I - t A - V)^{-1}.
\end{equation}
The power series expansion of \( G(\lambda) \) can be expressed as:
\begin{equation}
G(\lambda) = \sum_{n=0}^{\infty} \frac{A^n}{\lambda^{n+1}}.
\end{equation}
For \(\lambda = 0\), this simplifies to:
\begin{equation}
G(0) = -A^{-1}.
\end{equation}
The inverse matrix \( A^{-1} \) can be expressed as:
\begin{equation}
A^{-1} = \sum_{n=0}^{\infty} C_n A^n,
\end{equation}
where \( C_n \) are coefficients determined by the matrix \( A \). This expansion includes contributions from all possible walk lengths in the network.

To analyze contributions from different walk lengths, we consider the series expansion of \( G(\lambda) \). For even-length walks, we focus on the series with even powers of \( A \):
\begin{equation}
G_{\text{even}}(\lambda) = \sum_{n=0}^{\infty} \frac{A^{2n}}{\lambda^{2n+1}}.
\end{equation}
For \(\lambda = 0\), this becomes:
\begin{equation}
G_{\text{even}}(0) = -\left( \sum_{n=0}^{\infty} C_{2n} A^{2n} \right).
\end{equation}

The terms \( A^{2n} \) represent walks of length \( 2n \). If only even-length walks are present, the Green's function reflects these contributions.

For odd-length walks, we consider the series with odd powers of \( A \):
\begin{equation}
G_{\text{odd}}(\lambda) = \sum_{n=0}^{\infty} \frac{A^{2n+1}}{\lambda^{2n+2}}.
\end{equation}
For \(\lambda = 0\):
\begin{equation}
G_{\text{odd}}(0) = -\left( \sum_{n=0}^{\infty} C_{2n+1} A^{2n+1} \right).
\end{equation}

Here, the terms \( A^{2n+1} \) represent walks of length \( 2n+1 \). If only odd-length walks are present, Green's function captures these contributions.

When both even and odd-length walks are present, the Green's function includes contributions from all powers of \( A \):
\begin{equation}
G(\lambda) = \sum_{n=0}^{\infty} \frac{A^n}{\lambda^{n+1}}.
\end{equation}
For \(\lambda = 0\):
\begin{equation}
G(0) = -\left( \sum_{n=0}^{\infty} C_n A^n \right).
\end{equation}
This includes contributions from both even and odd-length walks:
\begin{equation}
G(0) = -\left( \sum_{n=0}^{\infty} C_{2n} A^{2n} + \sum_{n=0}^{\infty} C_{2n+1} A^{2n+1} \right).
\end{equation}

The presence of both even and odd-length contributions results in complex quantum interference effects. Optimizing thermoelectric effects in such networks may be achieved by designing graphs that support constructive interference, thereby enhancing specific walk lengths and improving overall transmission and performance.

\section{Non-linear transport and quantum devices under broken time-reversal symmetry}
 In a multiterminal interferometer magnetic field breaks time-reversal symmetry and can lead to nonreciprocal transport.
 For a three-terminal case the transmission probability $T_{L,R}(\omega, \phi)\neq T_{L,R}(\omega, -\phi)$. For inelastic scattering, this asymmetry in transmission gives rise to diode behavior.
  Even the symmetry of transmission probability is broken when a third terminal is introduced. This symmetry is also broken when many-body interactions are included. In these cases, $T_{LR}(\phi) \neq T_{LR}(-\phi)$. This transmission asymmetry can lead to diode behavior. It can also be exploited to enhance thermoelectric efficiency as discussed in the subsequent sections.
 \begin{figure}[h] 
  \centering
  \includegraphics[width=0.95\textwidth]{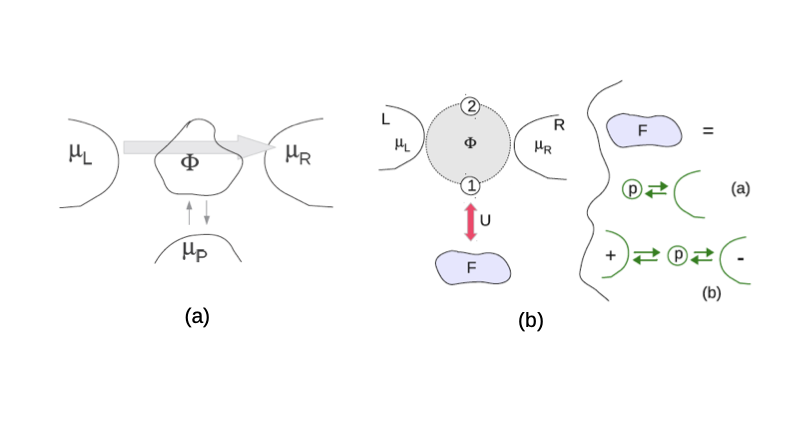}
  \caption{Reproduced from \cite{PhysRevB.90.235411_Bedkihal2014, Bedkihal2013, PhysRevB.88.155407}. (a) Scheme of Aharonov-Bohm diode using inelastic scattering. The horizontal arrow represents the
charge current I. The two parallel arrows represent (same magnitude) currents into and from the P terminal inducing elastic- and inelastic-scattering effects. The third terminal represents a probe reservoir.
(b) The diode behavior can also be observed when a
 Aharonov-Bohm interferometer is capacitively coupled to an equilibrium and the non-equilibrium fermionic
environment. The equilibrium case can be regarded as a microscopic model for a voltage probe. This environment consists of a quantum dot (labeled p)
itself hybridized with either (a) an equilibrium sea of noninteracting
electrons, or (b) two metals (±), possibly biased away from
equilibrium. In both cases dot 1 of the interferometer is coupled
capacitively (strength U) to dot p in the fermionic environment}
  \label{fig:Three_terminal}
\end{figure}
The Onsager-Casimir symmetry relations dictated reciprocal relations between linear response coefficients, \( L_{i,j} = L_{j,i} \). In the presence of a magnetic field \( \textbf{B} \), the reciprocity relation became \( L_{i,j}(\textbf{B}) = L_{j,i}(-\textbf{B}) \). This implied that the conductances (diagonal matrix elements) are even functions of the magnetic field. In a two-terminal Aharonov-Bohm interferometer, this symmetry is known as the “phase rigidity” of linear conductance \cite{yacoby1995}.

In the non-linear regime, Onsager-Casimir symmetries did not necessarily hold. A prominent example of this breakdown was the asymmetry of the differential conductance out-of-equilibrium. This effect was attributed to electron-electron interactions in the system, resulting in an asymmetric charge response under the reversal of a magnetic field, leading to a magnetoasymmetric differential conductance. Such interaction-induced asymmetry was observed recently in carbon nanotubes and semiconductor quantum dots \cite{wei2005phys, leturcq2006phys, leturcq2006physica, angers2007phys, sanchez2008phys, sanchez2013phys, hwang2013njp}.
Many-body interactions induced different types of phase-breaking processes in coherent transport. These included quasi-elastic scattering and inelastic scattering. Büttiker’s probes served as phenomenological tools to incorporate quasi-elastic and inelastic scattering effects. A systematic study of how many-body effects and spatial asymmetries affected magnetic field symmetries and magnetoasymmetries of charge and heat current beyond linear response was conducted.

Nonlinear transport measurements were performed recently on AB rings connected to two leads by Leturcq {\it et al.}, who reported that the even (odd) conductance terms were asymmetric (symmetric) in the magnetic field. It was also argued that these observations were insensitive to geometric asymmetries in the ring. Angers {\it et al.} also performed nonlinear transport measurements on GaAs/GaAlAs rings in a two-terminal configuration, reporting an antisymmetric second-order response coefficient under the reversal of the magnetic field, attributed to electron-electron interactions. Notably, in Leturcq \textit{et al.} \cite{leturcq2006phys} work, no particular symmetry of even coefficients was reported, whereas Angers \textit{et al.} \cite{angers2007phys} reported them to be antisymmetric under magnetic field reversal.

Magnetic field asymmetries of transport in mesoscopic conductors coupled to an environment were theoretically studied by Kang \textit{et al.} \cite{sanchez2008phys}. The model system used was a quantum dot conductor coupled to another conductor (treated as an environment) via Coulomb interaction, allowing energy exchange without particle exchange. The environment was driven out of equilibrium by applying a voltage bias. It was found that the interaction between the conductor and the environment caused magnetoasymmetry even in the linear regime if the environment was maintained out-of-equilibrium.

The necessary and sufficient conditions for diode behavior in a three-terminal AB interferometer were successfully obtained using B\"{u}ttiker probes beyond linear response regime \cite{Bedkihal2013, PhysRevB.90.235411_Bedkihal2014}. This study systematically examined four different probes (dephasing, voltage, temperature, and voltage-temperature), with numerical simulations establishing stable solutions for probe parameters under far-from-equilibrium conditions.
A comprehensive analysis was conducted on a generic model of the AB interferometer, incorporating elastic and inelastic effects as shown in Fig. \ref{fig:Three_terminal}:
\begin{itemize}
    \item The study rigorously demonstrated phase rigidity under voltage probes in the linear response regime.
    \item New sets of magnetic-field symmetry relations beyond linear response were derived specifically for symmetric AB interferometers.
    \item Spatial asymmetries were found to invalidate certain magnetic-field symmetries, although the study showed that generalized magnetic field-gate voltage symmetry relations, attributable to particle-hole symmetry, still hold.
    \item The necessary and sufficient conditions for diode behavior in multi-terminal AB interferometer were obtained using probe formalism.
    \item The operational characteristics of a double-dot interferometer, subject to inelastic effects, were explored, revealing its behavior as a charge rectifier under conditions of broken time-reversal symmetry.
\end{itemize}
Here we briefly summarise these results and discuss possible technological applications. For the sake of completeness, we briefly review B\"{u}ttiker probes \cite{Buttiker1986, jacquet2009stat, jacquet2012phys, damato1990phys, bergfield2013acs, saito2011}. These are phenomenological tools to incorporate the effects of many-body interactions and decoherence into otherwise coherent transport.\\
For a three-terminal AB interferometer as shown in Fig. \ref{fig:Three_terminal}, the particle currents from the left terminal ($I_L$) and the probe ($I_P$) can be expressed as
\begin{equation}\label{eq:IL}
    I_L(\phi) = \int_{-\infty}^{\infty} d\omega \bigg[
        T_{L,R}(\omega, \phi) f_L(\omega)
        - T_{R,L}(\omega, \phi) f_R(\omega)
        + T_{L,P}(\omega, \phi) f_L(\omega)
        - T_{P,L}(\omega, \phi) f_P(\omega)
    \bigg],
\end{equation}
\begin{equation}\label{eq:IP}
    I_P(\phi) = \int_{-\infty}^{\infty} d\omega \bigg[
        T_{P,L}(\omega, \phi) f_P(\omega)
        - T_{L,P}(\omega, \phi) f_L(\omega)
        + T_{P,R}(\omega, \phi) f_P(\omega)
        - T_{R,P}(\omega, \phi) f_R(\omega)
    \bigg].
\end{equation}
Similarly, we can write the heat currents from the left terminal ($Q_L$) and the probe ($Q_P$) as
\begin{equation}\label{eq:QL}
    Q_L(\phi) = \int_{-\infty}^{\infty} d\omega (\omega - \mu_L) \bigg[
        T_{L,R}(\omega, \phi) f_L(\omega)
        - T_{R,L}(\omega, \phi) f_R(\omega)
        + T_{L,P}(\omega, \phi) f_L(\omega)
        - T_{P,L}(\omega, \phi) f_P(\omega)
    \bigg],
\end{equation}
\begin{equation}\label{eq:QP}
    Q_P(\phi) = \int_{-\infty}^{\infty} d\omega (\omega - \mu_P) \bigg[
        T_{P,L}(\omega, \phi) f_P(\omega)
        - T_{L,P}(\omega, \phi) f_L(\omega)
        + T_{P,R}(\omega, \phi) f_P(\omega)
        - T_{R,P}(\omega, \phi) f_R(\omega)
    \bigg].
\end{equation}
Elastic dephasing effects can be incorporated with the dephasing probe. A particle entering the probe is incoherently re-emitted within a small energy interval $[\omega, \omega + d\omega]$ such that $d\omega \ll T\nu, \Delta\mu$ \cite{forster2007new}. It loses phase memory, but the change in energy is much smaller than the voltage bias and the temperature \cite{forster2007new}. Since scattering processes in each energy interval are independent, the distribution function of electrons in the probe reservoir is nonequilibrium. Elastic dephasing effects can thus be implemented by demanding that the energy-resolved particle current vanishes in the probe,
\begin{equation}
I_P(\omega) = 0 \quad \text{with} \quad I_P = \int I_P(\omega) d\omega.
\end{equation}
Using this condition, Eq. (\ref{eq:IP}) provides a closed form for the corresponding (flux-dependent) probe distribution, not necessarily in the form of a Fermi function as
\begin{equation}\label{eq:fP}
f_P(\phi) = \frac{T_{L,P} f_L + T_{R,P} f_R}{T_{P,L} + T_{P,R}},
\end{equation}
Dissipative inelastic effects can be introduced into the conductor using the voltage probe technique. The three reservoirs (L, R, P) are maintained at the same inverse temperature $\beta_a$, but the L and R chemical potentials are made distinct, $\mu_L \neq \mu_R$. In this case, the chemical potential $\mu_P$ of the probe P is evaluated by demanding that the net-total particle current flowing into the P reservoir diminishes \cite{jacquet2009stat,forster2007new, Bedkihal2013},
\begin{equation}\label{eq:IP0}
    I_P = 0.
\end{equation}
This choice allows dissipative energy exchange processes to occur within the probe. In the linear response regime, Eq. (\ref{eq:IP}) can be used to derive an analytic expression for $\mu_P$,
\begin{equation}
\mu_P(\phi) = \frac{\Delta\mu}{2} \frac{\int d\omega \frac{\partial f_a}{\partial \omega} (T_{L,P} - T_{R,P})}{\int d\omega \frac{\partial f_a}{\partial \omega} (T_{P,L} + T_{P,R})}.
\end{equation}
In the above equation, the derivative of the Fermi function is evaluated at equilibrium. In far-from-equilibrium situations, we obtain the unique \cite{jacquet2012phys, Bedkihal2013} chemical potential of the probe numerically by using the Newton-Raphson method as below,
\begin{equation}
\mu_P^{(k+1)} = \mu_P^{(k)} - I_P(\mu_P^{(k)}) \left[\frac{\partial I_P(\mu_P^{(k)})}{\partial \mu_P}\right]^{-1}.
\end{equation}
The current $I_P(\mu_P^{(k)})$ and its derivative are evaluated from Eq. (\ref{eq:IP}) using the probe (Fermi) distribution with $\mu_P^{(k)}$. Note that the self-consistent probe solution varies with the magnetic flux.

For a temperature probe, the three reservoirs L, R, and P are maintained at the same chemical potential $\mu_a$, but the temperatures at the L and R terminals are different, $T_L \neq T_R$. The probe temperature $T_P = \beta_P^{-1}$ is determined by requiring the net heat current at the probe to satisfy
\begin{equation}\label{eq:QP0}
    Q_P = 0.
\end{equation}
This constraint allows for charge leakage into the probe since we do not require Eq. (\ref{eq:IP0}) to hold. We can obtain the temperature $T_P$ numerically by following an iterative procedure,
\begin{equation}
T_P^{(k+1)} = T_P^{(k)} - \frac{Q_P(T_P^{(k)})}{\left[\frac{\partial Q_P(T_P^{(k)})}{\partial T_P}\right]}.
\end{equation}
Here, the probe temperature varies with the applied flux $\phi$.

\begin{figure}[h] 
  \centering
  \includegraphics[width=0.85\textwidth]{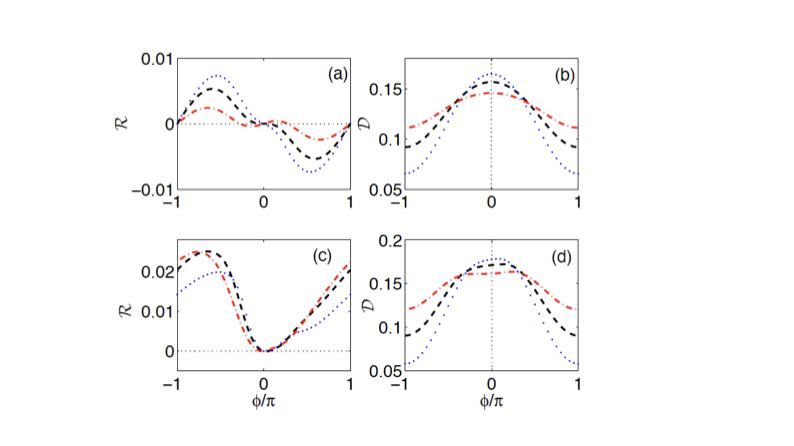}
  \caption{Reproduced from \cite{PhysRevB.90.235411_Bedkihal2014, Bedkihal2013, PhysRevB.88.155407}.Diode behavior in the presence of Buttiker voltage probe that induces inelastic scattering. Even (R) and odd
(D) conductance terms. (a)-(b) Spatially symmetric system, $\gamma_{L} =\gamma_{R}  = 0.05$. (c)-(d)
Spatially asymmetric junction, $\gamma_{L}= 0.05$  $\gamma_{R} = 0.2$ and $\gamma_{P} = 0.1$ (dot), $\gamma_{P} = 0.2$ (dashed
line) and $\gamma_{P} = 0.4$ (dashed-dotted). Light dotted lines represent symmetry lines. Other
parameters are $\Delta\mu = 0.4$, $\epsilon_{1} = \epsilon_{2} = 0.15$, $\beta_{a} = 50$. We can see that the symmetry of the odd conductance terms is weakly broken in both cases implying that the linear part of the conductance term dominates over higher order odd terms.}
  \label{fig:Diode}
\end{figure}

The voltage-temperature probe acts as an electron thermometer at weak coupling. This is achieved by setting the temperatures $T_L, T_R$ and the potentials $\mu_L, \mu_R$, then demanding that both
\begin{equation}\label{eq:IP0QP0}
    I_P = 0, \quad Q_P = 0.
\end{equation}
In other words, the charge and heat currents in the conductor satisfy $I_L = -I_R$ and $Q_L = -Q_R$, since neither charges nor heat are allowed to leak to the probe. Analytic results can be obtained in the linear response regime, see Refs. \cite{jacquet2009stat, PhysRevB.75.195110}. Beyond that, Eq. (\ref{eq:IP0QP0}) can be solved self-consistently, to provide $T_P$ and $\mu_P$. This can be done by using the two-dimensional Newton-Raphson method,
\begin{align}
\mu_P^{(k+1)} &= \mu_P^{(k)} - D^{-1}_{1,1} I_P(\mu_P^{(k)}, T_P^{(k)}) - D^{-1}_{1,2} Q_P(\mu_P^{(k)}, T_P^{(k)}), \\
T_P^{(k+1)} &= T_P^{(k)} - D^{-1}_{2,1} I_P(\mu_P^{(k)}, T_P^{(k)}) - D^{-1}_{2,2} Q_P(\mu_P^{(k)}, T_P^{(k)}),
\end{align}
where the Jacobian $D$ is re-evaluated at every iteration,
\begin{equation}
D(\mu_P, T_P) \equiv
\begin{pmatrix}
\frac{\partial I_P(\mu_P, T_P)}{\partial \mu_P} & \frac{\partial I_P(\mu_P, T_P)}{\partial T_P} \\
\frac{\partial Q_P(\mu_P, T_P)}{\partial \mu_P} & \frac{\partial Q_P(\mu_P, T_P)}{\partial T_P}
\end{pmatrix}.
\end{equation}
We emphasize that besides the case of the dephasing probe, the function $f_P(\phi)$ is forced to take the form of a Fermi-Dirac distribution function in the other probe models.
The following measures were used to quantify the magneto-asymmetries.  We now define several measures for quantifying phase symmetry in a voltage-biased three-terminal junction satisfying charge conservation $I(\phi)=I_L(\phi)=-I_R(\phi)$. Expanding the charge current in powers of the bias $\Delta\mu$ we write \cite{Bedkihal2013}
\begin{equation}
I(\phi) = G_1(\phi)\Delta\mu + G_2(\phi)(\Delta\mu)^2 + G_3(\phi)(\Delta\mu)^3 + \ldots,
\end{equation}
with $G_{n>1}$ as the nonlinear conductance coefficients. In this work, relations between two quantities are studied: a measure of the magnetic field asymmetry
\begin{equation}
\Delta I(\phi) \equiv \frac{1}{2} [I(\phi) - I(-\phi)],
\end{equation}
and the dc-rectification current,
\begin{equation}
R(\phi) \equiv \frac{1}{2} [I(\phi) + \bar{I}(\phi)] = G_2(\phi)(\Delta\mu)^2 + G_4(\phi)(\Delta\mu)^4 + \ldots,
\end{equation}
with $\bar{I}$ defined as the current obtained upon interchanging the chemical potentials of the two terminals. The behavior of odd conductance terms is also studied,
\begin{equation}
D(\phi) \equiv \frac{1}{2} [I(\phi) - \bar{I}(\phi)] = G_1(\phi)\Delta\mu + G_3(\phi)(\Delta\mu)^3 + \ldots.
\end{equation}
Notice that the rectification current $\mathcal{R}$ contains the sum of all even conductance terms, while $\mathcal{D}$ includes the sum of all odd conductance terms. For a two-terminal non-interacting system, $I(\phi) = I(-\phi)$, indicating that the current is always a symmetric function of the AB phase. Consequently, it can be shown that $\mathcal{R} = 0$. In other words, the non-interacting AB network cannot function as a diode due to the absence of even conductance terms.
In a three-terminal system within the linear response regime, it has been demonstrated that the linear conductance term $G_{1}(\phi) = G_{1}(-\phi)$ and $\mathcal{R}(\phi) = 0$, implying the absence of diode behavior in the linear response regime. This holds regardless of electron-electron and electron-phonon interactions. This result can be established by considering a specific scattering mechanism modeled using the phenomenological probe methods described above.
Under elastic dephasing effects, it was proven that
\begin{equation}
    \Delta{I} = 0,
\end{equation} and
\begin{equation}
    \mathcal{R} = 0,
\end{equation}
demonstrating that the current exhibits even symmetry with respect to the magnetic field.
Furthermore, inelastic effects were incorporated beyond linear response, and it was shown that in geometrically symmetric junctions,
\begin{align}
    \Delta I(\phi) &= \mathcal{R}(\phi), \\
    \mathcal{R}(\phi) &= -\mathcal{R}(-\phi), \\
    \mathcal{D}(\phi) &= \mathcal{D}(-\phi).
\end{align}
We can see that many-body interactions included in the form of inelastic scattering can lead to a diode behavior. The even conductance terms are anti-symmetric under the AB phase and the odd conductance terms are symmetric under the AB phase. The presence of even conductance terms can be attributed to many-body effects which modify the charge response of the system beyond linear regime thereby giving rise to diode behavior for $\phi\neq0$. Within B\"{u}ttiker probe formalism it can be shown that the above conditions hold for particle-hole symmetric case even if the set-up is geometrically asymmetric (left-right asymmetry). The anti-symmetry of the even conductance coefficient is strongly broken when the system is spatially asymmetric. However, the symmetry of $\mathcal{D}(\phi)$ is only weakly broken implying that the dominant contribution to this term comes from the linear term $G_{1}(\phi)$ compared to the higher order (odd) conductance terms. The results of B\"{u}ttiker probe methods were further compared to numerically exact path integral methods and it was found that in both cases the asymmetry
$\mathcal{D(\phi)}-\mathcal{D(-\phi)}\approx 1.5\times10^{-4}$. To conclude, AB networks can act as a diode when inelastic scattering effects are included beyond the linear regime and the time-reversal symmetry is broken i.e when $T_{LR}(\phi)\neq T_{LR}(-\phi)$.
Creating a pronounced transmission asymmetry can enhance diode effects. Interestingly, this diode behavior is not purely incoherent. Strong coupling to a Büttiker probe can suppress these effects, indicating that quantum interference is crucial for observing the diode behavior as shown in Fig. \ref{fig:Diode}.

The transition from quantum to classical physics is fundamentally characterized by the decoherence of quantum states.
Recent theoretical investigations have demonstrated that the transition regime facilitates a novel form of matter transport leading to a diode behavior \cite{PhysRevB.104.115413}. Nanoscale devices exhibit decoherence, modeled by random quantum jumps, which disrupts the unitary evolution of electron wave packets, leading to distinctive phenomena.
Noncentrosymmetric conductors, configured as AB rings at mesoscopic scales, function as two-terminal rectifiers with unique properties. The inelastic interactions between itinerant electrons and impurities, acting as electron trapping centers, induce a novel steady state characterized by partial charge separation between the leads or, in closed circuits, the emergence of persistent currents.
The interface between quantum and classical domains thus unveils a unique transport regime with significant implications for the advancement of mesoscopic electronic devices. This work complements prior theoretical investigations into diode behavior using Büttiker probes \cite{Bedkihal2013}.

Theoretical investigations have examined the behavior of a ballistic AB ring embedded in a normal metal (N) and superconducting (S) two-terminal setup. Utilizing current technologies, this device demonstrates potential as a hybrid quantum thermal device, functioning as both a quantum heat engine and a quantum thermal rectifier \cite{Blasi2023}. The interplay between single-particle quantum interferences within the AB ring and the superconducting properties enables its hybrid operation. The device achieves an efficiency of 55 percent of the Carnot limit as a quantum heat engine, and a thermal rectification factor reaching 350 percent. These findings suggest significant promise for future phase-coherent caloritronic nanodevices.

In recent studies, the thermoelectric transport properties of a ring structure influenced by an AB flux have been investigated. This setup includes a molecular bridge that interacts with both charge carriers and vibrational modes of the molecule \cite{entinwohlman2012, PhysRevResearch.6.013215}. The system is connected to three terminals: two electronic reservoirs providing charge carriers and a phonon bath that thermalizes molecular vibrations. The analysis focuses on deriving transport coefficients to second order in electron-vibration coupling. These coefficients establish relationships between temperature and chemical potential differences across the terminals and various charge and heat currents. At linear response, the coefficients adhere to the Onsager-Casimir relations. Notably, when the phonon bath is maintained at a temperature distinct from the electronic reservoirs, this setup allows for the conversion of heat currents between molecular vibrations and charge carriers into electric and/or additional heat currents. The study highlights how these transport coefficients, governed by electron-vibration coupling, exhibit changes in sign under interchange of electronic terminals and variations in the AB flux, indicating potential enhancements in thermoelectric conversion processes.

Double quantum dots have been studied as a practical realization of the two-impurity Kondo model, which features a non-Fermi-liquid quantum critical point (QCP) at specific parameter values \cite{PhysRevB.79.125110}. In a recent study \cite{PhysRevB.79.125110}, the investigation was generalized to a parallel configuration incorporating an AB flux. An exact universal result for the conductance $G_{V,T}$ at finite temperature and voltage was derived, covering the crossover from the QCP to the low-energy Fermi-liquid phase. Unlike the series configuration, the parallel setup typically exhibits an asymmetry $G_{V, T} \neq G_{-V, T}$, leading to current rectification arising from the interplay of the AB phase and the many-body interactions.

In the context of electron wave-packet rectification along quasi-one-dimensional chains, traditional methods involve nonlinear and longitudinal asymmetries confined within segments termed wave diodes \cite{li2014scientific}. However, recent investigations explore an alternative approach where spatial asymmetry perpendicular to the propagation direction, facilitated by an external magnetic field, can achieve rectification. This concept is exemplified by a nonlinear ring-shaped lattice divided into upper and lower halves (diode), coupled to elastic chains (leads). Despite mirror symmetry along the ring's vertical axis, asymmetry persists in the chain direction. Wave propagation along these diode paths is effectively modeled using a discrete Schrödinger equation with cubic nonlinearities. Notably, numerical simulations highlight the role of the AB effect in tuning the magnetic flux through the ring to control diode operation effectively.

Recent theoretical investigations have explored the transport signatures of quantum interference in highly symmetric double quantum dots arranged in a parallel geometry \cite{PhysRevB.99.125406}. It was demonstrated that even minor symmetry-breaking effects can significantly impact the current. Using a master equation approach, where quantum interference appears as nondiagonal elements in the density matrix of the double quantum dots, the study reveals insights into these phenomena.
By reformulating the equations as Bloch-like equations for a pseudospin associated with dot occupation, many results gain a physically intuitive understanding. In a perfectly symmetric setup with equal tunnel couplings and orbital energies, the system lacks a unique stationary-state density matrix. However, the introduction of minimal symmetry-breaking terms to the tunnel couplings or orbital energies stabilizes a stationary state, either with or without quantum interference, depending on the balance between these perturbations which may also arise by introducing AB flux.
These different states can correspond to significantly varied current levels. Therefore, controlling orbital energies or tunnel couplings, for example, through electrostatic gating, allows the double quantum dot to function as a highly sensitive electric switch. It is an outstanding problem to generalize this for heat current. It is further instructive to examine the effects of phonon modes on thermal switch operation. It is further instructive to generalize this work to include AB flux. The presence of multiple steady-states in the above system opens up novel possibilities of thermoelectric transport which remains largely unexplored.

Unlike electric currents, the control of heat flows presents significant challenges. Recent studies have shown that in mesoscopic conductors, manipulation of electronic thermal currents using a magnetic field through the AB effect is feasible \cite{balduque2023arxiv}. This method enhances the thermoelectric effect by regulating interference patterns and potentially allows for complete suppression of heat transport. In a three-terminal configuration, the asymmetry induced by the magnetic flux generates a non-local thermoelectric response, facilitating directed heat circulation. This capability opens avenues for developing efficient thermoelectric generators, thermal switches, thermal circulators, and nano-scale energy harvesters, offering effective thermal management with minimal disruption.
Recent advancements in manipulating thermal and thermoelectric effects in mesoscopic conductors using the AB effect and magnetic fields showed significant progress.  Additionally, the induction of flux-dependent destructive interference via the AB effect effectively suppressed both charge and heat currents, enabling efficient heat switches. Furthermore, the utilization of broken time-reversal symmetry in multiterminal setups resulted in asymmetric transport coefficients, facilitating practical applications such as thermal circulators and energy harvesters.
These studies examined the role of dephasing in two-terminal configurations and analyzed nonreciprocal transport in three-terminal setups. They explored the design and operational principles of thermal devices like thermal switches and circulators, particularly in all-thermal configurations where only heat was injected and controlled through magnetic flux. Comparative analyses with electrothermal configurations, involving charge flow, provided valuable insights and simpler analytical treatments under specific conditions.
Theoretical models, particularly those involving single-channel rings connected to two or three terminals, revealed complex interference patterns influenced by kinetic and magnetic phases. These patterns included resonances from circulating currents and regions of tunable destructive interference modulated by external factors such as gates and magnetic fields, interpreted through junction spectral properties and symmetries in transmission probabilities.
Practically, these systems demonstrated promising potential as thermoelectric generators, achieving high efficiency or power output under specific resonance conditions. Moreover, magnetic thermal switches exhibited robust efficiency across varying levels of dephasing.
Introducing a three-terminal geometry enhanced functional capabilities, particularly with heat injection from a third terminal, explored in both all-thermal and electro-thermal configurations. The spatial separation of cold terminal junctions introduced additional phases that prevented thermal switch effects across potential rings. However, efficient switching capabilities were achievable in intermediately coupled rings, leveraging nonreciprocal transport coefficients induced by magnetic fields to define efficient thermal circulators, especially in strongly coupled regimes.
Overall, these studies highlighted the role of kinetic and magnetic phases in breaking essential symmetries for advancing nonlocal thermoelectric engines. They emphasized the enhanced system performance due to the AB effect compared to systems reliant solely on kinetic phases, proposing optimal configurations grounded in reciprocity relations.
Considering both symmetric and asymmetric configurations, these findings suggested further optimization through junction distance asymmetries and the incorporation of energy-dependent scatterers, promising avenues for advancing nanoscale thermal management technologies.


\section{Thermoelectric Heat Engine: Bounds on Power and Efficiency}
In this section, we will discuss the thermoelectric devices that integrate heat and particle currents guided by local
gradients in temperature and chemical potential to generate electrical power or cooling \cite{ref1, ref2}. Compared to their cyclic counterparts, the advantage of such machines operating under periodic compression and expansion of a working fluid is that they work without moving parts. However, they have prevented wide-ranging applicability because of their low efficiency. However, it has been proved that utilizing proper energy filtering may lead to highly efficient thermoelectric heat engines \cite{Mahan1996}. They may even reach close to  Carnot efficiency \cite{ref4, ref5}. Thus, the challenge of finding better thermoelectric materials has attracted much scientific interest in recent decades.\\
\indent Recently, a new method to enhance the performance of thermoelectric engines is discovered by Benenti {\it et al.} \cite{ref6}. Utilizing a phenomenological framework of linear irreversible thermodynamics, their analysis within the phenomenological framework reveals that a magnetic field, which breaks time-reversal symmetry, could enhance thermoelectric efficiency significantly. It is even possible to construct devices that work close to Carnot efficiency while delivering finite power output. This is a promising observation that gives rise to the question of whether this possibility can be perceived in specific microscopic models.\\
\indent
Let us first introduce the concept of the phenomenological but powerful framework of irreversible thermodynamics as a toolbox for describing thermoelectric heat engines. In this methodology, the underlying thermoelectric transport processes in the nonequilibrium steady state can be expressed with a universal thermodynamic structure built from currents and affinities corresponding to temperature and chemical potential gradients. Thus the total rate of entropy production becomes a bilinear form in these quantities, even in the non-linear regime. Now, considering small gradients compared to their respective reference values, the currents can be linearized concerning the affinities. Two fundamental principles contrived the kinetic coefficients arising in this expansion. First, the second law constrains
the total entropy production rate, which is in quadratic form in the affinities through the linearization of the currents, to be non-negative. On the other hand, the microscopic reversibility guides to a reciprocal relation that provides the off-diagonal kinetic coefficients identical to time-reversal symmetric systems. This symmetry is, however, typically broken in the presence of a magnetic field. In this situation, the second law is effectively the only constraint.\\
\indent
With this general overview, we first discuss time-reversal symmetric thermoelectric engines in linear response limits. This is usually observed that the three possibly most important figures in the present context are maximum efficiency, efficiency at maximum power, and power at a given efficiency. They can usually be expressed in terms of a single dimensionless parameter, the thermoelectric figure of merit, function of the kinetic coefficients. Further, one may demonstrate an efficiency-dependent bound on power by utilizing the reciprocal relations.\\
\indent
Next, we move to investigate the systems with broken time-reversal symmetry. Following the analysis of Benenti \textit{et al.} we show how to express the previously introduced key figures in terms of two dimensionless parameters. The first one is just in terms of the generalized thermoelectric figure of merit, and the second one measures the magnetic-field-induced asymmetry between the off-diagonal kinetic coefficients. It is observed that one may recover the reversible transport process if this parameter deviates from its symmetric value $1$. It can be demonstrated that this interesting phenomenon can be related to reversible currents, which do not contribute to the total entropy production rate and they are unconstrained by the second law.\\
\indent
A promising platform for our investigations is provided by thermoelectric nano-devices which follow usual constraints on the kinetic coefficients beyond the second law, which can be modeled using the scattering approach to quantum transport. The basic idea behind our approach goes back to the pioneering work of Landauer \cite{ref9} and was later substantially reformed by Buttiker \cite{ref7, ref8}. A central region in an external magnetic field is connected to a set of n electronic reservoirs (terminals) via perfect, semi-infinite leads. Now, the kinetic coefficients that describe the coherent transport of non-interacting particles between the reservoirs of this multi-terminal geometry can be expressed in terms of quantum-mechanical transmission probabilities.\\
\indent
Considering a minimal two-terminal setup if one neglects interactions and dephasing, one can obtain identical off-diagonal kinetic coefficients, even in the presence of an external magnetic field. Thus, taking into account such a setup one may investigate the effect of non-symmetric kinetic coefficients within the phenomenological Landauer-Buttiker formalism by considering fictitious terminals whose temperature and chemical potential are adjusted such that they, on average, do not exchange any heat or particles with the real terminals. Consequently, these probe terminals do not contribute to the physical transport process but mimic the inelastic scattering process.\\
\indent
Hence we first investigate the most simple case of three terminals. Based on the multi-terminal Landauer formula, one can arrive at a universal bound on the kinetic coefficients, which is independent of model-specific details like the potential landscape inside the central region or the strength of the magnetic field. This result suggests that if we break the symmetry between the off-diagonal kinetic coefficients the total rate of entropy production should be larger than zero. This bound excludes the possibility of dissipationless transport generated solely by reversible currents. Furthermore, one may understand that this result does not follow either from Onsager's principle of micro-reversibility or from purely thermodynamic arguments involving only the second law. Rather it directly comes out from the law of current conservation, which should be regarded as the fundamental physical principle underlying these new bounds.\\
\indent
As a next step of the present study, one may consider more general models involving an arbitrary number of probe terminals. This generalized constraint on the kinetic coefficients depends only on the total number of terminals $n$. It reduces to the special case of $n=3$ i.e., for the three terminals model. It is further observed that this bound becomes successively weaker as one increases $n$. In the limit $n\rightarrow \infty$, it brings down to the same relation as required by the second law of thermodynamics only. The generalized constraint leads to a bound on efficiency that is stronger than the one imposed by the second law for any finite $n$. In particular, it averts the option of reversible transport. In the limit, $n\rightarrow \infty$, the situation reduces to the case originally discussed by Benenti {\it et al.} \cite{ref6}.
\subsection{Role of time-reversal symmetry}
\begin{figure}[ht!]
    \centering
    \includegraphics[width=0.4\linewidth]{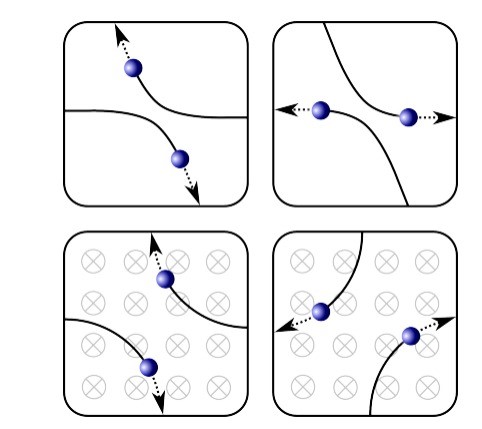}
    \caption{Reproduced from \cite{ref20}. Schematic of time-reversal symmetry breaking of a charged particle in the presence of a magnetic field indicated by the symbol \textcircled{$\times$}.}
    \label{time_reversal_breaking}
\end{figure}
In this subsection, we discuss the role of time-reversal symmetry. The microscopic equations of motion like Hamilton's equations in classical mechanics and Schrodinger's equations in the quantum mechanics regime are invariant under a simultaneous reversal of time and momenta. The constraint on the phenomenological laws, describing coupled transport processes on the macroscopic scale, is generated due to this time-reversal symmetry and it is established by the Onsager-Casimir reciprocity relation. In the context of thermoelectricity, it expresses that the off-diagonal kinetic coefficients $L_{pq}$ and $L_{qp}$  which are related to a temperature gradient to a particle current and a chemical potential gradient to a heat current, respectively should be identical.\\
\indent
Reciprocal relation has proved to be extremely useful in the sense that it leads to a unified theory to
treat the various thermoelectric effects such as the Peltier, the Seebeck, or the Thomson effect on an equal footing and demonstrate their interdependencies \cite{ref10, ref11}. In addition, it helps us to prove a quantitative relation between the efficiency and power output of thermoelectric heat engines. Particularly in the linear response regime, it can be shown that a quadratic function of efficiency bounds power. It vanishes at the Carnot value $\eta_{C}$ and reaches its maximum at the Curzon-Ahlborn value $\eta_{CA}=\frac{\eta_C}{2}$.\\
\indent
This result must, however, be reassessed in the presence of an external magnetic field ${\bf B}$. The microscopic time-reversal symmetry is broken by this ${\bf B}$ and the reciprocal relation between the kinetic coefficients $L_{pq}$ and $L_{qp}$ is also impeded. We should mention here that we consider a local standpoint, where ${\bf B}$ enters the microscopic equations of motion as a fixed parameter. Indeed, time-reversal symmetry is not broken in an extended system where the sources of the magnetic field are included as a part of the system and the reciprocal relation can be recovered as $L_{pq}({\bf B}) = L_{qp}(-{\bf B})$. This relation is not of interest since it involves the kinetic coefficients of two distinct setups, which differ in the direction of the magnetic field. This has been explained in the figure (\ref{time_reversal_breaking}).\\
\indent
In Fig. \ref{time_reversal_breaking},  the collision of two identically charged particles in the absence of an external field is shown in the upper left panel. One may observe that the particles can follow their original trajectories if their momenta are reversed after the scattering event, i.e., the process appears as
running reverse in time as shown in the upper right panel. In the lower-left panel, the particles move in circles due to the external magnetic field as indicated by the symbol \textcircled{$\times$}. This Lorentz force changes its direction as the momenta of the particles are reversed and they depart from their original trajectories as shown in the lower right panel. As a consequence, the magnetic field
breaks the symmetry between the two directions of time.\\
\indent
For a multiterminal set-up and interacting systems, the transmission probability $T_{LR}({\bf B})\neq T_{LR}(-{\bf B})$. This asymmetry provides an additional pathway to enhance thermoelectric efficiency and power output.
In an interesting paper, Benenti {\it et al.} \cite{ref6} mentioned that this additional freedom might open the doorway to enhance the performance of thermoelectric engines in such a significant manner that even devices delivering finite power close to Carnot efficiency seem to be achievable \cite{ref12,ref13}. The result has two immediate consequences. First, it has the potential to improve the efficiency of thermoelectric devices in practice. Secondly, it establishes the first significant opportunity to overcome the longstanding difficulty between high efficiency and large power output. The examination of these features as well as the search for microscopic accomplishments are the major aims of this section.\\
\begin{figure}[ht!]
    \centering
    \includegraphics[width=0.6\linewidth]{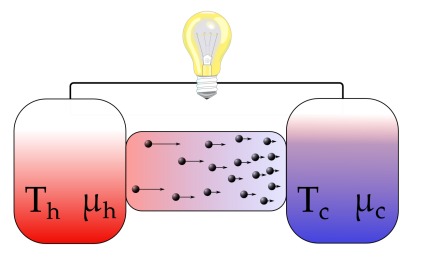}
    \caption{Reproduced from \cite{ref20}. Schematic of a two-terminal heat engine. Two real reservoirs at different temperatures $T_h>T_c$ and chemical potentials $\mu_h<\mu_c$ are connected to a conductor that can maintain a constant steady state heat and particle current $J_q$ and $J_p$, respectively. The output power for the same is $P=(\mu_c-\mu_h)J_p$. }
    \label{fig:two_terminal}
\end{figure}
\subsection{Framework}
In this subsection, we introduce the framework of linear irreversible thermodynamics that is used as a basic tool for describing thermoelectric devices. Considering time-reversal symmetric systems first, we briefly analyze the standard method that steers us to universal bounds on the efficiency and power of thermoelectric heat engines. We then demonstrate how these bounds are affected by an external magnetic field.
For our convenience, we have drawn a schematic setup of thermoelectric transport (shown in Fig. \ref{fig:two_terminal} ), and all our discussion is centered around such a setup. Considering two particle reservoirs, at temperatures $T_c$ and $T_h>T_c$ and chemical potentials $\mu_c$ and $\mu_h < \mu_c$, the exchange of heat and particles are allowed by connecting them by a conductor. As a consequence constant heat and particle currents, $J_q$ and $J_p$, start to flow from the hot reservoir to the cold reservoir as soon as the steady state is reached. If we apply a constant magnetic field ${\bf B}$ to the conductor the time-reversal symmetry is broken for this setup.
Now, the temperature difference $\Delta T = (T_h -T_c)>0$ and the chemical potential difference $\Delta \mu = (\mu_h-\mu_c)< 0$ are small compared to their respective reference values $T=\frac{T_h+T_c}{2}$ and $\mu =\frac{\mu_c+\mu_h}{2}$, respectively. In the linear response regime, the particle and heat current are related to the affinities $\mathcal{F}_p=\frac{\Delta \mu}{T}$ and $\mathcal{F}_q=\frac{\Delta T}{T^2}$ through the phenomenological equations :
\begin{eqnarray}
J_p &=& L_{pp}\mathcal{F}_p + L_{pq}\mathcal{F}_q, \\
J_q &=& L_{qp}\mathcal{F}_p + L_{qq}\mathcal{F}_q.
\end{eqnarray}
The entropy production rate associated with the setup is given by
\begin{equation}
\dot{S}=\mathcal{F}_pJ_p+\mathcal{F}_qJ_q= L_{pp}\mathcal{F}_p^2+L_{qq}\mathcal{F}_q^2+(L_{pq}+L_{qp})\mathcal{F}_p\mathcal{F}_q.
\end{equation}
One may understand that the kinetic coefficients $L_{ij},(i,j=p,q)$ are the main quantities of the present formalism and they are constrained by two fundamental relations. First, the second law of thermodynamics $\dot{S}\geq 0$ leads to :
\begin{equation}\label{Onsager_bound1}
L_{pp},L_{qq}\geq 0,
\end{equation}
and
\begin{equation}\label{Onsager_bound2}
L_{pp}L_{qq}-\frac{(L_{pq}+L_{qp})^2}{4}\geq 0.
\end{equation}
Secondly, the Onsager's theorem implies the reciprocal relation
\begin{equation}\label{Casimir_reln}
L_{pq}({\bf B})=L_{qp}(-{\bf B}).
\end{equation}
This is to notify you that these constraints Eq. (\ref{Onsager_bound1}), Eq. (\ref{Onsager_bound2}), and Eq. (\ref{Casimir_reln}) are the only conditions related to the off-diagonal kinetic coefficients. We can choose the affinities $\mathcal{F}_p$ and $\mathcal{F}_q$ in such a manner that $J_p, J_q>0$ and the setup can act as a heat engine. Thus the power output of the heat engine can be defined as
\begin{equation}\label{eq:power}
    P=-\Delta \mu J_p = -T_c\mathcal{F}_p(L_{pp}\mathcal{F}_p+L_{pq}\mathcal{F}_q),
\end{equation}
and the efficiency is given by
\begin{equation}\label{eq:effc}
    \eta = \frac{P}{J_q}=-\frac{T_c\mathcal{F}_p(L_{pp}\mathcal{F}_p+L_{pq}\mathcal{F}_q)}{(L_{qp}\mathcal{F}_p+L_{qq}\mathcal{F}_q)}.
\end{equation}
Now utilizing the entropy production rate and the second law of thermodynamics one can easily show that the efficiency $\eta$ is universally bounded by the Carnot efficiency $\eta_c =1-\frac{T_c}{T_h}$ and it becomes $\eta_c\approx\frac{\Delta T}{T_c}=T_c\mathcal{F}_q$ in the linear response regime.
\begin{figure}[t!]
    \centering
    \includegraphics[width=0.4\linewidth]{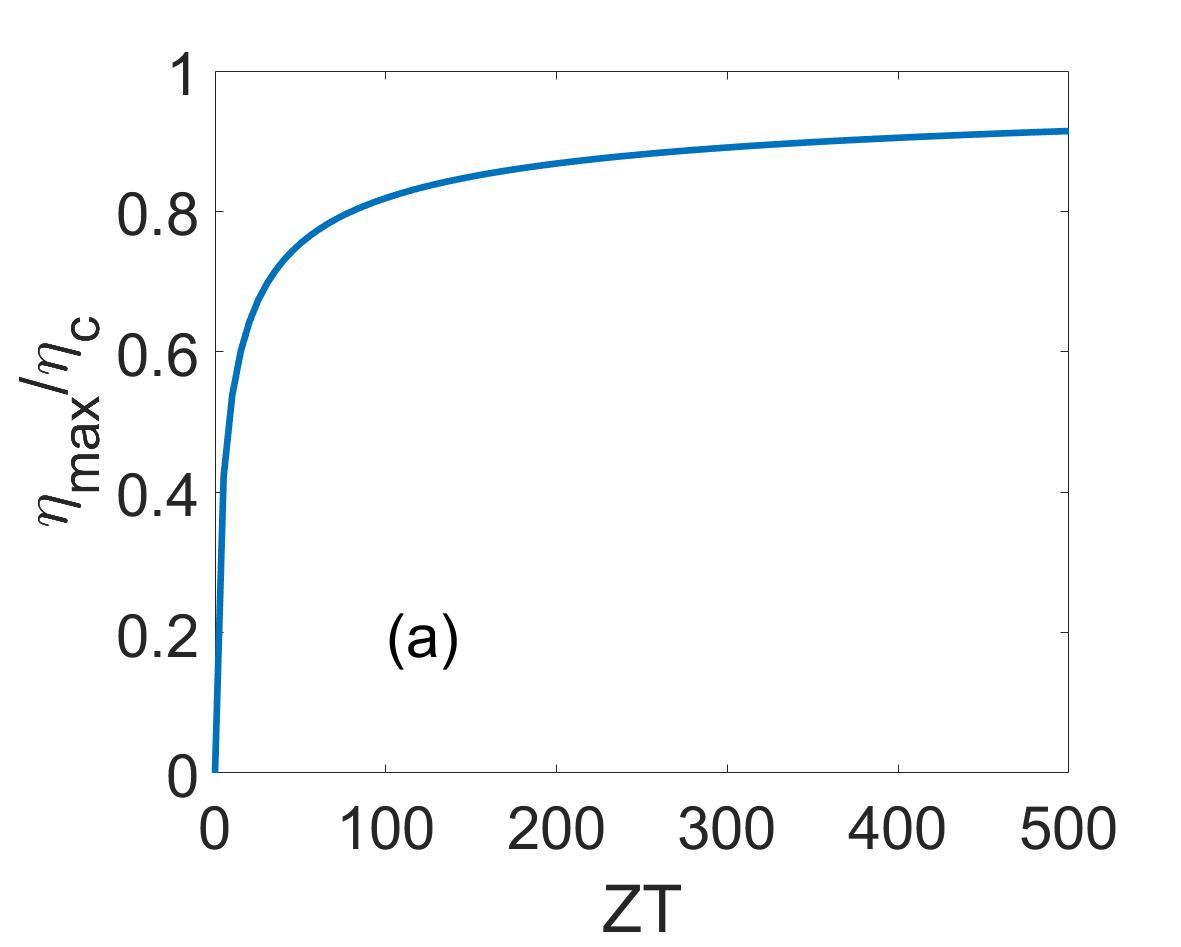}
    \includegraphics[width=0.4\linewidth]{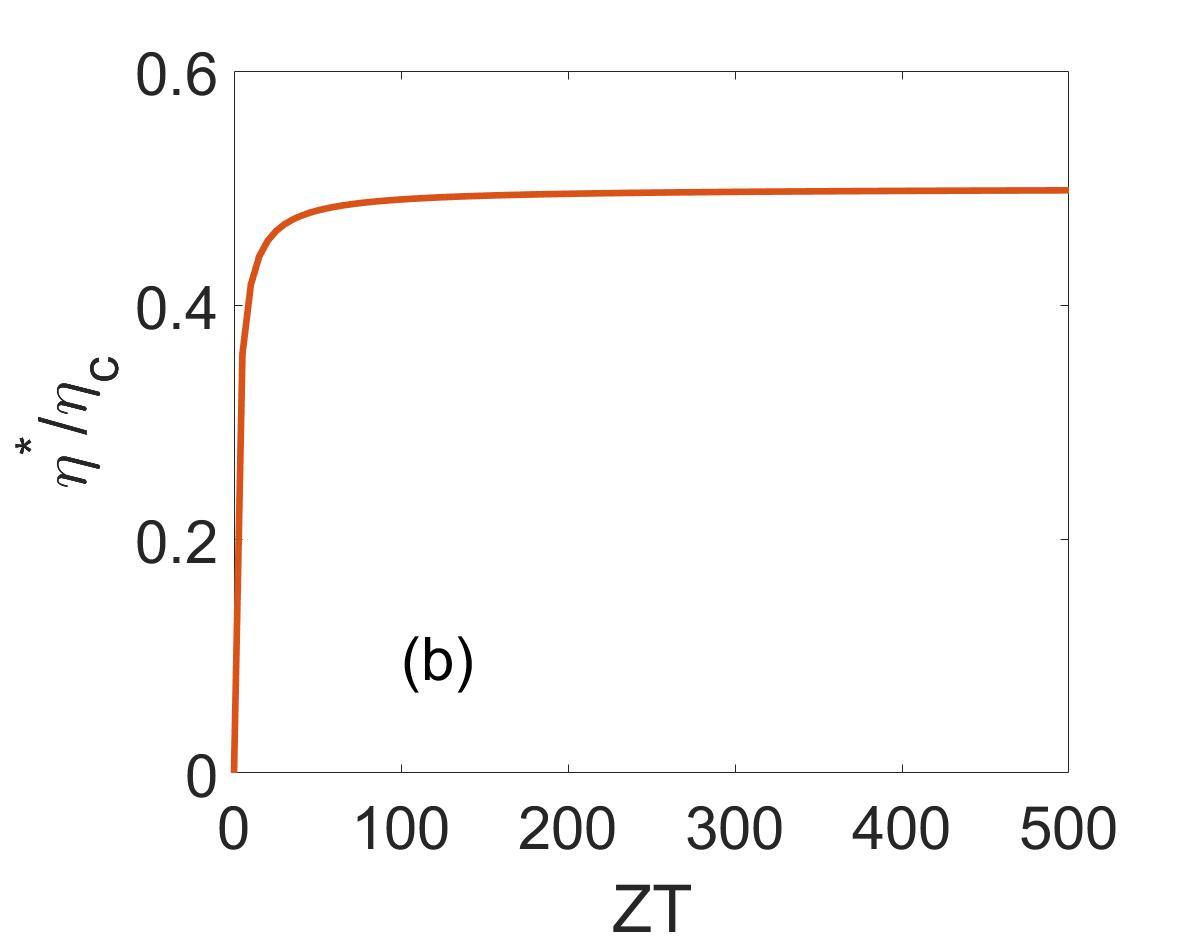}
    \caption{Plot of bound on (a) maximum efficiency, $\eta_{max}$ and (b) efficiency at maximum power, $\eta^*$ in units of Carnot efficiency $\eta_C$ as a function of $ZT$ for a time-reversal symmetric heat engine.}
    \label{fig:nZT}
\end{figure}
\subsubsection{Time-reversal symmetric engines}
\begin{figure}[t!]
    \centering
    \includegraphics[width=0.6\linewidth]{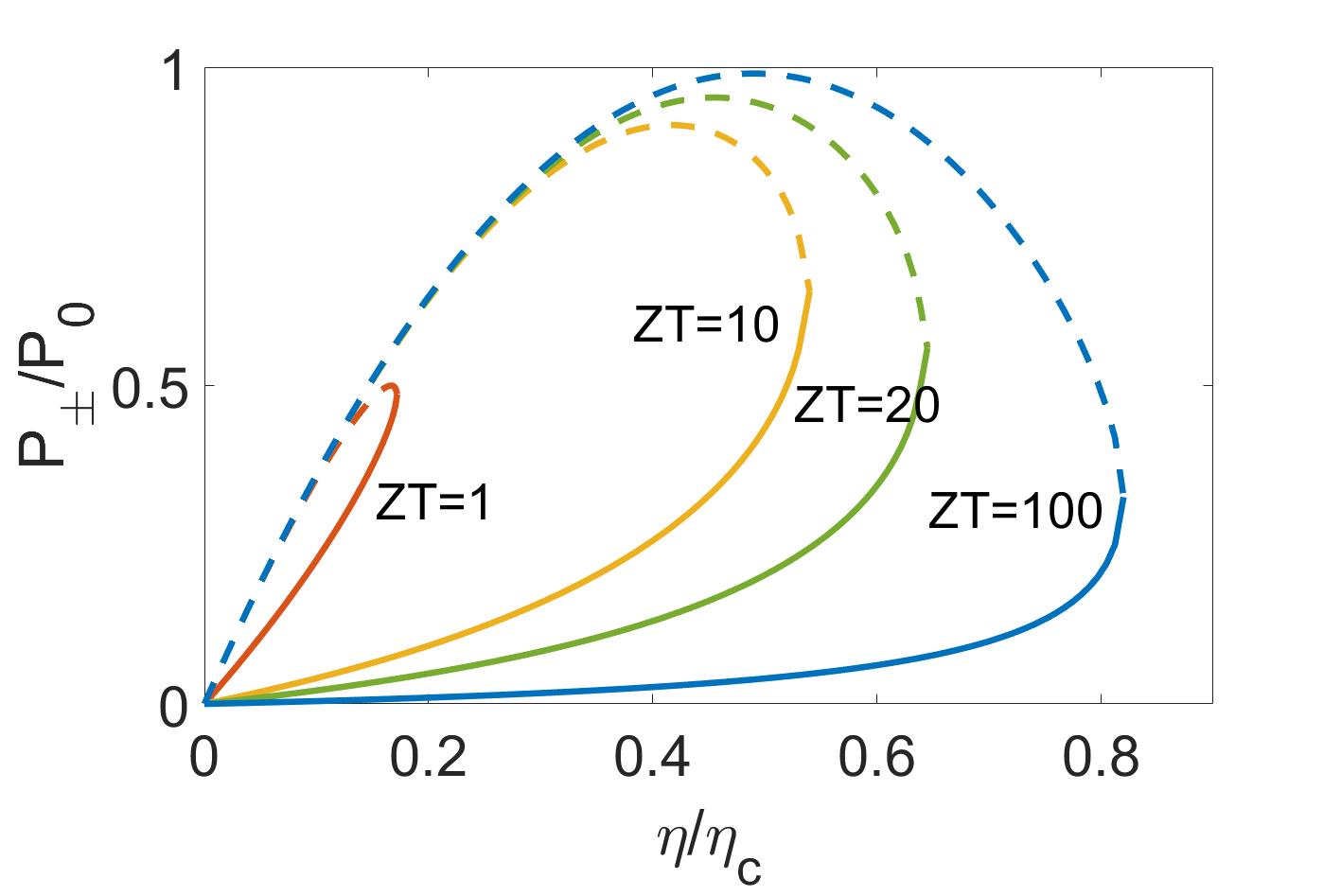}
    \caption{Reproduced from \cite{ref20}. Plot of normalized output power $\frac{P_{\pm}}{P_0}$ of a time reversal symmetric thermoelectric heat engine as a function of normalized efficiency $\tilde{\eta}=\frac{\eta}{\eta_C}$ ($\eta_C$ = Carnot efficiency) for different values of $ZT$. The dashed lines represent the upper bound branch $\frac{P_+}{P_0}$ and the solid lines correspond to the lower bound branch $\frac{P_-}{P_0}$.}
    \label{fig:Pplus1}
\end{figure}

At the outset, if we switch off the magnetic field the Onsager's reciprocal relation becomes $L_{pq}=L_{qp}$. Due to this symmetry, the characterization of a time-reversal thermoelectric heat engine can be determined by three kinetic coefficients $L_{pp}, L_{qq}$, and $L_{pq}$. The thermoelectric figure of merit of such a device is defined as \cite{ref6,Bell}:
\begin{equation}\label{eq:ZTB0}
    ZT = \frac{L_{pq}^2}{L_{pp}L_{qq}-L_{pq}^2},
\end{equation}
and it must be non-negative which comes out from the second law of thermodynamics. The next task is to find the bound on the efficiency which can be obtained by optimizing Eq. (\ref{eq:power}) and Eq. (\ref{eq:effc}) concerning $\mathcal{F}_p$ constraining $P>0$ and the maximum efficiency is found to be \cite{ref6}
\begin{equation}
\eta_{max}(ZT)= \eta_c \frac{\sqrt{ZT+1}-1}{\sqrt{ZT+1}+1},
\end{equation}
and the efficiency at maximum power is given by
\begin{equation}
\eta^{*}(ZT)=\frac{\eta_C}{2}\frac{ZT}{ZT+2}.
\end{equation}
One may observe that both these functions $\eta_{max}(ZT)$ and $\eta^{*}(ZT)$ increase monotonically with $ZT$ and they reach their respective upper bounds $\eta_c$ and $\eta_c/2$, respectively in the limit $ZT \rightarrow \infty$ (see Fig. \ref{fig:nZT}). Now, the assessment of the performance of a heat engine can only be completed if we can characterize the output power performance. Introducing a normalized efficiency $\tilde{\eta}=\frac{\eta}{\eta_c}$, one can obtain the affinity \cite{ref20}
\begin{equation}
    \mathcal{F}_p = -\mathcal{F}_q\frac{L_{qq}}{L_{pq}}\Big\lbrack\frac{ZT(1+\tilde{\eta})}{1+ZT}-\sqrt{\frac{ZT^2(1+\tilde{\eta})^2}{4(1+ZT)^2}-\frac{ZT\tilde{\eta}}{1+ZT}}\Big\rbrack.
\end{equation}
The power at a given efficiency can be written as \cite{ref20}
\begin{equation}
    P_{\pm}(ZT,\tilde{\eta})=4P_0\tilde{\eta}\Big\lbrack\frac{2+ZT(1-\tilde{\eta})}{2(1+ZT)}\pm\sqrt{\frac{ZT^2(1+\tilde{\eta})^2}{4(1+ZT)^2}-\frac{ZT\tilde{\eta}}{1+ZT}} \Big\rbrack,
\end{equation}
where the standard power $P_0 = \frac{T_cL_{qq}\mathcal{F}_{q}}{4}$, and it is constrained by
\begin{equation}
\frac{4\tilde{\eta}}{(\tilde{\eta}-1)^2}\leq ZT <\infty,
\end{equation}
which is equivalent to $\eta\leq \eta_{max}(ZT)$. The two branches as denoted by $P_{\pm}(ZT,\tilde{\eta})$ are the two admissible values of output power $P$ for a fixed value of $\tilde{\eta}$ and $ZT$. These two branches are shown in Fig. \ref{fig:Pplus1}. If we fix $\eta = \eta^{*}$, one may observe that the upper branch values $P_+(ZT,\tilde{\eta})$ are always either greater than or equal to the lower branch values $P_-(ZT,\tilde{\eta})$ and its maximum is given by
\begin{equation}
P_{max}(ZT)=P_0\frac{ZT}{1+ZT},
\end{equation}
and the power output at maximum efficiency is given by
\begin{equation}
P^{*}(ZT)=\frac{4P_0}{\sqrt{ZT+1}}\frac{\sqrt{ZT+1}-1}{\sqrt{ZT+1}+1}.
\end{equation}
One may observe that the $P^{*}(ZT)$ decays to zero in a linear manner as the efficiency $\eta$ approaches $\eta_C$. This result confirms that high efficiency is obtained at the cost of low output power. In addition, as we increase $ZT \rightarrow \infty$, the two branches of the available output power become:
\begin{eqnarray}\label{eq:Ppm}
P_+(\tilde{\eta})&=&4P_0\tilde{\eta}(1-\tilde{\eta}) \nonumber, \\
P_-(\tilde{\eta})&=& 0.
\end{eqnarray}
Considering the above expression we may conclude that the quadratic function of efficiency as in Eq. (\ref{eq:Ppm}) can be considered as a universal bound on power in the sense that it is applicable for any time-reversal-symmetric thermoelectric heat engine in the linear response regime.
\subsubsection{Extensive changes for broken time-reversal symmetry}
\begin{figure}[h!]
    \centering
    \includegraphics[width=0.4\linewidth]{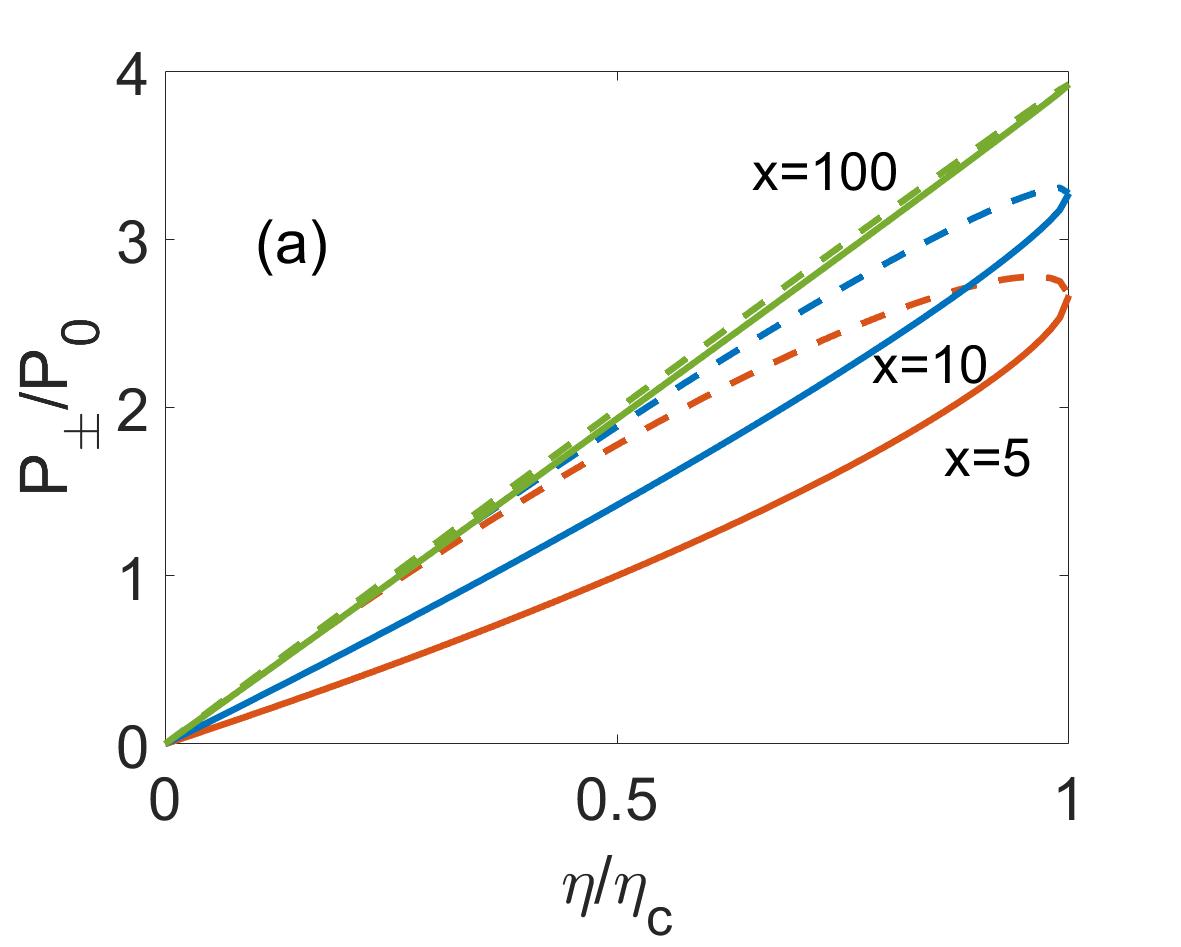}
    \includegraphics[width=0.4\linewidth]{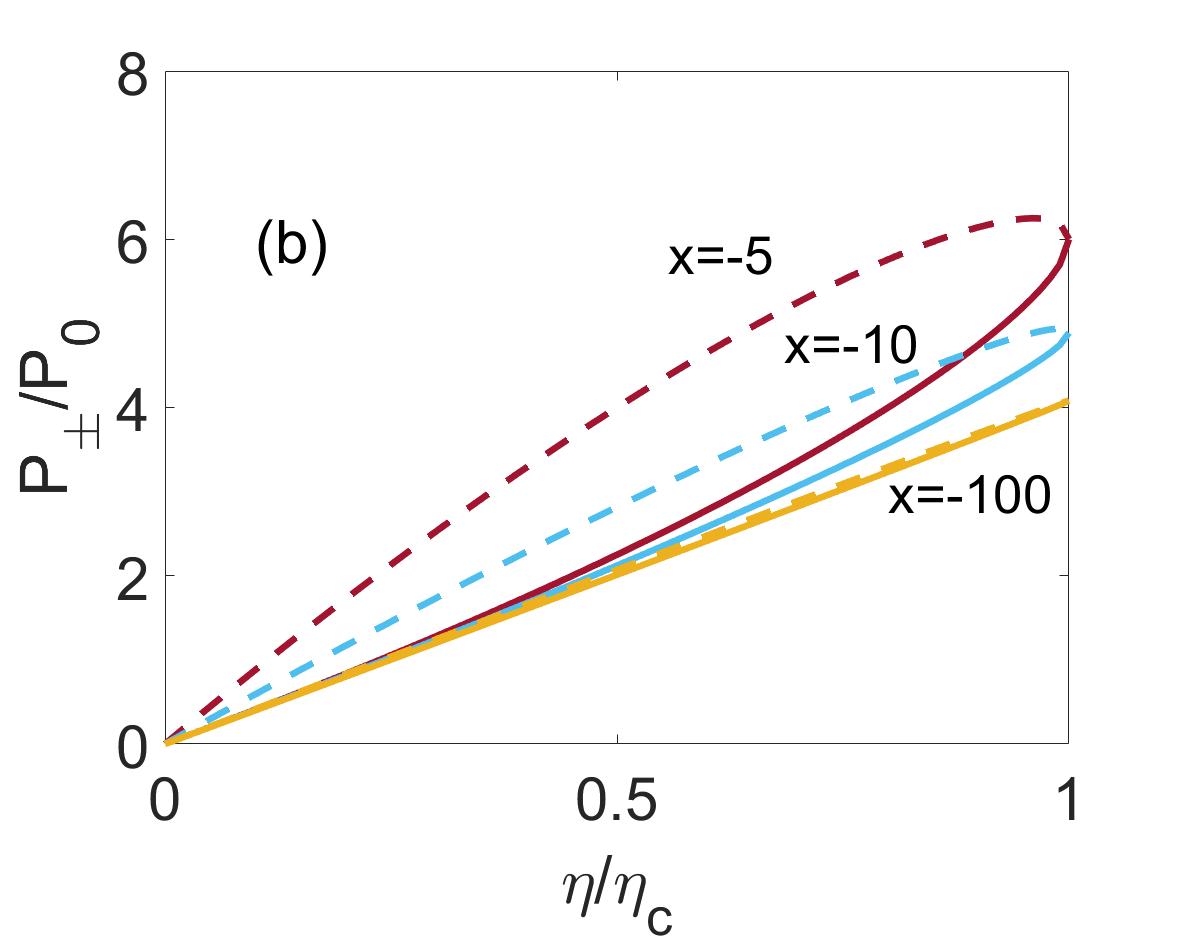}
    \caption{Reproduced from \cite{ref6}. Plot of normalized output power $\frac{P_{\pm}}{P_0}$ of a thermoelectric heat engine with a broken time-reversal symmetry as a function of normalized efficiency $\tilde{\eta}=\frac{\eta}{\eta_C}$ ($\eta_C$ = Carnot efficiency) for different values of the asymmetry parameter with (a) $x>0$ and (b) $x<0$.The dashed lines represent the upper bound branch $\frac{P_+}{P_0}$ and the solid lines correspond to the lower bound branch $\frac{P_-}{P_0}$.}
    \label{fig:Pplus2}
\end{figure}

We already observed that Onsager's reciprocal relation (Eq. (\ref{Casimir_reln})) does not introduce any new kind of restrictions on the kinetic coefficients as that of the second law (Eq.(\ref{Onsager_bound1}) and Eq.(\ref{Onsager_bound2})). Contrary to a symmetric setup, one needs two dimensionless parameters to characterize and quantify the performance of a thermoelectric heat engine with broken time-reversal symmetry. One can introduce two dimensionless parameters following Ref. (\cite{ref6}):
\begin{eqnarray}
x &=& \frac{L_{pq}}{L_{qp}} \nonumber, \\
y &=& \frac{L_{pq}L_{qp}}{L_{pp}L_{qq}-L_{pq}L_{qp}}.
\end{eqnarray}
One may notice that in the presence of ${\bf B}$, $x$ is a measure of the asymmetry between the kinetic coefficients $L_{pq}$ and $L_{qp}$, while $y$ is a generalization of the figure of merit as introduced in Eq. (\ref{eq:ZTB0}) for ${\bf B}=0$. Now, these parameters are inter-related via a few inequalities following from Eq. (\ref{Onsager_bound2}) as mentioned below \cite{ref6}
\begin{eqnarray}
h(x)\leq y \leq 0  \qquad \text{for} \qquad x < 0          \nonumber, \\
0 \leq y \leq h(x) \qquad \text{for} \qquad x \geq 0,
\end{eqnarray}
where, $h(x) = \frac{4x}{(x-1)^2}$ which follows from the second law.
\begin{figure}[t!]
    \centering
    \includegraphics[width=0.4\linewidth]{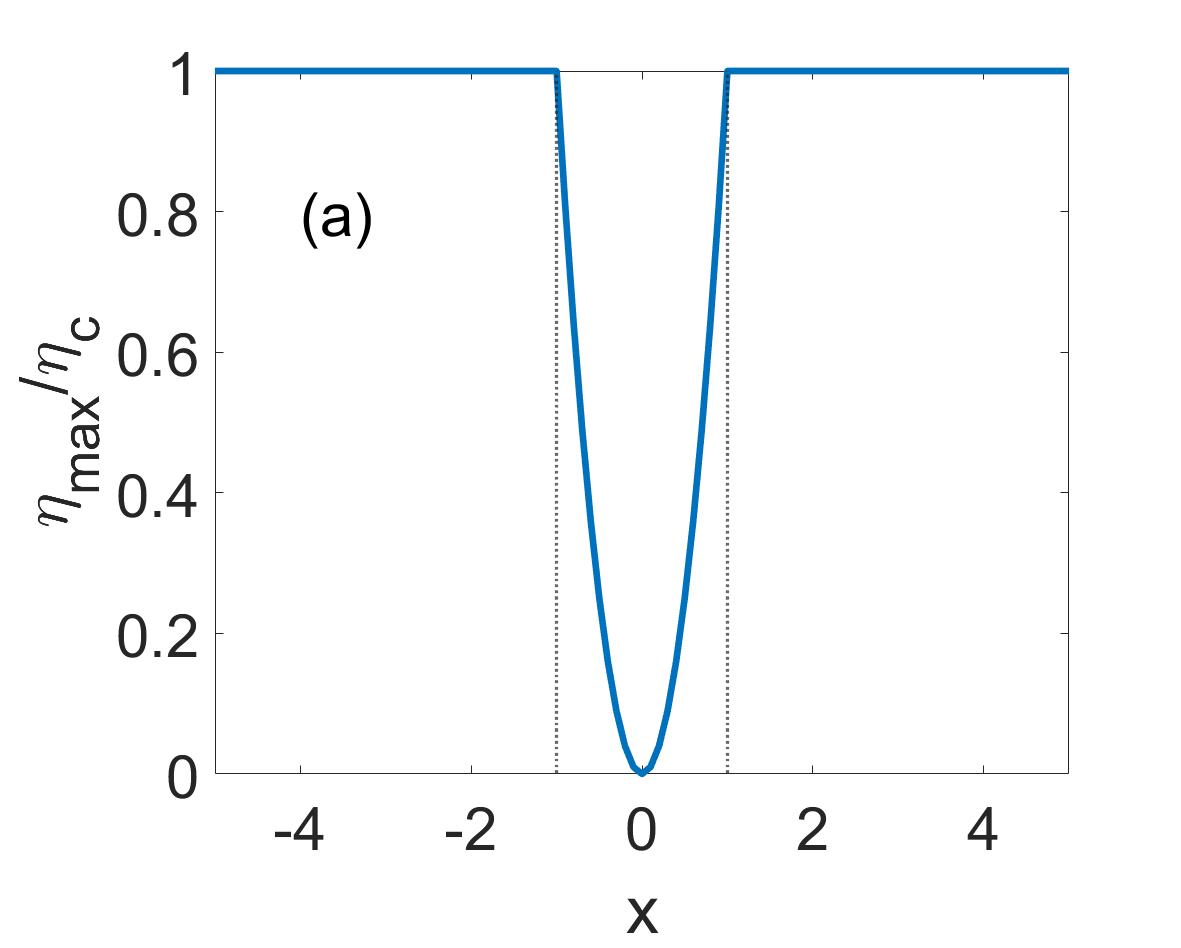}
    \includegraphics[width=0.4\linewidth]{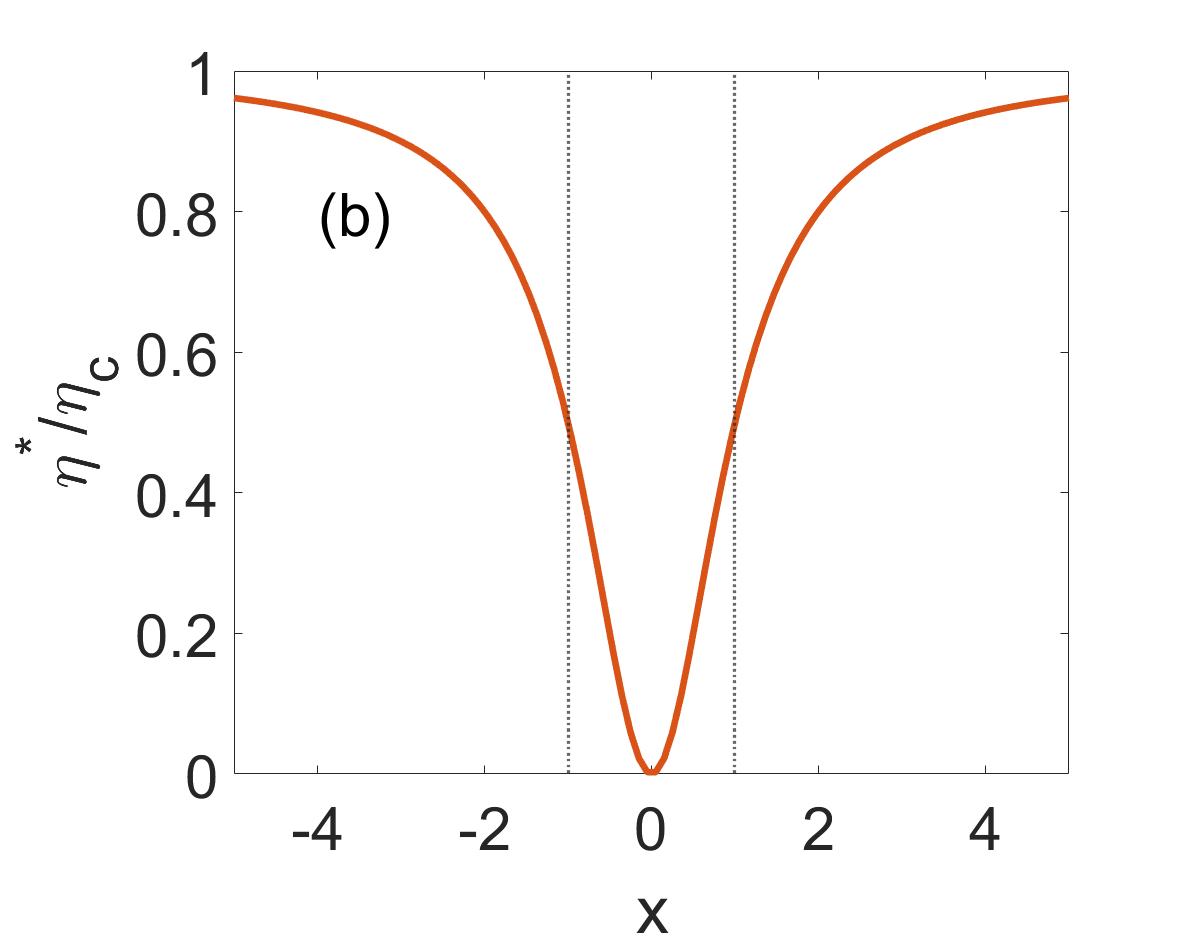}
    \caption{Reproduced from \cite{ref6}. The upper bound on (a) maximum efficiency, $\eta_{max}$ and (b) efficiency at maximum power, $\eta^*$ in units of Carnot efficiency $\eta_C$ as a function of asymmetry parameter $x$ for a two-terminal heat engine with broken time-reversal symmetry. The vertical dotted lines denote $|x|=1$.}
    \label{fig:etax}
\end{figure}
Thus, the maximum efficiency in terms of the dimensionless parameters $x$ and $y$ is given by
\begin{equation}
\eta_{max}(x,y)= \eta_C x \frac{\sqrt{y+1}-1}{\sqrt{y+1}+1},
\end{equation}
while the efficiency at maximum power is given by
\begin{equation}
\eta^*(x,y)= \eta_C\frac{xy}{4+2y}.
\end{equation}
For all $x \in \mathbb{R}$, these two functions are monotonically increasing functions of $|y|$. Now setting $y=h(x)$ one can obtain the upper bounds :
\begin{eqnarray}
\eta_{max}(x)&=& \eta_C  \qquad \text{for} \qquad |x|\geq 1 \nonumber, \\
\eta_{max}(x)&=& \eta_C x^2 \qquad \text{for} \qquad |x|< 1,
\end{eqnarray}
and
\begin{equation}
\eta^*(x)= \eta_C \frac{x^2}{1+x^2}.
\end{equation}
These two functions are plotted in Fig. \ref{fig:etax}. One can observe that the maximum efficiency can reach $\eta_C$ for any $|x|\geq 1$. Moreover in the regime $|x|>1$, as $x$ becomes larger than 1, the efficiency at maximum power can exceed its symmetric bound $\eta_C/2$ that can only occur for $|x|= 1$. This also suggests that the performance of a thermoelectric heat engine with broken time-reversal symmetry is much better compared to its symmetric counterparts.\\
\indent
Next, our discussion will be centered around the profound changes observed in the bound on power due to the presence of a magnetic field. Following the similar analysis as shown for the time-reversal symmetric thermoelectric heat engine, one can obtain the power output \cite{ref20}
\begin{equation}
\frac{P_{\pm}(x,y,\tilde{\eta})}{P_0}=4\tilde{\eta}\Big\lbrack \frac{x(y+2)-y\tilde{\eta}}{2x(1+y)}\pm\sqrt{\frac{y^2(x+\tilde{\eta})^2}{4x^2(1+y)^2}-\frac{y\tilde{\eta}}{x(1+y)}} \Big\rbrack,
\end{equation}
where $x\in \mathbb{R}$, and $0\leq \eta \leq \eta_{max}(x)$, with
\begin{eqnarray}
\frac{4x\tilde{\eta}}{(x-\tilde{\eta})^2}\leq y\leq h(x) \qquad \text{for} x\geq 0 \nonumber, \\
h(x)\leq y \leq \frac{4x\tilde{\eta}}{(x-\tilde{\eta})^2} \qquad \text{for} x<0.
\end{eqnarray}
Again setting $y=h(x)$ one can obtain the upper bound
\begin{equation}
P_+(x,\tilde{\eta})= 4P_0\tilde{\eta}\frac{(x^2-1)+2(1-\tilde{\eta})+2\sqrt{(x^2-\tilde{\eta})(1-\tilde{\eta})}}{(1+x)^2}.
\end{equation}
On the other hand, the lower bound is given by
\begin{equation}
P_-(x,\tilde{\eta})= 4P_0\tilde{\eta}\frac{(x^2-1)+2(1-\tilde{\eta})-2\sqrt{(x^2-\tilde{\eta})(1-\tilde{\eta})}}{(1+x)^2}.
\end{equation}
These two functions are plotted in Fig. \ref{fig:Pplus2}. One may observe that for $x>1$ and $x\leq -1$, the Carnot efficiency can be achieved at finite power :
\begin{equation}
P_+(x,\eta_C)=P_-(x,\eta_C)=4P_0\frac{x-1}{x+1}.
\end{equation}
This surprising result can be explained as follows. The total rate of entropy production indicates that the currents $J_p$ and $J_q$ can be split into a reversible and irreversible part:
\begin{eqnarray}
J_i^{rev}&=&\frac{L_{ij}-L_{ji}}{2}\mathcal{F}_j \nonumber,\\
J_i^{irr}&=&L_{ii}\mathcal{F}_i+\frac{L_{ij}+L_{ji}}{2}\mathcal{F}_j,
\end{eqnarray}
and it is well understood that $J_i^{rev}$ vanishes for the unbroken reversible symmetry case i.e. for ${\bf B}=0$. As soon as one breaks the time-reversal symmetry the reversible current can in principle increase to arbitrarily large. Since there are no further general relations between $L_{ij}({\bf B})$ and $L_{ji}({\bf B})$ are known, except Eqs. (\ref{Onsager_bound2}) and (\ref{Casimir_reln}), such unconstrained reversible currents ultimately give rise to the possibility of dissipationless transport. This enables one to achieve Carnot efficiency at finite power as first discussed by Benenti {\it et al.} \cite{ref6}.
In conclusion, it can be observed that when time-reversal symmetry is broken, both the maximum efficiency and the efficiency at maximum power are influenced by factors beyond the figure of merit alone $ZT$. To determine them fully, two more parameters are needed: (i) an asymmetry parameter $x$ and (ii) another parameter $y$ related to the figure of merit and it reduces to $ZT$  in the symmetric limit $x=1$. We have also observed that it is possible to overcome the Curzon-Ahlborn limit within linear response for $|x|>1$. One may even reach the Carnot efficiency for an increasingly smaller and smaller figure of merit $y$ as one increases $|x|$. From a practical point of view, one should
note that the scattering matrix follows $S({\bf B})=S(-{\bf B})$ for the noninteracting case with $x=1$ and it is a consequence of the symmetry properties of the scattering matrix \cite{ref6}. Further, the Onsager-Casimir symmetry relations do not inflict the symmetry of the Seebeck coefficient under the exchange of ${\bf B}$ to ${-\bf B}$. Therefore this symmetry is broken in the presence of the
electron-phonon and electron-electron interactions. In a two-terminal purely metallic mesoscopic system \cite{ref21} the Seebeck coefficient is always found to be an even function of the magnetic field, but Andreev interferometer experiments \cite{ref22} and recent theoretical studies \cite{ref23} have proved that systems in contact with a superconductor or with a heat bath can exhibit asymmetric thermopowers. It isn't easy to find realistic setups with $x$ that are significantly different from unity while approaching Carnot efficiency.
\subsection{Mutiterminal setup}
\begin{figure}[b!]
    \centering
    \includegraphics[width=0.6\linewidth]{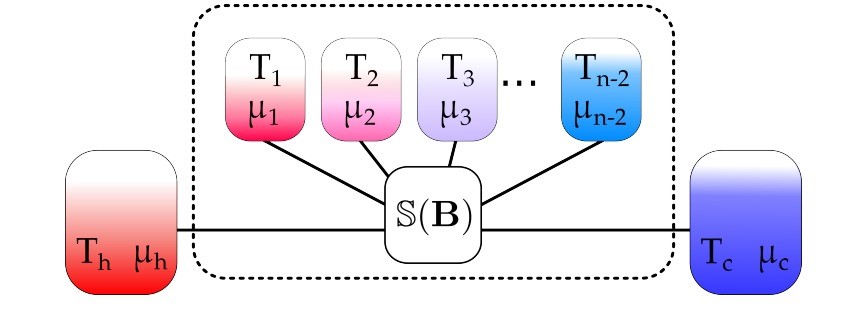}
    \caption{Reproduced from \cite{ref20}. Schematic of a multi-terminal irreversible heat engine with broken time-reversal symmetry. The system is coupled with two real reservoirs maintained at different temperatures $T_h>T_c$ and chemical potentials $\mu_h<\mu_c$. On the other hand, the terminals $1,2,\cdots,n-2$ are considered as probes. In the case of probe terminals the time-reversal symmetry is broken i.e $T_{LR}(\phi)\neq T_{LR}(-\phi)$}
    \label{fig:multi_terminal}
\end{figure}
From the discussion of the previous subsection, one can understand that there are no other possible constraints on the kinetic coefficients except the second law and Onsager's reciprocal relations for two terminal setups. One needs to modify the phenomenological setup discussed in the previous subsection to explore more possibilities of the restrictions on the kinetic coefficients. We need to mention that our discussion is limited to those thermoelectric nano-devices, which can be described within the Landauer-Buttiker approach. If one considers inelastic scattering events, which are crucial to break the symmetry between the off-diagonal kinetic coefficients, can be incorporated phenomenologically in terms of probe terminals. The most simple model of this type which has a non-symmetric matrix of kinetic coefficients involves one probe terminal along with the two real terminals. For this minimal setup, Brandner {\it et al.} \cite{ref19} have shown that current conservation steers an additional constraint on the kinetic coefficients, much stronger than the second law. If one breaks the timer-reversal symmetry, the efficiency of the three-terminal thermoelectric heat engine is considerably smaller than the Carnot value due to this additional constraint coming out from the current conservation. One can generalize this observation for the three-terminal case and Brandner {\it et al.} \cite{ref19} proved that similar kinds of results hold for models with an arbitrary number of terminals. They have shown that the corresponding bounds become consecutively weaker as this number of probes increases. If one increases this probe number to the limit of infinitely large, one may land up with the situation studied by Benenti {\it et al.} \cite{ref6}, where the second law is the only constraint. Naturally, this leads to the question of whether not only efficiency but also power can be bounded. Brandner {\it et al.} \cite{ref18} proved that such bounds exist for the power also at least for the multiterminal setup. In this subsection, we will summarize the results obtained by Brandner {\it et al.} \cite{ref19, ref18} for the multiterminal setup.\\
\indent
One may find a multiterminal model setup as shown schematically in Fig. \ref{fig:multi_terminal}. Here, A central region is pierced by a constant magnetic field ${\bf B}$ and is connected to $n$ independent electronic reservoirs (terminals) characterized by their respective chemical potential $\mu_{\alpha}$ and temperature $T_{\alpha}$, where $\alpha=h,1,2,3,\cdots,n-2,c$.
Out of these $n$ number of electronic reservoirs, two reservoirs (h and c) are real terminals and others are probe terminals. Considering such multi-terminal geometry and utilizing the linear irreversible thermodynamics by fixing reference temperature $T_c$ and chemical potential $\mu_c$, one can define the affinities :
$\mathcal{F}_p^{\alpha}=\frac{\mu_{\alpha}-\mu_c}{T_c}=\frac{\Delta\mu_{\alpha}}{T_c}$ and $\mathcal{F}_q^{\alpha}=\frac{T_{\alpha}-T_c}{T_c^2}=\frac{\Delta T_{\alpha}}{T_c^2}$ for $\alpha = h, 1, 2, 3,\cdots n-2, c$. Denoting $J_p^{\alpha}$ and $J_q^{\alpha}$ as a particle and heat current leaving the reservoir $\alpha$, respectively, one can write in the linear response regime ($\Delta\mu_{\alpha}<< \mu_c, \Delta T_{\alpha}<<T_c$) :
\begin{equation}
\vect{J}= \mathbb{L}^n \vect{\mathcal{F}},
\end{equation}
where the current vector
\begin{equation}
\vect{J}=
\begin{pmatrix}
\vect{J}_h\\
\vect{J}_1\\
\vdots \\
\vect{J}_{n-2}
\end{pmatrix},
\end{equation}
and the affinity vector
\begin{equation}
\vect{\mathcal{F}}=
\begin{pmatrix}
\vect{\mathcal{F}}_h\\
\vect{\mathcal{F}}_1\\
\vdots \\
\vect{\mathcal{F}}_{n-2}
\end{pmatrix},
\end{equation}
and the matrix of kinetic coefficients :
\begin{equation}
\mathbb{L}^n =
\begin{pmatrix}
\mathbb{L}_{h,h}^n & \cdots & \mathbb{L}_{h, n-2}^n\\
\vdots & \ddots & \vdots \\
\mathbb{L}_{n-2, h}^n & \cdots & \mathbb{L}_{n-2, n-2}^n
\end{pmatrix}.
\end{equation}
Considering the present multiterminal setup one can perceive a thermoelectric heat engine employing
the terminals $1,2,\cdots, n-2$ as probes that can mimic inelastic scattering. We need to utilize the probe conditions in such a way that they do not exchange any net quantities with the real terminals and these conditions enable us to determine the temperature and chemical potential of the corresponding reservoirs. This constraint reads as :
\begin{equation}
\begin{pmatrix}
\vect{J}_1\\
\vdots\\
\vect{J}_{n-2}
\end{pmatrix}
=0.
\end{equation}
With this one may end up with a reduced system :
\begin{equation}
\begin{pmatrix}
J_p\\
J_q
\end{pmatrix}
=\mathbb{L}^{'}
\begin{pmatrix}
\mathcal{F}_p\\
\mathcal{F}_q
\end{pmatrix},
\end{equation}
where one may identify particle current $J_p^h=J_p$ and heat current $J_q^h=J_q$ leaving the hot reservoir and the affinities are chosen in such a way $\mathcal{F}_p=\mathcal{F}_p^h=\frac{\Delta\mu_h}{T_c}<0$ and $\mathcal{F}_q=\mathcal{F}_q^h=\frac{\Delta T_h}{T_c^2}>0$ that the multiterminal setup works as a heat engine. The reduced effective kinetic matrix is identified as :
\begin{equation}
\mathbb{L^{'}}= \begin{pmatrix} L_{pp}^{'} & L_{pq}^{'}\\
L_{qp}^{'} & L_{qq}^{'}\end{pmatrix}.
\end{equation}
\subsubsection{Minimal three-terminal model}
In general, a three-terminal system can be characterized by the following equation :
\begin{equation}\label{eq:72}
\begin{pmatrix}
\vect{J}^h\\
\vect{J}^p
\end{pmatrix}
=\mathbb{L}\begin{pmatrix} \mathcal{\vect{F}}^h\\
 \mathcal{\vect{F}}^p\end{pmatrix},
\end{equation}
where the kinetic coefficient matrix is given by:
\begin{equation}
\mathbb{L}=\begin{pmatrix}
\mathbb{L}_{hh} & \mathbb{L}_{hP}\\
\mathbb{L}_{Ph} & \mathbb{L}_{PP}
\end{pmatrix},
\end{equation}
and $\mathbb{L}_{i,j}$ ($i,j= h, P$) are $2\times2$ matrices of kinetic coefficients. Now, employing the additional probe condition $\vect{J}^P=0$ one can eliminate $\vect{\mathcal{F}}^P$ and reduce Eq. (\ref{eq:72}) to a system of two equations :
\begin{equation}
\vect{J}^h = \mathbb{L}^{'}\vect{\mathcal{F}}^h,
\end{equation}
where the effective matrix of kinetic coefficients is given by :
\begin{equation}
\mathbb{L}^{'}=\mathbb{L}_{hh}-\mathbb{L}_{hP}\mathbb{L}_{PP}^{-1}\mathbb{L}_{Ph}=\begin{pmatrix}L_{pp}^{'} & L_{pq}^{'}\\
L_{qp}^{'}& L_{qq}^{'}
\end{pmatrix}.
\end{equation}
In the linear response regimes, the closed form expression for the $2\times2$ block matrices $\mathcal{L}_{i,j}$ ($i,j = h,P$) are given by \cite{ref19}
\begin{equation}
\mathbb{L}_{i,j}=\frac{2\pi e^2T_c}{\hbar}\int_{-\infty}^{\infty}dE F(E) \begin{pmatrix} E & \frac{E-\mu_c}{e}\\
 \frac{E-\mu_c}{e}& \Big( \frac{E-\mu_c}{e}\Big)^2 \end{pmatrix} (\delta_{i,j}-|S_{i,j}(E,{\bf B})|^2),
\end{equation}
where $F(E)$ is the negative derivative of the Fermi function, $S_{i,j}(E,{\bf B})\neq S_{j,i}(E,{\bf B})$ are the matrix elements of the $3\times 3$ scattering matrix $\mathbb{S}(E,{\bf B})$  for the central scattering region. The current conservation leads to the fact that $\mathbb{S}(E,{\bf B})$ to be unitary, while time reversal invariance enforces the condition $\mathbb{S}(E,{\bf B})=\mathbb{S}(E,-{\bf B})$. The reduced matrix of kinetic coefficients acquires a nonvanishing asymmetry part. Now the constrain on the asymmetry part of the reduced kinetic coefficient matrix leads to \cite{ref19}
\begin{equation}
L_{pp}^{'}L_{qq}^{'}+L_{pq}^{'}L_{qp}^{'}-L_{pq}^{'2}-L_{qp}^{'2} \geq 0,
\end{equation}
which can be rewritten as follows :
\begin{equation}\label{eq:new_bound}
L_{pp}^{'}L_{qq}^{'}-\frac{(L_{pq}^{'}+L_{qp}^{'})^2}{4} \geq \frac{3(L_{pq}^{'}-L_{qp}^{'})^2}{4}.
\end{equation}
On the other hand, the second law implies
\begin{equation}
L_{pp}^{'}L_{qq}^{'}-\frac{(L_{pq}^{'}+L_{qp}^{'})^2}{4} \geq 0.
\end{equation}
One may easily understand from the above discussion that the additional constraint due to current conservation leads to a tighter bound on the effective kinetic coefficients. Thus, the reversible currents associated with the asymmetric part of the reduced kinetic matrix come at the expense of a tight lower bound on the Onsager coefficients. It makes a stronger bound on the entropy production rate than that of the bare second law. \\
Now, this new bound obtained in Eq. (\ref{eq:new_bound}) has an immense effect on the performance of the three-terminal thermoelectric heat engine. Considering the dimensionless parameters $x=\frac{{L}_{pq}^{'}}{{L}_{qp}^{'}}$ and $y = \frac{L_{pq}^{'}L_{qp}^{'}}{Det(\mathbb{L}^{'})}$, one may find the two important quantities to quantify the performance of the thermoelectric heat engine :
the maximum efficiency
\begin{equation}
\eta_{max}= \eta_C x \frac{\sqrt{y+1}-1}{\sqrt{y+1}+1},
\end{equation}
and the efficiency at maximum power is given by
\begin{equation}
\eta(P_{max})= \eta_C \frac{xy}{(4+2y)},
\end{equation}
where $\eta_C = 1- \frac{T_c}{T_h}$. By setting $y=h(x)=\frac{x}{(x-1)^2}$, which is obtained from the Onsager bounds in Eq. (\ref{eq:new_bound}), we can derive the upper bound on maximum efficiency as
\begin{equation}
    \eta_{max}^*=\eta_Cx\frac{\sqrt{(x-1)x+1}-|x-1|}{\sqrt{(x-1)x+1}+|x-1|},
\end{equation}
and the upper bound on the efficiency at maximum power as
\begin{equation}
    \eta^*(P_{max})=\frac{\eta_C}{2}\frac{x^2}{2x^2-3x+2}.
\end{equation}

\begin{figure}
    \centering
    \includegraphics[width=0.4\linewidth]{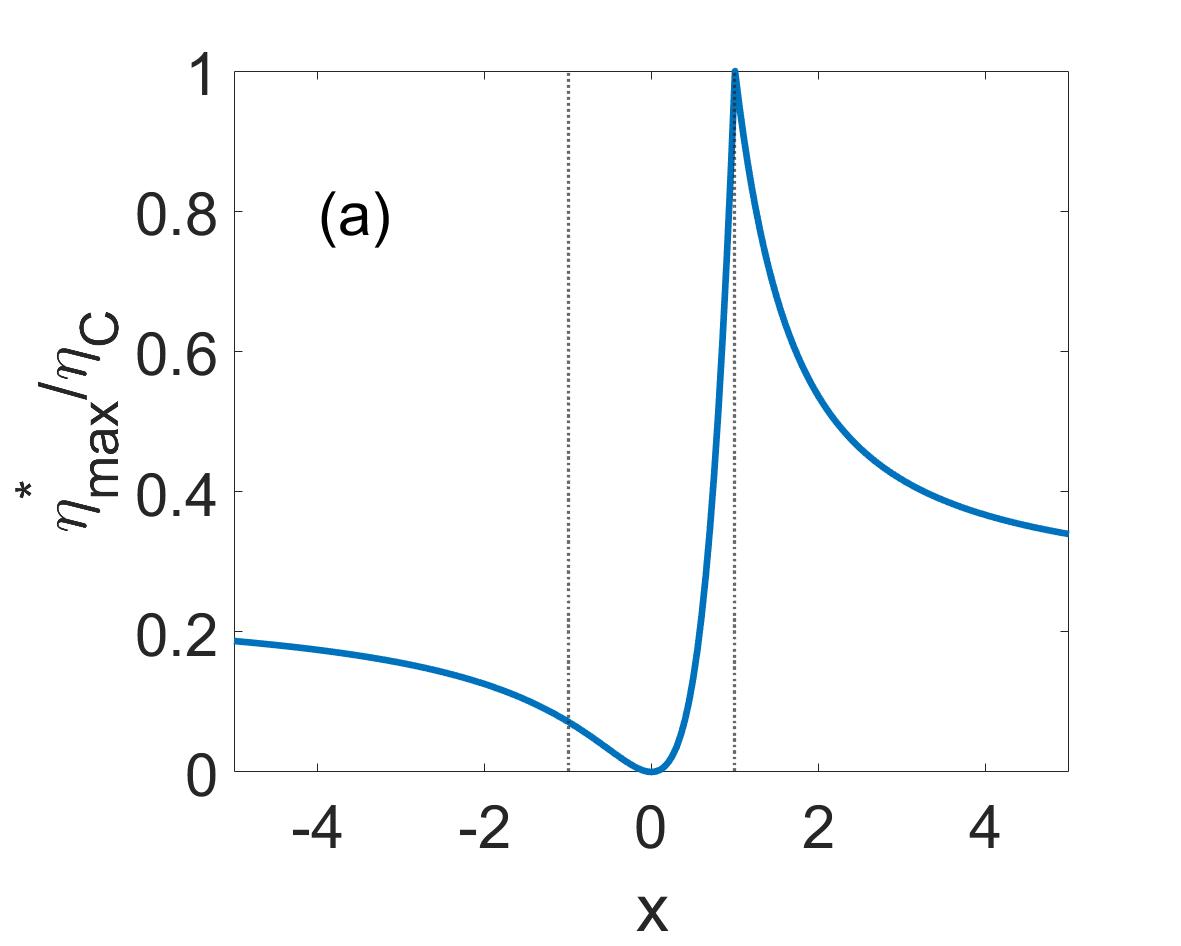}
    \includegraphics[width=0.4\linewidth]{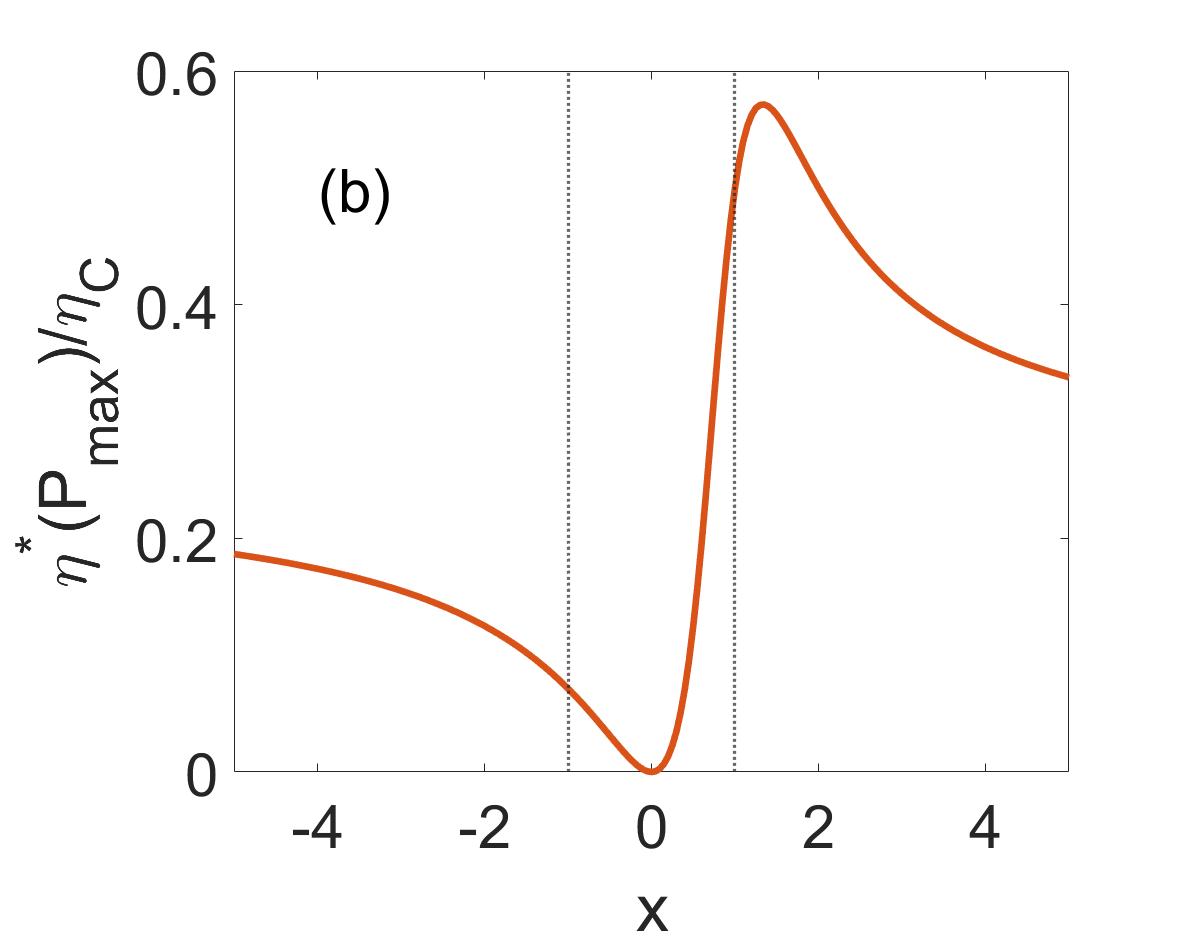}
    \caption{Reproduced from \cite{ref18}. Upper bound on (a) maximum efficiency, $\eta_{max}^*$ and (b) efficiency at maximum power, $\eta^*(P_{max})$ in units of Carnot efficiency $\eta_C$ as a function of asymmetry parameter $x$ for a three-terminal heat engine with broken time-reversal symmetry. The vertical dotted lines denote $|x|=1$.}
    \label{fig:nmaxb3t}
\end{figure}
Figure \ref{fig:nmaxb3t} shows the plots for the upper bound on maximum efficiency and efficiency at maximum power for a three-terminal heat engine with broken time-reversal symmetric. We notice that the Carnot efficiency can be reached only at $x=1$ and the maximum efficiency decays rapidly as the asymmetry parameter deviates from its symmetric value $x=1$. For efficiency at maximum power, the Curzon-Ahlborn limit $\eta_{CA}=\eta_C/2$ is reached for $x=1$ and this limit can be overcome for $x$ values slightly larger than 1.
In the limit $x\rightarrow\infty$, maximum efficiency and efficiency at maximum power, both reach $\frac{\eta_C}{4}$. This result implies that from the maximal attainable efficiency perspective, the reversible currents' thermodynamic cost is larger than the benefit they bring.

\subsection{Minimally Nonlinear Heat Engine}
\begin{figure}[t!]
    \centering
    \includegraphics[width=0.6\linewidth]{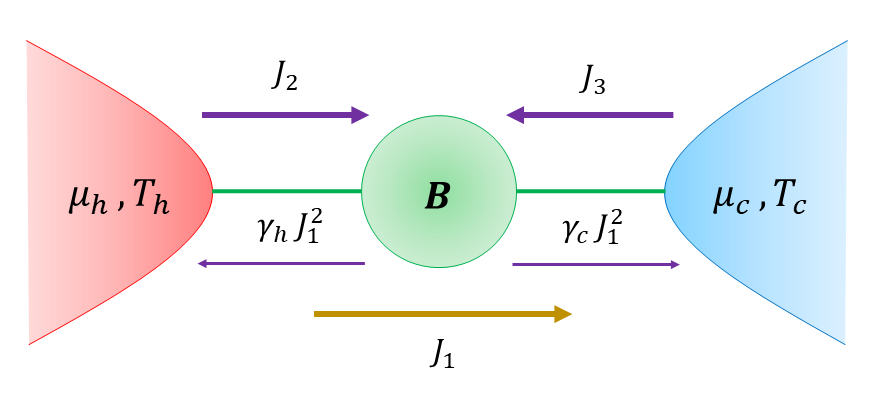}
    \caption{Schematic of a minimally nonlinear irreversible heat engine with broken time-reversal symmetry. The system is coupled with two reservoirs maintained at different temperatures $T_h>T_c$ and chemical potentials $\mu_h<\mu_c$. The figure shows the dissipation to the hot and cold reservoirs as $\gamma_h J_1^2$ and $\gamma_c J_1^2$, respectively. Here $\gamma_h$ and $\gamma_c$ represent the strength of dissipation to the hot and cold reservoirs, respectively.}
    \label{mnl_model}
\end{figure}

The heat dissipation effects at the nanoscale are inevitable and can influence a thermoelectric device's performance. It is necessary to investigate these effects to enhance the performance of the nanoscale thermoelectric heat engine. Izumida and Okuda proposed a minimally nonlinear irreversible model that accounts for the possible thermal dissipation effects due to the interaction between the system and the reservoirs \cite{Izumida1, Izumida2}. This model incorporates the Onsager relations with an additional second-order dissipation term and is thus called the extended Onsager relations. Let us redefine the linear Onsager relations between the thermodynamic fluxes and forces for a two-terminal system as follows:
\begin{equation}
    J_1=L_{11}X_1+L_{12}X_2,
\end{equation}
\begin{equation}\label{heat_current}
    J_2=L_{21}X_1+L_{22}X_2,
\end{equation}
where $J_1$ and $J_2$ are the particles and the heat currents, respectively. $L_{ij}$ are the Onsager coefficients satisfying the reciprocal relation $L_{12}({\bf B})=L_{21}(-{\bf B})$. The thermodynamic forces $X_1=\Delta\mu/T_c$ and $X_2=1/T_c-1/T_h$ ($\approx\Delta T/T^2$ for $T_c\approx T_h\approx T$ in the linear response regime). Here $\Delta\mu=\mu_h-\mu_c<0$ and $\Delta T=T_h-T_c>0$. \\
\indent Izumida {\it et al.} \cite{Izumida1, Izumida2} propose a minimally nonlinear irreversible heat engine model to incorporate dissipation into the system. A nonlinear dissipation term $-\gamma_hJ_1^2$ in the Onsager relation for heat current in Eq. (\ref{heat_current}) and the new expressions are called the extended Onsager relations \cite{Izumida1}:
\begin{equation}\label{mnl_particle}
    J_1=L_{11}X_1+L_{12}X_2,
\end{equation}
\begin{equation}\label{mnl_heat}
    J_2=L_{21}X_1+L_{22}X_2-\gamma_hJ_1^2.
\end{equation}
Here $\gamma_h$ is the strength of dissipation to the hot reservoir and is assumed to be positive constant $\gamma_h>0$. Unlike the linear response regime, the thermodynamic forces $X_1$ and $X_2$ are not restricted to small values. Also, the dissipation is assumed to be weak so the constraints on the Onsager coefficients still hold for the extended Onsager relations. The heat engine described by Eq. (\ref{mnl_particle}) and Eq. (\ref{mnl_heat}) is called a minimally nonlinear heat engine, since we assume that no other higher-order nonlinear terms contribute to the entropy production. A schematic model of a two-terminal minimally nonlinear heat engine is shown in Fig. \ref{mnl_model}.\\
For a fixed value of $X_2$, $J_1$ and $X_1$ are uniquely related through Eq. (\ref{mnl_particle}). So, one can choose $J_1$ as the control parameter instead of $X_1$. The heat current from the cold reservoir, $J_3=P-J_2$, where $P=-J_1X_1T_c>0$ is the output power of the heat engine. So, we can write $J_2$ and $J_3$ in terms of $J_1$ for a fixed $X_2$ as follows:
\begin{equation}\label{J2}
    J_2=\frac{L_{21}}{L_{11}}J_1+\bigg(\frac{L_{11}L_{22}-L_{12}L_{21}}{L_{11}}\bigg)X_2-\gamma_hJ_1^2,
\end{equation}
\begin{equation}\label{J3}
    J_3=\frac{L_{12}\eta_C-L_{21}}{L_{11}}J_1-\bigg(\frac{L_{11}L_{22}-L_{12}L_{21}}{L_{11}}\bigg)X_2-\gamma_cJ_1^2.
\end{equation}
Here $\eta_C=1-T_c/T_h=X_2T_c$ is the Carnot efficiency. The dissipation strength to the cold reservoir denoted by $\gamma_c$ is defined as
\begin{equation}
    \gamma_c=\frac{T_c}{L_{11}}-\gamma_h>0.
\end{equation}
Here $\gamma_h$ and $\gamma_c$ are assumed to be positive constants. The dissipation terms $\gamma_hJ_1^2$ and $\gamma_cJ_1^2$ turn out to be the inevitable power loss terms due to the finite time operation of the heat engine which converted into heat and then transferred to the hot and cold reservoir, respectively. A dimensionless parameter $\alpha=1/(1+\gamma_c/\gamma_h)$ representing the dissipation ratio is defined in Ref. \cite{entropy_Liu} to characterize the extent of dissipation to the respective reservoirs. In the asymmetric dissipation limit $\gamma_c/\gamma_h\to\infty$ and $\gamma_c/\gamma_h\to 0$, we have $\alpha=0$ and $\alpha=1$, respectively. Thus, $0<\alpha<1$ and $\alpha=0$ (i.e., $\gamma_h=0$) recovers the linear response relations.\\
\indent
In the minimally nonlinear irreversible heat engines, the dissipation is assumed to be weak, thus the bounds on the Onsager coefficients due to the positivity of the entropy production rate obtained in Eq. (\ref{Onsager_bound1}) and Eq. (\ref{Onsager_bound2}) for still hold as mentioned in Ref. \cite{Izumida1, Izumida2, Long1, Long2, Iyyappan1, Ponmurugan, Bai, Zhang2}. However, Liu {\it et al.} in Ref. \cite{entropy_Liu} argued that direct use Eq. (\ref{Onsager_bound2}) would yield nonphysical, negative entropy production rate in the nonlinear response regime $\alpha\ne 0$, since the nonlinear dissipative terms $\gamma_hJ_1^2$ and $\gamma_cJ_1^2$ also contribute to the entropy production rate, $\dot{\sigma}=J_1X_1+J_2X_2$ as
\begin{equation}\label{entropy_rate}
  \dot{\sigma}=\frac{(L_{21}-L_{12})X_2}{L_{11}}J_1+\bigg(\frac{L_{11}L_{22}-L_{12}L_{21}}{L_{11}}\bigg)X_2^2+\frac{\gamma_h}{T_h}J_1^2+\frac{\gamma_c}{T_c}J_1^2.
\end{equation}
Using Eq. (\ref{entropy_rate}), Liu {\it et al.} obtained the bounds on the Onsager coefficients in the minimally nonlinear regime as follows \cite{entropy_Liu}:
\begin{equation}\label{newbound}
  L_{11}\ge0,\,\ L_{22}\ge0,\,\ L_{11}L_{22}-L_{11}L_{22}\alpha\eta_C-(L_{12}+L_{21})^2/4+L_{12}L_{21}\alpha\eta_C\ge0.
\end{equation}
The performance of a thermoelectric heat engine is characterized by the output power and the heat-to-work conversion efficiency. The output power can be expressed in terms of $J_1$ as
\begin{equation}\label{power}
    P=\frac{L_{12}}{L_{11}}\eta_CJ_1-\frac{T_c}{L_{11}}J_1^2.
\end{equation}
One can notice that the dissipative terms in Eq. (\ref{J2}) and Eq. (\ref{J3}) do not contribute to the output power at all. The first term in Eq. (\ref{power}) describes the power generation by the heat engine due to the temperature difference, whereas the second term denotes the inevitable power dissipation due to the finite time operation of the heat engine. The second term can also be seen as the effect of Joule heating if we consider $T_c/L_{11}$ and $J_1$ as resistance and electric current, respectively \cite{Izumida1, Izumida2}.\\
\indent The heat-to-work conversion efficiency at a given output power for the minimally nonlinear heat engine is defined as follows:
\begin{equation}\label{effc}
  \eta=\frac{P}{J_2}=\frac{L_{12}\eta_CJ_1-T_cJ_1^2}{L_{21}J_1+(L_{11}L_{22}-L_{12}L_{21})X_2-\alpha T_cJ_1^2}.
\end{equation}
We notice that the dissipative term contributes to the denominator of the efficiency and by increasing the dissipation strength $\alpha$ one can enhance the efficiency of the heat engine as compared to the linear response case ($\alpha=0$).\\
\indent As discussed earlier, the time-reversal symmetry of the system is broken in the presence of a magnetic field and the Onsager coefficients satisfy the reciprocal relation also called as Onsager-Casimir relation defined in Eq. (\ref{Casimir_reln}). With broken time-reversal symmetry, the performance of a heat engine is characterized by the asymmetry parameter $x$ and the figure of merit $y$ defined as follows:
\begin{equation}\label{xy}
  x=\frac{L_{12}}{L_{21}},\,\ y=\frac{L_{12}L_{21}}{L_{11}L_{22}-L_{12}L_{21}}.
\end{equation}
There is no restriction on the attainable values of asymmetry parameter $x$. However, the third inequality in Eq. (\ref{newbound}) impose an upper bound the values of $y$ given as follows \cite{entropy_Liu}:
\begin{equation}\label{yupper}
  \begin{aligned}
    h(x)\le y\le0,\,\ \mathrm{for}\,\ x\le0,\\
    0\le y\le h(x),\,\ \mathrm{for}\,\ x\ge0,
  \end{aligned}
\end{equation}
where the bound function $h(x)$ is defined as \cite{entropy_Liu}
\begin{equation}\label{hx}
  h(x)=\frac{4(1-\alpha\eta_C)x}{(x-1)^2}.
\end{equation}
For the time-reversal symmetric case ($x=1$), $h(x)\to\infty$, thus $y=ZT$ is called the dimensionless figure of merit and there is no upper bound on the attainable value of $ZT$.
Although most of the papers discussed the minimally nonlinear irreversible heat engine for the time-reversal symmetric case \cite{Izumida1, Izumida2, Long1, Long2, Iyyappan1, Ponmurugan, Bai}, two recent papers \cite{Zhang2, entropy_Liu} discussed the optimization of the minimally nonlinear heat engine in the broken time-reversal symmetry case.\\
\indent To optimize the performance of the heat engine, it is necessary to investigate the maximum output power, the efficiency at maximum power, and the maximum efficiency generated by it. The maximum output power can be obtained by setting $\partial P/\partial J_1=0$, then we have
\begin{equation}\label{xpmax}
  J_1^*=\frac{L_{12}\eta_C}{2T_c},\,\ X_1^*=-\frac{L_{12}X_2}{2L_{11}}.
\end{equation}
Using the above expression, the maximum output power is obtained as \cite{Zhang2, entropy_Liu}
\begin{equation}\label{pmax}
  P_{max}=\frac{\eta_CL_{12}^2}{4L_{11}}X_2.
\end{equation}
Substituting Eq. (\ref{xpmax}) in Eq. (\ref{effc}), we obtain the efficiency at maximum power in terms of $x$ and $y$ as \cite{entropy_Liu}
\begin{equation}\label{npmax}
  \eta(P_{max})=\eta_C\frac{xy}{4+y(2-\alpha\eta_Cx)}.
\end{equation}
The upper bound on efficiency at maximum output power is achieved at $y=h(x)$ \cite{entropy_Liu},
\begin{equation}\label{npmaxb}
  \eta^*(P_{max})=\eta_C\frac{1-\alpha\eta_C}{(\alpha\eta_C-x^{-1})^2-\alpha\eta_C+1}.
\end{equation}
For $x=1$, the upper bound on efficiency at maximum power can be written as
\begin{equation}\label{npmaxbx1}
  \eta^*(P_{max})=\frac{\eta_C}{2-\alpha\eta_C}.
\end{equation}
For $0\le\alpha\le1$, Eq. (\ref{npmaxbx1}) lies between $\eta_C/2\le\eta^*(P_{max})\le\eta_C/(2-\eta_C)$. We observe that the Curzon-Ahlborn (CA) limit $\eta^*(P_{max})=\eta_{CA}=\eta_C/2$ as discussed in Ref. \cite{Broeck} is no longer the upper bound for efficiency at maximum power when nonlinearity is introduced to the system. Equation (\ref{npmaxbx1}) is also obtained in \cite{Izumida1, Izumida2, Long2, Iyyappan1} and the validity of this model is examined with the low dissipation Carnot engine in the tight-coupling limit (i.e., $det({\bf L})=0$) \cite{lowCarnot}. The broken time-reversal symmetry with $x\ne1$ can further help in overcoming the CA efficiency and the Carnot efficiency can be reached, $\eta^*(P_{max})=\eta_C$ when $x\to\infty$ \cite{ref6, entropy_Liu}. Liu {\it et al.} show that for $x>0$, the efficiency at maximum power reaches the Carnot efficiency faster in the nonlinear regime due to the presence of more fluctuations than in the linear response regime. The nonlinearity enhances the performance of a heat engine by enhancing the efficiency at maximum power. \\
\indent The optimization of the performance of the heat engine in the minimally nonlinear regime is done by investing the relative power gain and relative efficiency gain by the heat engine in Ref. \cite{Long2, Zhang2}. Zhang {\it et al.} also observe that the nonlinear dissipative term enhances the performance of the heat engine. They discuss that the efficiency increases with increasing the dissipation strength for the lower voltage bias, however, the efficiency reduces for higher dissipation when we move towards high voltage bias.\\
\indent The minimally nonlinear heat engine can be investigated for an Aharonov-Bohm interferometer that can incorporate broken time-reversal symmetry in the system. Further, the minimally nonlinear heat engine model can be extended to a three-terminal model. A third terminal can be a Buttiker probe which introduces inelastic scattering effects into the system. A three-terminal system having the third terminal as the Buttiker probe with broken time-reversal symmetry is discussed in the linear response regime \cite{Saito, balachandran2013, ref18, RongZhang, Zahra}. One can investigate the minimally nonlinear model for the three-terminal system and investigate how the upper bound on efficiency at maximum power is modified for the three-terminal case. Enhancing the efficiency and power output of minimally nonlinear heat engines using coherent control is an open question, and the interplay between minimal nonlinearity and that induced by inelastic scattering beyond the linear regime remains less explored.

\section{Conclusions and Outlook}
Engineering the transmission function through quantum interference effects within quantum-dot networks, and between these networks and the continuum, is a promising approach for enhancing efficiency and power output. This review presents a scattering theory formalism that relates transmission features to transport coefficients. Examples discussed include quantum coherent control in thermoelectric heat engines using triple-dot Aharonov-Bohm interferometers and Mach-Zehnder interferometers. The review also covers coherent control of heat currents in quantum-dot Aharonov-Bohm interferometers and explores its applications in designing nanoscale devices.
We also reviewed the conditions for achieving even, odd, and combined quantum walks within Green's function formalism that may be useful to tailor maximal constructive quantum interference effects favoring enhanced efficiency and power.
The review also addresses the effects of scattering, providing examples of diode behavior under broken time-reversal symmetry and decoherence. We presented an overview of B{\"u}ttiker probe methods to incorporate various phase-breaking processes in otherwise coherent transport. Within this formalism, we discussed the coherent diode behavior and presented sufficient conditions for it.
We further extended our review to include bounds on efficiency and power in systems with and without broken time-reversal symmetric systems.
Designing sharp features in transmission by tailoring quantum transport pathways and or breaking time-reversal symmetry using probe terminals offers a promising route to enhance the efficiency of thermoelectric heat engines.

 We provided an overview of quantum coherent control in quantum-dot Aharonov-Bohm networks and discussed recent advancements in enhancing power efficiency through interference effects. We introduced another paradigm of breaking time-reversal symmetry within a three-terminal setup to enhance efficiency and power. Most works follow either of these methods to explore thermoelectric transport. However, could combining these two approaches offer a promising platform for designing efficient thermoelectric engines? For instance, by introducing a probe reservoir and engineering sharp Fano-type transmission from the source and drain terminals to the probe terminal, could we achieve higher power output? By exploiting the interplay of Joule heating, scattering by the probe, and quantum interference effects within the conductor, might we enhance power efficiency in deeply nonlinear regime?

The topology of the quantum dot network is crucial to engineer optimal transmissions. In the case of the molecular graphs, it has been found that the quantum interference effects are sensitive to the position of the source-drain terminals.
Given the quantum interference effects arising from the presence of both the even and the odd-length contributions, can thermoelectric effects in such networks be optimized by designing graphs that support solely constructive interference? Would enhancing specific walk lengths through graph design lead to improved overall transmission and performance?
The quantum coherent control of nonlinear thermoelectric effects using non-Euclidean graphs is a promising future direction. Non-Euclidean graphs can exhibit various features, including knots and twists. These lattices can be constructed using superconducting qubits within a circuit QED framework, providing the flexibility to design artificial structures similar to graphene. Creating a lattice structure through a network of interconnected dots presents significant challenges. However, this approach could lead to configurations that optimize power efficiency and potentially suppress phonon effects by leveraging topological phenomena.

In conclusion, quantum coherent control of nonlinear thermoelectricity presents compelling opportunities for designing artificial quantum-dot networks with tailored transmission functions. These functions arise from complex interference patterns, which are particularly evident in structures such as triple quantum dots. Our analysis suggests that optimal efficiency and power output can be achieved when the inter-dot tunneling strength is comparable to the dot-lead hybridization. This insight highlights the potential of leveraging complex networks and graph-based structures as highly efficient thermoelectric heat engines.

Moreover, the introduction of a multi-terminal configuration could provide additional asymmetry parameters, potentially leading to further performance optimization. Notably, the non-reciprocal outcomes observed when measurements are introduced in otherwise coherent conductors can be harnessed to enhance thermoelectric heat engine designs. These findings underscore the significant role that the delicate interplay between coherence and decoherence can play in enhancing energy transfer in such nanoscale systems.
\section*{Acknowledgements}
J.B. acknowledges the financial support from IIT Bhubaneswar. M.B. acknowledges support from the Science and Engineering Research Board (SERB), India under SERB-MATRICS scheme Grant No. MTR/2021/000566, and SERB-CRG scheme Grant No. CRG/2020/001768.
\appendix
\section{Appendix A: Aharonov-Bohm Effect}
Since our review paper is based on the Aharonov-Bohm effect, a brief detail on this effect is described as follows. The AB effect manifests as a fundamental phenomenon in quantum mechanics, where a magnetic vector potential influences the phase of a charged particle's wavefunction, even in regions where the magnetic field is zero \cite{aharonov1959}. This influence becomes evident through interference effects.
\begin{figure}[h] 
  \centering
  \includegraphics[width=0.7\textwidth]{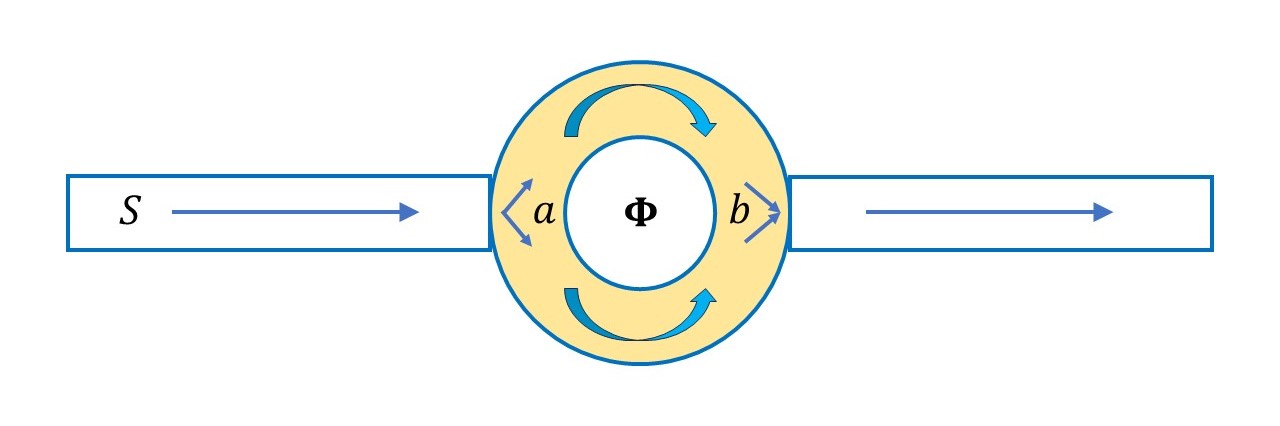}
  \caption{Aharonov-Bohm interferometer with magnetic flux $\Phi$.}
  \label{fig:AB}
\end{figure}

Consider the schematic depiction in Fig. \ref{fig:AB}, illustrating an AB interferometer setup, where $S$ denotes an electron source and the bold arrows delineate two distinct paths. A uniform magnetic field $\mathbf{B}$, perpendicular to the plane of the interferometer, is introduced, potentially generated by an infinitely long thin cylinder. In this setup, the vector potential $\mathbf{A}(r)$ takes the form:
\begin{equation}
\mathbf{A}(r) =
\begin{cases}
Br\hat{\phi} & : r < R, \\
\frac{BR^2}{2r}\hat{\phi} & : r > R.
\end{cases}
\end{equation}
Here, $B$ represents the magnetic field's magnitude, $\hat{\phi}$ denotes a unit vector along the $z$-axis, $R$ signifies the cylinder's radius, and $r$ stands for the radial coordinate. When an electron traverses the lower arm, it aligns with the vector potential's direction, while in the upper arm, it moves in opposition. Consequently, the two paths acquire opposite phases, leading to a discernible phase difference. The phase acquired along a given path from $r_a$ to $r_b$ (designated as points ``$a$" and ``$b$" in Fig. \ref{fig:AB}) is provided by:
\begin{equation}
\phi = \oint_{r_a}^{r_b} \mathbf{A}(r) \cdot d\mathbf{r}.
\end{equation}
If the two paths enclose an area $S$, the resultant net phase difference $\Delta \phi$ is expressed as:
\begin{equation}
\Delta \phi = 2\pi \oint_{S} \mathbf{B} \cdot d\mathbf{S} = 2\pi \frac{\Phi}{\Phi_0},
\end{equation}
where $\mathbf{B} = \nabla \times \mathbf{A}$. The second integral is derived from Stokes’ theorem, and $\Phi = \oint_{S} \mathbf{B}(r') \cdot d\mathbf{S} = BS$ with $\Phi_0 = \frac{h}{e}$. The phase difference remains invariant irrespective of the chosen gauge for $\mathbf{A}$.\\
\bibliography{Review}

\begin{thebibliography}{137}%
\makeatletter
\providecommand \@ifxundefined [1]{%
 \@ifx{#1\undefined}
}%
\providecommand \@ifnum [1]{%
 \ifnum #1\expandafter \@firstoftwo
 \else \expandafter \@secondoftwo
 \fi
}%
\providecommand \@ifx [1]{%
 \ifx #1\expandafter \@firstoftwo
 \else \expandafter \@secondoftwo
 \fi
}%
\providecommand \natexlab [1]{#1}%
\providecommand \enquote  [1]{``#1''}%
\providecommand \bibnamefont  [1]{#1}%
\providecommand \bibfnamefont [1]{#1}%
\providecommand \citenamefont [1]{#1}%
\providecommand \href@noop [0]{\@secondoftwo}%
\providecommand \href [0]{\begingroup \@sanitize@url \@href}%
\providecommand \@href[1]{\@@startlink{#1}\@@href}%
\providecommand \@@href[1]{\endgroup#1\@@endlink}%
\providecommand \@sanitize@url [0]{\catcode `\\12\catcode `\$12\catcode
  `\&12\catcode `\#12\catcode `\^12\catcode `\_12\catcode `\%12\relax}%
\providecommand \@@startlink[1]{}%
\providecommand \@@endlink[0]{}%
\providecommand \url  [0]{\begingroup\@sanitize@url \@url }%
\providecommand \@url [1]{\endgroup\@href {#1}{\urlprefix }}%
\providecommand \urlprefix  [0]{URL }%
\providecommand \Eprint [0]{\href }%
\providecommand \doibase [0]{https://doi.org/}%
\providecommand \selectlanguage [0]{\@gobble}%
\providecommand \bibinfo  [0]{\@secondoftwo}%
\providecommand \bibfield  [0]{\@secondoftwo}%
\providecommand \translation [1]{[#1]}%
\providecommand \BibitemOpen [0]{}%
\providecommand \bibitemStop [0]{}%
\providecommand \bibitemNoStop [0]{.\EOS\space}%
\providecommand \EOS [0]{\spacefactor3000\relax}%
\providecommand \BibitemShut  [1]{\csname bibitem#1\endcsname}%
\let\auto@bib@innerbib\@empty
\bibitem [{\citenamefont {Haack}\ and\ \citenamefont
  {Giazotto}(2019)}]{study1}%
  \BibitemOpen
  \bibfield  {author} {\bibinfo {author} {\bibfnamefont {G.}~\bibnamefont
  {Haack}}\ and\ \bibinfo {author} {\bibfnamefont {F.}~\bibnamefont
  {Giazotto}},\ }\bibfield  {title} {\bibinfo {title} {Efficient and tunable
  aharonov-bohm quantum heat engine},\ }\href@noop {} {\bibfield  {journal}
  {\bibinfo  {journal} {Phys. Rev. B}\ }\textbf {\bibinfo {volume} {100}},\
  \bibinfo {pages} {235442} (\bibinfo {year} {2019})}\BibitemShut {NoStop}%
\bibitem [{\citenamefont {Thingna}\ \emph {et~al.}(2020)\citenamefont
  {Thingna}, \citenamefont {Manzano},\ and\ \citenamefont {Cao}}]{study2}%
  \BibitemOpen
  \bibfield  {author} {\bibinfo {author} {\bibfnamefont {J.}~\bibnamefont
  {Thingna}}, \bibinfo {author} {\bibfnamefont {D.}~\bibnamefont {Manzano}},\
  and\ \bibinfo {author} {\bibfnamefont {J.}~\bibnamefont {Cao}},\ }\bibfield
  {title} {\bibinfo {title} {Magnetic field induced symmetry breaking in
  nonequilibrium quantum networks},\ }\href@noop {} {\bibfield  {journal}
  {\bibinfo  {journal} {New J. Phys.}\ }\textbf {\bibinfo {volume} {22}},\
  \bibinfo {pages} {083026} (\bibinfo {year} {2020})}\BibitemShut {NoStop}%
\bibitem [{\citenamefont {Samuelsson}\ \emph {et~al.}(2017)\citenamefont
  {Samuelsson}, \citenamefont {Kheradsoud},\ and\ \citenamefont
  {Sothmann}}]{study3}%
  \BibitemOpen
  \bibfield  {author} {\bibinfo {author} {\bibfnamefont {P.}~\bibnamefont
  {Samuelsson}}, \bibinfo {author} {\bibfnamefont {S.}~\bibnamefont
  {Kheradsoud}},\ and\ \bibinfo {author} {\bibfnamefont {B.}~\bibnamefont
  {Sothmann}},\ }\bibfield  {title} {\bibinfo {title} {Optimal quantum
  interference thermoelectric heat engine with edge states},\ }\href
  {https://doi.org/10.1103/PhysRevLett.118.256801} {\bibfield  {journal}
  {\bibinfo  {journal} {Phys. Rev. Lett.}\ }\textbf {\bibinfo {volume} {118}},\
  \bibinfo {pages} {256801} (\bibinfo {year} {2017})}\BibitemShut {NoStop}%
\bibitem [{\citenamefont {Taniguchi}(2020)}]{study4}%
  \BibitemOpen
  \bibfield  {author} {\bibinfo {author} {\bibfnamefont {N.}~\bibnamefont
  {Taniguchi}},\ }\bibfield  {title} {\bibinfo {title} {Quantum control of
  nonlinear thermoelectricity at the nanoscale},\ }\href@noop {} {\bibfield
  {journal} {\bibinfo  {journal} {Phys. Rev. B}\ }\textbf {\bibinfo {volume}
  {101}},\ \bibinfo {pages} {115404} (\bibinfo {year} {2020})}\BibitemShut
  {NoStop}%
\bibitem [{\citenamefont {Jasleen~Kaur}(2022)}]{study5}%
  \BibitemOpen
  \bibfield  {author} {\bibinfo {author} {\bibfnamefont {M.~B.}\ \bibnamefont
  {Jasleen~Kaur}, \bibfnamefont {Aritra~Ghosh}},\ }\bibfield  {title} {\bibinfo
  {title} {Quantum counterpart of energy equipartition theorem for fermionic
  systems},\ }\href@noop {} {\bibfield  {journal} {\bibinfo  {journal} {Journal
  of Statistical Mechanics: Theory and Experiment}\ }\textbf {\bibinfo {volume}
  {2022}},\ \bibinfo {pages} {053105} (\bibinfo {year} {2022})}\BibitemShut
  {NoStop}%
\bibitem [{\citenamefont {Asam~Rajesh}(2015)}]{study6}%
  \BibitemOpen
  \bibfield  {author} {\bibinfo {author} {\bibfnamefont {M.~B.}\ \bibnamefont
  {Asam~Rajesh}},\ }\href@noop {} {\bibfield  {journal} {\bibinfo  {journal}
  {Phys. Rev. A}\ }\textbf {\bibinfo {volume} {92}},\ \bibinfo {pages} {012105}
  (\bibinfo {year} {2015})}\BibitemShut {NoStop}%
\bibitem [{\citenamefont {Jasleen~Kaur}\ and\ \citenamefont
  {Bandyopadhyay}(2023)}]{study7}%
  \BibitemOpen
  \bibfield  {author} {\bibinfo {author} {\bibfnamefont {A.~G.}\ \bibnamefont
  {Jasleen~Kaur}}\ and\ \bibinfo {author} {\bibfnamefont {M.}~\bibnamefont
  {Bandyopadhyay}},\ }\bibfield  {title} {\bibinfo {title} {Partition of
  kinetic energy and magnetic moment in dissipative diamagnetism},\ }\href@noop
  {} {\bibfield  {journal} {\bibinfo  {journal} {Physica A :Statistical
  Mechanics and its Applications}\ }\textbf {\bibinfo {volume} {625}},\
  \bibinfo {pages} {128993} (\bibinfo {year} {2023})}\BibitemShut {NoStop}%
\bibitem [{\citenamefont {Svilans}\ \emph {et~al.}(2016)\citenamefont
  {Svilans}, \citenamefont {Leijnse},\ and\ \citenamefont {Linke}}]{expt1}%
  \BibitemOpen
  \bibfield  {author} {\bibinfo {author} {\bibfnamefont {A.}~\bibnamefont
  {Svilans}}, \bibinfo {author} {\bibfnamefont {M.}~\bibnamefont {Leijnse}},\
  and\ \bibinfo {author} {\bibfnamefont {H.}~\bibnamefont {Linke}},\ }\bibfield
   {title} {\bibinfo {title} {Experiments on the thermoelectric properties of
  quantum dots},\ }\href
  {https://doi.org/https://doi.org/10.1016/j.crhy.2016.08.002} {\bibfield
  {journal} {\bibinfo  {journal} {Comptes Rendus Physique}\ }\textbf {\bibinfo
  {volume} {17}},\ \bibinfo {pages} {1096} (\bibinfo {year}
  {2016})}\BibitemShut {NoStop}%
\bibitem [{\citenamefont {Josefsson}\ \emph {et~al.}(2018)\citenamefont
  {Josefsson}, \citenamefont {Svilans}, \citenamefont {Burke}, \citenamefont
  {Hoffmann}, \citenamefont {Fahlvik}, \citenamefont {Thelander}, \citenamefont
  {Leijnse},\ and\ \citenamefont {Linke}}]{expt2}%
  \BibitemOpen
  \bibfield  {author} {\bibinfo {author} {\bibfnamefont {M.}~\bibnamefont
  {Josefsson}}, \bibinfo {author} {\bibfnamefont {A.}~\bibnamefont {Svilans}},
  \bibinfo {author} {\bibfnamefont {A.~M.}\ \bibnamefont {Burke}}, \bibinfo
  {author} {\bibfnamefont {E.~A.}\ \bibnamefont {Hoffmann}}, \bibinfo {author}
  {\bibfnamefont {S.}~\bibnamefont {Fahlvik}}, \bibinfo {author} {\bibfnamefont
  {C.}~\bibnamefont {Thelander}}, \bibinfo {author} {\bibfnamefont
  {M.}~\bibnamefont {Leijnse}},\ and\ \bibinfo {author} {\bibfnamefont
  {H.}~\bibnamefont {Linke}},\ }\bibfield  {title} {\bibinfo {title} {A
  quantum-dot heat engine operating close to the thermodynamic efficiency
  limits},\ }\href@noop {} {\bibfield  {journal} {\bibinfo  {journal} {Nature
  Nanotechnology}\ }\textbf {\bibinfo {volume} {13}},\ \bibinfo {pages} {920}
  (\bibinfo {year} {2018})}\BibitemShut {NoStop}%
\bibitem [{\citenamefont {Dorsch}\ \emph {et~al.}(2021)\citenamefont {Dorsch},
  \citenamefont {Svilans}, \citenamefont {Josefsson}, \citenamefont
  {Goldozian}, \citenamefont {Kumar}, \citenamefont {Thelander}, \citenamefont
  {Wacker},\ and\ \citenamefont {Burke}}]{expt3}%
  \BibitemOpen
  \bibfield  {author} {\bibinfo {author} {\bibfnamefont {S.}~\bibnamefont
  {Dorsch}}, \bibinfo {author} {\bibfnamefont {A.}~\bibnamefont {Svilans}},
  \bibinfo {author} {\bibfnamefont {M.}~\bibnamefont {Josefsson}}, \bibinfo
  {author} {\bibfnamefont {B.}~\bibnamefont {Goldozian}}, \bibinfo {author}
  {\bibfnamefont {M.}~\bibnamefont {Kumar}}, \bibinfo {author} {\bibfnamefont
  {C.}~\bibnamefont {Thelander}}, \bibinfo {author} {\bibfnamefont
  {A.}~\bibnamefont {Wacker}},\ and\ \bibinfo {author} {\bibfnamefont
  {A.}~\bibnamefont {Burke}},\ }\bibfield  {title} {\bibinfo {title} {Heat
  driven transport in serial double quantum dot devices},\ }\href@noop {}
  {\bibfield  {journal} {\bibinfo  {journal} {Nano Lett.}\ }\textbf {\bibinfo
  {volume} {21}},\ \bibinfo {pages} {988} (\bibinfo {year} {2021})}\BibitemShut
  {NoStop}%
\bibitem [{\citenamefont {Kobayashi}\ \emph {et~al.}(2002)\citenamefont
  {Kobayashi}, \citenamefont {Aikawa}, \citenamefont {Katsumoto},\ and\
  \citenamefont {Iye}}]{AB1}%
  \BibitemOpen
  \bibfield  {author} {\bibinfo {author} {\bibfnamefont {K.}~\bibnamefont
  {Kobayashi}}, \bibinfo {author} {\bibfnamefont {H.}~\bibnamefont {Aikawa}},
  \bibinfo {author} {\bibfnamefont {S.}~\bibnamefont {Katsumoto}},\ and\
  \bibinfo {author} {\bibfnamefont {Y.}~\bibnamefont {Iye}},\ }\bibfield
  {title} {\bibinfo {title} {Tuning of the fano effect through a quantum dot in
  an aharonov-bohm interferometer},\ }\href@noop {} {\bibfield  {journal}
  {\bibinfo  {journal} {Phys. Rev. Lett.}\ }\textbf {\bibinfo {volume} {88}},\
  \bibinfo {pages} {256806} (\bibinfo {year} {2002})}\BibitemShut {NoStop}%
\bibitem [{\citenamefont {Kang}\ and\ \citenamefont {Cho}(2004)}]{AB2}%
  \BibitemOpen
  \bibfield  {author} {\bibinfo {author} {\bibfnamefont {K.}~\bibnamefont
  {Kang}}\ and\ \bibinfo {author} {\bibfnamefont {S.~Y.}\ \bibnamefont {Cho}},\
  }\bibfield  {title} {\bibinfo {title} {Tunable molecular resonances of a
  double quantum dot aharonov–bohm interferometer},\ }\href@noop {}
  {\bibfield  {journal} {\bibinfo  {journal} {J. Phys. Cond. Matt.}\ }\textbf
  {\bibinfo {volume} {16}},\ \bibinfo {pages} {117} (\bibinfo {year}
  {2004})}\BibitemShut {NoStop}%
\bibitem [{\citenamefont {Gomez-Silva}\ \emph {et~al.}(2012)\citenamefont
  {Gomez-Silva}, \citenamefont {Ávalos Ovando}, \citenamefont {de~Guevara},\
  and\ \citenamefont {Orellana}}]{AB3}%
  \BibitemOpen
  \bibfield  {author} {\bibinfo {author} {\bibfnamefont {G.}~\bibnamefont
  {Gomez-Silva}}, \bibinfo {author} {\bibfnamefont {O.}~\bibnamefont {Ávalos
  Ovando}}, \bibinfo {author} {\bibfnamefont {M.~L.~L.}\ \bibnamefont
  {de~Guevara}},\ and\ \bibinfo {author} {\bibfnamefont {P.~A.}\ \bibnamefont
  {Orellana}},\ }\bibfield  {title} {\bibinfo {title} {Enhancement of
  thermoelectric efficiency and violation of the wiedemann-franz law due to
  fano effect},\ }\href@noop {} {\bibfield  {journal} {\bibinfo  {journal} {J.
  Appl. Phys.}\ }\textbf {\bibinfo {volume} {111}},\ \bibinfo {pages} {053704}
  (\bibinfo {year} {2012})}\BibitemShut {NoStop}%
\bibitem [{\citenamefont {Washburn}\ and\ \citenamefont
  {Webb}(1986)}]{washburn1986}%
  \BibitemOpen
  \bibfield  {author} {\bibinfo {author} {\bibfnamefont {S.}~\bibnamefont
  {Washburn}}\ and\ \bibinfo {author} {\bibfnamefont {R.}~\bibnamefont
  {Webb}},\ }\bibfield  {title} {\bibinfo {title} {Aharonov-bohm effect in
  disordered metals},\ }\href@noop {} {\bibfield  {journal} {\bibinfo
  {journal} {Advances in Physics}\ }\textbf {\bibinfo {volume} {35}},\ \bibinfo
  {pages} {375} (\bibinfo {year} {1986})}\BibitemShut {NoStop}%
\bibitem [{\citenamefont {Yacoby}\ \emph {et~al.}(1995)\citenamefont {Yacoby},
  \citenamefont {Heiblum}, \citenamefont {Mahalu},\ and\ \citenamefont
  {Shtrikman}}]{yacoby1995}%
  \BibitemOpen
  \bibfield  {author} {\bibinfo {author} {\bibfnamefont {A.}~\bibnamefont
  {Yacoby}}, \bibinfo {author} {\bibfnamefont {M.}~\bibnamefont {Heiblum}},
  \bibinfo {author} {\bibfnamefont {D.}~\bibnamefont {Mahalu}},\ and\ \bibinfo
  {author} {\bibfnamefont {H.}~\bibnamefont {Shtrikman}},\ }\bibfield  {title}
  {\bibinfo {title} {Coherence and phase sensitive measurements in a quantum
  dot},\ }\href@noop {} {\bibfield  {journal} {\bibinfo  {journal} {Phys. Rev.
  Lett.}\ }\textbf {\bibinfo {volume} {74}},\ \bibinfo {pages} {4047} (\bibinfo
  {year} {1995})}\BibitemShut {NoStop}%
\bibitem [{\citenamefont {Holleitner}\ \emph {et~al.}(2001)\citenamefont
  {Holleitner}, \citenamefont {Decker}, \citenamefont {Quin}, \citenamefont
  {Eberl},\ and\ \citenamefont {Blick}}]{holleitner2001}%
  \BibitemOpen
  \bibfield  {author} {\bibinfo {author} {\bibfnamefont {A.~W.}\ \bibnamefont
  {Holleitner}}, \bibinfo {author} {\bibfnamefont {C.~E.}\ \bibnamefont
  {Decker}}, \bibinfo {author} {\bibfnamefont {H.}~\bibnamefont {Quin}},
  \bibinfo {author} {\bibfnamefont {K.}~\bibnamefont {Eberl}},\ and\ \bibinfo
  {author} {\bibfnamefont {R.~H.}\ \bibnamefont {Blick}},\ }\bibfield  {title}
  {\bibinfo {title} {Dynamic phase coherent electron systems},\ }\href@noop {}
  {\bibfield  {journal} {\bibinfo  {journal} {Phys. Rev. Lett.}\ }\textbf
  {\bibinfo {volume} {87}},\ \bibinfo {pages} {256802} (\bibinfo {year}
  {2001})}\BibitemShut {NoStop}%
\bibitem [{\citenamefont {Verduijn}\ \emph {et~al.}(2013)\citenamefont
  {Verduijn}, \citenamefont {Agundez}, \citenamefont {Blaauboer},\ and\
  \citenamefont {Rogge}}]{verduijn2013}%
  \BibitemOpen
  \bibfield  {author} {\bibinfo {author} {\bibfnamefont {J.}~\bibnamefont
  {Verduijn}}, \bibinfo {author} {\bibfnamefont {R.}~\bibnamefont {Agundez}},
  \bibinfo {author} {\bibfnamefont {M.}~\bibnamefont {Blaauboer}},\ and\
  \bibinfo {author} {\bibfnamefont {S.}~\bibnamefont {Rogge}},\ }\bibfield
  {title} {\bibinfo {title} {Controlling quantum interference in aharonov-bohm
  rings with local electric fields},\ }\href@noop {} {\bibfield  {journal}
  {\bibinfo  {journal} {New J. Phys.}\ }\textbf {\bibinfo {volume} {15}},\
  \bibinfo {pages} {033020} (\bibinfo {year} {2013})}\BibitemShut {NoStop}%
\bibitem [{\citenamefont {Hackenbroich}(2001)}]{hackenbroich2001}%
  \BibitemOpen
  \bibfield  {author} {\bibinfo {author} {\bibfnamefont {G.}~\bibnamefont
  {Hackenbroich}},\ }\bibfield  {title} {\bibinfo {title} {Phase coherence in
  quantum dots: A review},\ }\href@noop {} {\bibfield  {journal} {\bibinfo
  {journal} {Phys. Rep.}\ }\textbf {\bibinfo {volume} {343}},\ \bibinfo {pages}
  {463} (\bibinfo {year} {2001})}\BibitemShut {NoStop}%
\bibitem [{\citenamefont {Schuster}\ \emph {et~al.}(1997)\citenamefont
  {Schuster}, \citenamefont {Buks}, \citenamefont {Heiblum}, \citenamefont
  {Mahalu}, \citenamefont {Umansky},\ and\ \citenamefont
  {Shtrikman}}]{schuster1997}%
  \BibitemOpen
  \bibfield  {author} {\bibinfo {author} {\bibfnamefont {R.}~\bibnamefont
  {Schuster}}, \bibinfo {author} {\bibfnamefont {E.}~\bibnamefont {Buks}},
  \bibinfo {author} {\bibfnamefont {M.}~\bibnamefont {Heiblum}}, \bibinfo
  {author} {\bibfnamefont {D.}~\bibnamefont {Mahalu}}, \bibinfo {author}
  {\bibfnamefont {V.}~\bibnamefont {Umansky}},\ and\ \bibinfo {author}
  {\bibfnamefont {H.}~\bibnamefont {Shtrikman}},\ }\bibfield  {title} {\bibinfo
  {title} {Phase measurement in a quantum dot via a double-slit interference
  experiment},\ }\href@noop {} {\bibfield  {journal} {\bibinfo  {journal}
  {Nature}\ }\textbf {\bibinfo {volume} {385}},\ \bibinfo {pages} {417}
  (\bibinfo {year} {1997})}\BibitemShut {NoStop}%
\bibitem [{\citenamefont {König}\ and\ \citenamefont
  {Gefen}(2001)}]{konig2001}%
  \BibitemOpen
  \bibfield  {author} {\bibinfo {author} {\bibfnamefont {J.}~\bibnamefont
  {König}}\ and\ \bibinfo {author} {\bibfnamefont {Y.}~\bibnamefont {Gefen}},\
  }\bibfield  {title} {\bibinfo {title} {Quantum interference and aharonov-bohm
  oscillations in strongly interacting quantum dots},\ }\href@noop {}
  {\bibfield  {journal} {\bibinfo  {journal} {Phys. Rev. Lett.}\ }\textbf
  {\bibinfo {volume} {86}},\ \bibinfo {pages} {3855} (\bibinfo {year}
  {2001})}\BibitemShut {NoStop}%
\bibitem [{\citenamefont {Li}\ \emph {et~al.}(2009{\natexlab{a}})\citenamefont
  {Li}, \citenamefont {Li}, \citenamefont {Zhang},\ and\ \citenamefont
  {Gurvitz}}]{li2009}%
  \BibitemOpen
  \bibfield  {author} {\bibinfo {author} {\bibfnamefont {F.}~\bibnamefont
  {Li}}, \bibinfo {author} {\bibfnamefont {X.-Q.}\ \bibnamefont {Li}}, \bibinfo
  {author} {\bibfnamefont {W.-M.}\ \bibnamefont {Zhang}},\ and\ \bibinfo
  {author} {\bibfnamefont {S.~A.}\ \bibnamefont {Gurvitz}},\ }\bibfield
  {title} {\bibinfo {title} {Aharonov-bohm effect in open quantum systems},\
  }\href@noop {} {\bibfield  {journal} {\bibinfo  {journal} {Euro. Phys.
  Lett.}\ }\textbf {\bibinfo {volume} {88}},\ \bibinfo {pages} {37001}
  (\bibinfo {year} {2009}{\natexlab{a}})}\BibitemShut {NoStop}%
\bibitem [{\citenamefont {Tokura}\ \emph {et~al.}(2007)\citenamefont {Tokura},
  \citenamefont {Nakano},\ and\ \citenamefont {Kubo}}]{tokura2007}%
  \BibitemOpen
  \bibfield  {author} {\bibinfo {author} {\bibfnamefont {Y.}~\bibnamefont
  {Tokura}}, \bibinfo {author} {\bibfnamefont {H.}~\bibnamefont {Nakano}},\
  and\ \bibinfo {author} {\bibfnamefont {T.}~\bibnamefont {Kubo}},\ }\bibfield
  {title} {\bibinfo {title} {Control of quantum dynamics via an aharonov-bohm
  ring},\ }\href@noop {} {\bibfield  {journal} {\bibinfo  {journal} {New J.
  Phys.}\ }\textbf {\bibinfo {volume} {9}},\ \bibinfo {pages} {113} (\bibinfo
  {year} {2007})}\BibitemShut {NoStop}%
\bibitem [{\citenamefont {Liu}\ \emph {et~al.}(2007{\natexlab{a}})\citenamefont
  {Liu}, \citenamefont {Chen}, ,\ and\ \citenamefont {Yang}}]{liu2007}%
  \BibitemOpen
  \bibfield  {author} {\bibinfo {author} {\bibfnamefont {Y.-S.}\ \bibnamefont
  {Liu}}, \bibinfo {author} {\bibfnamefont {H.}~\bibnamefont {Chen}}, ,\ and\
  \bibinfo {author} {\bibfnamefont {X.-F.}\ \bibnamefont {Yang}},\ }\bibfield
  {title} {\bibinfo {title} {Aharonov-bohm interference in mesoscopic
  systems},\ }\href@noop {} {\bibfield  {journal} {\bibinfo  {journal} {J.
  Phys.: Condens. Matter}\ }\textbf {\bibinfo {volume} {19}},\ \bibinfo {pages}
  {246201} (\bibinfo {year} {2007}{\natexlab{a}})}\BibitemShut {NoStop}%
\bibitem [{\citenamefont {Boese}\ \emph {et~al.}(2002)\citenamefont {Boese},
  \citenamefont {Hofstetter},\ and\ \citenamefont {Schoeller}}]{boese2002}%
  \BibitemOpen
  \bibfield  {author} {\bibinfo {author} {\bibfnamefont {D.}~\bibnamefont
  {Boese}}, \bibinfo {author} {\bibfnamefont {W.}~\bibnamefont {Hofstetter}},\
  and\ \bibinfo {author} {\bibfnamefont {H.}~\bibnamefont {Schoeller}},\
  }\bibfield  {title} {\bibinfo {title} {Influence of spin and charge on the
  aharonov-bohm effect},\ }\href@noop {} {\bibfield  {journal} {\bibinfo
  {journal} {Phys. Rev. B}\ }\textbf {\bibinfo {volume} {66}},\ \bibinfo
  {pages} {125315} (\bibinfo {year} {2002})}\BibitemShut {NoStop}%
\bibitem [{\citenamefont {Rai}\ and\ \citenamefont {Galperin}(2012)}]{rai2012}%
  \BibitemOpen
  \bibfield  {author} {\bibinfo {author} {\bibfnamefont {D.}~\bibnamefont
  {Rai}}\ and\ \bibinfo {author} {\bibfnamefont {M.}~\bibnamefont {Galperin}},\
  }\bibfield  {title} {\bibinfo {title} {Quantum interference and
  thermoelectric effects in molecular junctions},\ }\href@noop {} {\bibfield
  {journal} {\bibinfo  {journal} {Phys. Rev. B}\ }\textbf {\bibinfo {volume}
  {86}},\ \bibinfo {pages} {045420} (\bibinfo {year} {2012})}\BibitemShut
  {NoStop}%
\bibitem [{\citenamefont {Sigrist}\ \emph {et~al.}(2006)\citenamefont
  {Sigrist}, \citenamefont {Ihn}, \citenamefont {Ensslin}, \citenamefont
  {Loss}, \citenamefont {Reinwald},\ and\ \citenamefont
  {Wegscheider}}]{sigrist2006}%
  \BibitemOpen
  \bibfield  {author} {\bibinfo {author} {\bibfnamefont {M.}~\bibnamefont
  {Sigrist}}, \bibinfo {author} {\bibfnamefont {T.}~\bibnamefont {Ihn}},
  \bibinfo {author} {\bibfnamefont {K.}~\bibnamefont {Ensslin}}, \bibinfo
  {author} {\bibfnamefont {D.}~\bibnamefont {Loss}}, \bibinfo {author}
  {\bibfnamefont {M.}~\bibnamefont {Reinwald}},\ and\ \bibinfo {author}
  {\bibfnamefont {W.}~\bibnamefont {Wegscheider}},\ }\bibfield  {title}
  {\bibinfo {title} {Universal conductance fluctuations in an aharonov-bohm
  ring with a quantum dot},\ }\href@noop {} {\bibfield  {journal} {\bibinfo
  {journal} {Phys. Rev. Lett.}\ }\textbf {\bibinfo {volume} {96}},\ \bibinfo
  {pages} {036804} (\bibinfo {year} {2006})}\BibitemShut {NoStop}%
\bibitem [{\citenamefont {Golosov}\ and\ \citenamefont
  {Gefen}(2007)}]{golosov2007}%
  \BibitemOpen
  \bibfield  {author} {\bibinfo {author} {\bibfnamefont {D.~I.}\ \bibnamefont
  {Golosov}}\ and\ \bibinfo {author} {\bibfnamefont {Y.}~\bibnamefont
  {Gefen}},\ }\bibfield  {title} {\bibinfo {title} {Interplay of phase
  coherence and charging energy in a quantum dot aharonov-bohm
  interferometer},\ }\href@noop {} {\bibfield  {journal} {\bibinfo  {journal}
  {New J. Phys.}\ }\textbf {\bibinfo {volume} {9}},\ \bibinfo {pages} {120}
  (\bibinfo {year} {2007})}\BibitemShut {NoStop}%
\bibitem [{\citenamefont {Hofstetter}\ \emph {et~al.}(2001)\citenamefont
  {Hofstetter}, \citenamefont {König},\ and\ \citenamefont
  {Schoeller}}]{hofstetter2001}%
  \BibitemOpen
  \bibfield  {author} {\bibinfo {author} {\bibfnamefont {W.}~\bibnamefont
  {Hofstetter}}, \bibinfo {author} {\bibfnamefont {J.}~\bibnamefont {König}},\
  and\ \bibinfo {author} {\bibfnamefont {H.}~\bibnamefont {Schoeller}},\
  }\bibfield  {title} {\bibinfo {title} {Kondo effect in quantum dots with
  aharonov-bohm oscillations},\ }\href@noop {} {\bibfield  {journal} {\bibinfo
  {journal} {Phys. Rev. Lett.}\ }\textbf {\bibinfo {volume} {87}},\ \bibinfo
  {pages} {156803} (\bibinfo {year} {2001})}\BibitemShut {NoStop}%
\bibitem [{\citenamefont {Sun}\ and\ \citenamefont {Guo}(2002)}]{sun2002}%
  \BibitemOpen
  \bibfield  {author} {\bibinfo {author} {\bibfnamefont {Q.-F.}\ \bibnamefont
  {Sun}}\ and\ \bibinfo {author} {\bibfnamefont {H.}~\bibnamefont {Guo}},\
  }\bibfield  {title} {\bibinfo {title} {Kondo effect and aharonov-bohm
  oscillations in quantum dots},\ }\href@noop {} {\bibfield  {journal}
  {\bibinfo  {journal} {Phys. Rev. B}\ }\textbf {\bibinfo {volume} {66}},\
  \bibinfo {pages} {155308} (\bibinfo {year} {2002})}\BibitemShut {NoStop}%
\bibitem [{\citenamefont {Malecki}\ and\ \citenamefont
  {Affleck}(2010)}]{malecki2010}%
  \BibitemOpen
  \bibfield  {author} {\bibinfo {author} {\bibfnamefont {J.}~\bibnamefont
  {Malecki}}\ and\ \bibinfo {author} {\bibfnamefont {I.}~\bibnamefont
  {Affleck}},\ }\bibfield  {title} {\bibinfo {title} {Kondo effect in an
  aharonov-bohm ring},\ }\href@noop {} {\bibfield  {journal} {\bibinfo
  {journal} {Phys. Rev. B}\ }\textbf {\bibinfo {volume} {82}},\ \bibinfo
  {pages} {165426} (\bibinfo {year} {2010})}\BibitemShut {NoStop}%
\bibitem [{\citenamefont {Hod}\ \emph {et~al.}(2006)\citenamefont {Hod},
  \citenamefont {Baer},\ and\ \citenamefont {Rabani}}]{hod2006}%
  \BibitemOpen
  \bibfield  {author} {\bibinfo {author} {\bibfnamefont {O.}~\bibnamefont
  {Hod}}, \bibinfo {author} {\bibfnamefont {R.}~\bibnamefont {Baer}},\ and\
  \bibinfo {author} {\bibfnamefont {E.}~\bibnamefont {Rabani}},\ }\bibfield
  {title} {\bibinfo {title} {Interplay of phase and electron correlations in
  molecular conductance},\ }\href@noop {} {\bibfield  {journal} {\bibinfo
  {journal} {Phys. Rev. Lett.}\ }\textbf {\bibinfo {volume} {97}},\ \bibinfo
  {pages} {266803} (\bibinfo {year} {2006})}\BibitemShut {NoStop}%
\bibitem [{\citenamefont {Imry}(2002)}]{imry2002}%
  \BibitemOpen
  \bibfield  {author} {\bibinfo {author} {\bibfnamefont {Y.}~\bibnamefont
  {Imry}},\ }\href@noop {} {\emph {\bibinfo {title} {Introduction to Mesoscopic
  Physics}}},\ \bibinfo {edition} {2nd}\ ed.\ (\bibinfo  {publisher} {Oxford
  University Press},\ \bibinfo {year} {2002})\BibitemShut {NoStop}%
\bibitem [{\citenamefont {Li}\ \emph {et~al.}(2009{\natexlab{b}})\citenamefont
  {Li}, \citenamefont {Jiao}, \citenamefont {Luo}, \citenamefont {Li},\ and\
  \citenamefont {Gurvitz}}]{li2009a}%
  \BibitemOpen
  \bibfield  {author} {\bibinfo {author} {\bibfnamefont {F.}~\bibnamefont
  {Li}}, \bibinfo {author} {\bibfnamefont {H.-J.}\ \bibnamefont {Jiao}},
  \bibinfo {author} {\bibfnamefont {J.-Y.}\ \bibnamefont {Luo}}, \bibinfo
  {author} {\bibfnamefont {X.-Q.}\ \bibnamefont {Li}},\ and\ \bibinfo {author}
  {\bibfnamefont {S.~A.}\ \bibnamefont {Gurvitz}},\ }\bibfield  {title}
  {\bibinfo {title} {Quantum interference and phase control in aharonov-bohm
  rings},\ }\href@noop {} {\bibfield  {journal} {\bibinfo  {journal} {Physica
  E}\ }\textbf {\bibinfo {volume} {41}},\ \bibinfo {pages} {1707} (\bibinfo
  {year} {2009}{\natexlab{b}})}\BibitemShut {NoStop}%
\bibitem [{\citenamefont {Kashcheyevs}\ \emph {et~al.}(2006)\citenamefont
  {Kashcheyevs}, \citenamefont {Aharony},\ and\ \citenamefont
  {Entin-Wohlman}}]{kashcheyevs2006}%
  \BibitemOpen
  \bibfield  {author} {\bibinfo {author} {\bibfnamefont {V.}~\bibnamefont
  {Kashcheyevs}}, \bibinfo {author} {\bibfnamefont {A.}~\bibnamefont
  {Aharony}},\ and\ \bibinfo {author} {\bibfnamefont {O.}~\bibnamefont
  {Entin-Wohlman}},\ }\bibfield  {title} {\bibinfo {title} {Quantum
  interference in mesoscopic rings with magnetic fields},\ }\href@noop {}
  {\bibfield  {journal} {\bibinfo  {journal} {Phys. Rev. B}\ }\textbf {\bibinfo
  {volume} {73}},\ \bibinfo {pages} {125338} (\bibinfo {year}
  {2006})}\BibitemShut {NoStop}%
\bibitem [{\citenamefont {Entin-Wohlman}\ and\ \citenamefont
  {Aharony}(2012)}]{entinwohlman2012}%
  \BibitemOpen
  \bibfield  {author} {\bibinfo {author} {\bibfnamefont {O.}~\bibnamefont
  {Entin-Wohlman}}\ and\ \bibinfo {author} {\bibfnamefont {A.}~\bibnamefont
  {Aharony}},\ }\bibfield  {title} {\bibinfo {title} {Aharonov-bohm
  interference in molecular junctions},\ }\href@noop {} {\bibfield  {journal}
  {\bibinfo  {journal} {Phys. Rev. B}\ }\textbf {\bibinfo {volume} {85}},\
  \bibinfo {pages} {085401} (\bibinfo {year} {2012})}\BibitemShut {NoStop}%
\bibitem [{\citenamefont {Aharony}\ and\ \citenamefont
  {Entin-Wohlman}(2005)}]{aharony2005}%
  \BibitemOpen
  \bibfield  {author} {\bibinfo {author} {\bibfnamefont {A.}~\bibnamefont
  {Aharony}}\ and\ \bibinfo {author} {\bibfnamefont {O.}~\bibnamefont
  {Entin-Wohlman}},\ }\bibfield  {title} {\bibinfo {title} {Interference
  effects in quantum dots and molecular junctions},\ }\href@noop {} {\bibfield
  {journal} {\bibinfo  {journal} {Phys. Rev. B}\ }\textbf {\bibinfo {volume}
  {72}},\ \bibinfo {pages} {073311} (\bibinfo {year} {2005})}\BibitemShut
  {NoStop}%
\bibitem [{\citenamefont {Rai}\ \emph {et~al.}(2011)\citenamefont {Rai},
  \citenamefont {Hod},\ and\ \citenamefont {Nitzan}}]{rai2011}%
  \BibitemOpen
  \bibfield  {author} {\bibinfo {author} {\bibfnamefont {D.}~\bibnamefont
  {Rai}}, \bibinfo {author} {\bibfnamefont {O.}~\bibnamefont {Hod}},\ and\
  \bibinfo {author} {\bibfnamefont {A.}~\bibnamefont {Nitzan}},\ }\bibfield
  {title} {\bibinfo {title} {Magnetic effects in molecular junctions: The role
  of quantum interference},\ }\href@noop {} {\bibfield  {journal} {\bibinfo
  {journal} {J. Phys. Chem. Lett.}\ }\textbf {\bibinfo {volume} {2}},\ \bibinfo
  {pages} {2118} (\bibinfo {year} {2011})}\BibitemShut {NoStop}%
\bibitem [{\citenamefont {Kubala}\ and\ \citenamefont
  {K\"onig}(2002)}]{kubala2002}%
  \BibitemOpen
  \bibfield  {author} {\bibinfo {author} {\bibfnamefont {B.}~\bibnamefont
  {Kubala}}\ and\ \bibinfo {author} {\bibfnamefont {J.}~\bibnamefont
  {K\"onig}},\ }\bibfield  {title} {\bibinfo {title} {Quantum transport through
  single molecules: Influence of the electronic environment},\ }\href@noop {}
  {\bibfield  {journal} {\bibinfo  {journal} {Physical Review B}\ }\textbf
  {\bibinfo {volume} {65}},\ \bibinfo {pages} {245301} (\bibinfo {year}
  {2002})}\BibitemShut {NoStop}%
\bibitem [{\citenamefont {Sztenkiel}\ and\ \citenamefont
  {Swirkowicz}(2007)}]{sztenkiel2007}%
  \BibitemOpen
  \bibfield  {author} {\bibinfo {author} {\bibfnamefont {D.}~\bibnamefont
  {Sztenkiel}}\ and\ \bibinfo {author} {\bibfnamefont {R.}~\bibnamefont
  {Swirkowicz}},\ }\bibfield  {title} {\bibinfo {title} {J. phys.: Condens.
  matter 19 386224 (2007)},\ }\href@noop {} {\bibfield  {journal} {\bibinfo
  {journal} {Journal of Physics: Condensed Matter}\ }\textbf {\bibinfo {volume}
  {19}},\ \bibinfo {pages} {386224} (\bibinfo {year} {2007})}\BibitemShut
  {NoStop}%
\bibitem [{\citenamefont {Liu}\ \emph {et~al.}(2007{\natexlab{b}})\citenamefont
  {Liu}, \citenamefont {Chen},\ and\ \citenamefont {Yang}}]{liu22007}%
  \BibitemOpen
  \bibfield  {author} {\bibinfo {author} {\bibfnamefont {Y.-S.}\ \bibnamefont
  {Liu}}, \bibinfo {author} {\bibfnamefont {H.}~\bibnamefont {Chen}},\ and\
  \bibinfo {author} {\bibfnamefont {X.-F.}\ \bibnamefont {Yang}},\ }\bibfield
  {title} {\bibinfo {title} {J. phys.: Condens. matter 19 246201 (2007)},\
  }\href@noop {} {\bibfield  {journal} {\bibinfo  {journal} {Journal of
  Physics: Condensed Matter}\ }\textbf {\bibinfo {volume} {19}},\ \bibinfo
  {pages} {246201} (\bibinfo {year} {2007}{\natexlab{b}})}\BibitemShut
  {NoStop}%
\bibitem [{\citenamefont {Loss}\ and\ \citenamefont
  {Sukhorukov}(2000)}]{loss2000}%
  \BibitemOpen
  \bibfield  {author} {\bibinfo {author} {\bibfnamefont {D.}~\bibnamefont
  {Loss}}\ and\ \bibinfo {author} {\bibfnamefont {E.}~\bibnamefont
  {Sukhorukov}},\ }\bibfield  {title} {\bibinfo {title} {Phys. rev. lett. 84,
  1035, (2000)},\ }\href@noop {} {\bibfield  {journal} {\bibinfo  {journal}
  {Physical Review Letters}\ }\textbf {\bibinfo {volume} {84}},\ \bibinfo
  {pages} {1035} (\bibinfo {year} {2000})}\BibitemShut {NoStop}%
\bibitem [{\citenamefont {Leturcq}\ \emph
  {et~al.}(2006{\natexlab{a}})\citenamefont {Leturcq}, \citenamefont {Sanchez},
  \citenamefont {Götz}, \citenamefont {Ihn}, \citenamefont {Ensslin},
  \citenamefont {Driscoll},\ and\ \citenamefont {Gossard}}]{leturcq2006phys}%
  \BibitemOpen
  \bibfield  {author} {\bibinfo {author} {\bibfnamefont {R.}~\bibnamefont
  {Leturcq}}, \bibinfo {author} {\bibfnamefont {D.}~\bibnamefont {Sanchez}},
  \bibinfo {author} {\bibfnamefont {G.}~\bibnamefont {Götz}}, \bibinfo
  {author} {\bibfnamefont {T.}~\bibnamefont {Ihn}}, \bibinfo {author}
  {\bibfnamefont {K.}~\bibnamefont {Ensslin}}, \bibinfo {author} {\bibfnamefont
  {D.~C.}\ \bibnamefont {Driscoll}},\ and\ \bibinfo {author} {\bibfnamefont
  {A.~C.}\ \bibnamefont {Gossard}},\ }\href@noop {} {\bibfield  {journal}
  {\bibinfo  {journal} {Phys. Rev. Lett.}\ }\textbf {\bibinfo {volume} {96}},\
  \bibinfo {pages} {126801} (\bibinfo {year} {2006}{\natexlab{a}})}\BibitemShut
  {NoStop}%
\bibitem [{\citenamefont {Leturcq}\ \emph
  {et~al.}(2006{\natexlab{b}})\citenamefont {Leturcq}, \citenamefont
  {Bianchetti}, \citenamefont {Götz}, \citenamefont {Ihn}, \citenamefont
  {Ensslin}, \citenamefont {Driscoll},\ and\ \citenamefont
  {Gossard}}]{leturcq2006physica}%
  \BibitemOpen
  \bibfield  {author} {\bibinfo {author} {\bibfnamefont {R.}~\bibnamefont
  {Leturcq}}, \bibinfo {author} {\bibfnamefont {R.}~\bibnamefont {Bianchetti}},
  \bibinfo {author} {\bibfnamefont {G.}~\bibnamefont {Götz}}, \bibinfo
  {author} {\bibfnamefont {T.}~\bibnamefont {Ihn}}, \bibinfo {author}
  {\bibfnamefont {K.}~\bibnamefont {Ensslin}}, \bibinfo {author} {\bibfnamefont
  {D.~C.}\ \bibnamefont {Driscoll}},\ and\ \bibinfo {author} {\bibfnamefont
  {A.~C.}\ \bibnamefont {Gossard}},\ }\href@noop {} {\bibfield  {journal}
  {\bibinfo  {journal} {Physica E}\ }\textbf {\bibinfo {volume} {35}},\
  \bibinfo {pages} {327} (\bibinfo {year} {2006}{\natexlab{b}})}\BibitemShut
  {NoStop}%
\bibitem [{\citenamefont {Sanchez}\ and\ \citenamefont
  {Kang}(2008)}]{sanchez2008phys}%
  \BibitemOpen
  \bibfield  {author} {\bibinfo {author} {\bibfnamefont {D.}~\bibnamefont
  {Sanchez}}\ and\ \bibinfo {author} {\bibfnamefont {K.}~\bibnamefont {Kang}},\
  }\href@noop {} {\bibfield  {journal} {\bibinfo  {journal} {Phys. Rev. Lett.}\
  }\textbf {\bibinfo {volume} {100}},\ \bibinfo {pages} {036806} (\bibinfo
  {year} {2008})}\BibitemShut {NoStop}%
\bibitem [{\citenamefont {Bedkihal}\ \emph
  {et~al.}(2013{\natexlab{a}})\citenamefont {Bedkihal}, \citenamefont
  {Bandyopadhyay},\ and\ \citenamefont {Segal}}]{Bedkihal2013}%
  \BibitemOpen
  \bibfield  {author} {\bibinfo {author} {\bibfnamefont {S.}~\bibnamefont
  {Bedkihal}}, \bibinfo {author} {\bibfnamefont {M.}~\bibnamefont
  {Bandyopadhyay}},\ and\ \bibinfo {author} {\bibfnamefont {D.}~\bibnamefont
  {Segal}},\ }\bibfield  {title} {\bibinfo {title} {The probe technique far
  from equilibrium: Magnetic field symmetries of nonlinear transport},\ }\href
  {https://doi.org/10.1140/epjb/e2013-40971-7} {\bibfield  {journal} {\bibinfo
  {journal} {Eur. Phys. J. B}\ }\textbf {\bibinfo {volume} {86}},\ \bibinfo
  {pages} {506} (\bibinfo {year} {2013}{\natexlab{a}})}\BibitemShut {NoStop}%
\bibitem [{\citenamefont {Bedkihal}\ and\ \citenamefont
  {Segal}(2014)}]{PhysRevB.90.235411_Bedkihal2014}%
  \BibitemOpen
  \bibfield  {author} {\bibinfo {author} {\bibfnamefont {S.}~\bibnamefont
  {Bedkihal}}\ and\ \bibinfo {author} {\bibfnamefont {D.}~\bibnamefont
  {Segal}},\ }\bibfield  {title} {\bibinfo {title} {Magnetotransport in
  aharonov-bohm interferometers: Exact numerical simulations},\ }\href
  {https://doi.org/10.1103/PhysRevB.90.235411} {\bibfield  {journal} {\bibinfo
  {journal} {Phys. Rev. B}\ }\textbf {\bibinfo {volume} {90}},\ \bibinfo
  {pages} {235411} (\bibinfo {year} {2014})}\BibitemShut {NoStop}%
\bibitem [{\citenamefont {Mahan}\ and\ \citenamefont {Sofo}(1996)}]{Mahan1996}%
  \BibitemOpen
  \bibfield  {author} {\bibinfo {author} {\bibfnamefont {G.~D.}\ \bibnamefont
  {Mahan}}\ and\ \bibinfo {author} {\bibfnamefont {J.~O.}\ \bibnamefont
  {Sofo}},\ }\bibfield  {title} {\bibinfo {title} {The best thermoelectric},\
  }\href@noop {} {\bibfield  {journal} {\bibinfo  {journal} {Proc. Natl Acad.
  Sci. USA}\ }\textbf {\bibinfo {volume} {93}},\ \bibinfo {pages} {7436}
  (\bibinfo {year} {1996})}\BibitemShut {NoStop}%
\bibitem [{\citenamefont {Humphrey}\ \emph {et~al.}(2002)\citenamefont
  {Humphrey}, \citenamefont {Newbury}, \citenamefont {Taylor},\ and\
  \citenamefont {Linke}}]{ref4}%
  \BibitemOpen
  \bibfield  {author} {\bibinfo {author} {\bibfnamefont {T.~E.}\ \bibnamefont
  {Humphrey}}, \bibinfo {author} {\bibfnamefont {R.}~\bibnamefont {Newbury}},
  \bibinfo {author} {\bibfnamefont {P.~R.}\ \bibnamefont {Taylor}},\ and\
  \bibinfo {author} {\bibfnamefont {H.}~\bibnamefont {Linke}},\ }\bibfield
  {title} {\bibinfo {title} {Reversible quantum brownian heat engines for
  electrons},\ }\href@noop {} {\bibfield  {journal} {\bibinfo  {journal} {Phys.
  Rev. Lett.}\ }\textbf {\bibinfo {volume} {89}},\ \bibinfo {pages} {116801}
  (\bibinfo {year} {2002})}\BibitemShut {NoStop}%
\bibitem [{\citenamefont {Humphrey}\ and\ \citenamefont {Linke}(2005)}]{ref5}%
  \BibitemOpen
  \bibfield  {author} {\bibinfo {author} {\bibfnamefont {T.~E.}\ \bibnamefont
  {Humphrey}}\ and\ \bibinfo {author} {\bibfnamefont {H.}~\bibnamefont
  {Linke}},\ }\bibfield  {title} {\bibinfo {title} {Reversible thermoelectric
  nanomaterials},\ }\href@noop {} {\bibfield  {journal} {\bibinfo  {journal}
  {Phys. Rev. Lett.}\ }\textbf {\bibinfo {volume} {94}},\ \bibinfo {pages}
  {096601} (\bibinfo {year} {2005})}\BibitemShut {NoStop}%
\bibitem [{\citenamefont {Sánchez}\ \emph {et~al.}(2013)\citenamefont
  {Sánchez}, \citenamefont {Sothmann}, \citenamefont {Jordan},\ and\
  \citenamefont {Büttiker}}]{engine4}%
  \BibitemOpen
  \bibfield  {author} {\bibinfo {author} {\bibfnamefont {R.}~\bibnamefont
  {Sánchez}}, \bibinfo {author} {\bibfnamefont {B.}~\bibnamefont {Sothmann}},
  \bibinfo {author} {\bibfnamefont {A.~N.}\ \bibnamefont {Jordan}},\ and\
  \bibinfo {author} {\bibfnamefont {M.}~\bibnamefont {Büttiker}},\ }\bibfield
  {title} {\bibinfo {title} {Correlations of heat and charge currents in
  quantum-dot thermoelectric engines},\ }\href
  {https://doi.org/10.1088/1367-2630/15/12/125001} {\bibfield  {journal}
  {\bibinfo  {journal} {New J. Phys.}\ }\textbf {\bibinfo {volume} {15}},\
  \bibinfo {pages} {125001} (\bibinfo {year} {2013})}\BibitemShut {NoStop}%
\bibitem [{\citenamefont {Benenti}\ \emph {et~al.}(2017)\citenamefont
  {Benenti}, \citenamefont {Casati}, \citenamefont {Saito},\ and\ \citenamefont
  {Whitney}}]{engine5}%
  \BibitemOpen
  \bibfield  {author} {\bibinfo {author} {\bibfnamefont {G.}~\bibnamefont
  {Benenti}}, \bibinfo {author} {\bibfnamefont {G.}~\bibnamefont {Casati}},
  \bibinfo {author} {\bibfnamefont {K.}~\bibnamefont {Saito}},\ and\ \bibinfo
  {author} {\bibfnamefont {R.~S.}\ \bibnamefont {Whitney}},\ }\bibfield
  {title} {\bibinfo {title} {Fundamental aspects of steady-state conversion of
  heat to work at the nanoscale},\ }\href
  {https://doi.org/https://doi.org/10.1016/j.physrep.2017.05.008} {\bibfield
  {journal} {\bibinfo  {journal} {Physics Reports}\ }\textbf {\bibinfo {volume}
  {694}},\ \bibinfo {pages} {1} (\bibinfo {year} {2017})}\BibitemShut {NoStop}%
\bibitem [{\citenamefont {Kennes}\ \emph {et~al.}(2013)\citenamefont {Kennes},
  \citenamefont {Schuricht},\ and\ \citenamefont {Meden}}]{engine6}%
  \BibitemOpen
  \bibfield  {author} {\bibinfo {author} {\bibfnamefont {D.~M.}\ \bibnamefont
  {Kennes}}, \bibinfo {author} {\bibfnamefont {D.}~\bibnamefont {Schuricht}},\
  and\ \bibinfo {author} {\bibfnamefont {V.}~\bibnamefont {Meden}},\ }\bibfield
   {title} {\bibinfo {title} {Efficiency and power of a thermoelectric quantum
  dot device},\ }\href {https://doi.org/10.1209/0295-5075/102/57003} {\bibfield
   {journal} {\bibinfo  {journal} {Europhysics Letters}\ }\textbf {\bibinfo
  {volume} {102}},\ \bibinfo {pages} {57003} (\bibinfo {year}
  {2013})}\BibitemShut {NoStop}%
\bibitem [{\citenamefont {Mazza}\ \emph {et~al.}(2014)\citenamefont {Mazza},
  \citenamefont {Bosisio}, \citenamefont {Benenti}, \citenamefont
  {Giovannetti}, \citenamefont {Fazio},\ and\ \citenamefont
  {Taddei}}]{engine7}%
  \BibitemOpen
  \bibfield  {author} {\bibinfo {author} {\bibfnamefont {F.}~\bibnamefont
  {Mazza}}, \bibinfo {author} {\bibfnamefont {R.}~\bibnamefont {Bosisio}},
  \bibinfo {author} {\bibfnamefont {G.}~\bibnamefont {Benenti}}, \bibinfo
  {author} {\bibfnamefont {V.}~\bibnamefont {Giovannetti}}, \bibinfo {author}
  {\bibfnamefont {R.}~\bibnamefont {Fazio}},\ and\ \bibinfo {author}
  {\bibfnamefont {F.}~\bibnamefont {Taddei}},\ }\bibfield  {title} {\bibinfo
  {title} {Thermoelectric efficiency of three-terminal quantum thermal
  machines},\ }\href {https://doi.org/10.1088/1367-2630/16/8/085001} {\bibfield
   {journal} {\bibinfo  {journal} {New J. Phys.}\ }\textbf {\bibinfo {volume}
  {16}},\ \bibinfo {pages} {085001} (\bibinfo {year} {2014})}\BibitemShut
  {NoStop}%
\bibitem [{\citenamefont {Whitney}(2014)}]{PhysRevLett.112.130601}%
  \BibitemOpen
  \bibfield  {author} {\bibinfo {author} {\bibfnamefont {R.~S.}\ \bibnamefont
  {Whitney}},\ }\bibfield  {title} {\bibinfo {title} {Most efficient quantum
  thermoelectric at finite power output},\ }\href
  {https://doi.org/10.1103/PhysRevLett.112.130601} {\bibfield  {journal}
  {\bibinfo  {journal} {Phys. Rev. Lett.}\ }\textbf {\bibinfo {volume} {112}},\
  \bibinfo {pages} {130601} (\bibinfo {year} {2014})}\BibitemShut {NoStop}%
\bibitem [{\citenamefont {Whitney}(2015)}]{engine9}%
  \BibitemOpen
  \bibfield  {author} {\bibinfo {author} {\bibfnamefont {R.~S.}\ \bibnamefont
  {Whitney}},\ }\bibfield  {title} {\bibinfo {title} {Finding the quantum
  thermoelectric with maximal efficiency and minimal entropy production at
  given power output},\ }\href {https://doi.org/10.1103/PhysRevB.91.115425}
  {\bibfield  {journal} {\bibinfo  {journal} {Phys. Rev. B}\ }\textbf {\bibinfo
  {volume} {91}},\ \bibinfo {pages} {115425} (\bibinfo {year}
  {2015})}\BibitemShut {NoStop}%
\bibitem [{\citenamefont {Allahverdyan}\ \emph {et~al.}(2013)\citenamefont
  {Allahverdyan}, \citenamefont {Hovhannisyan}, \citenamefont {Melkikh},\ and\
  \citenamefont {Gevorkian}}]{engine10}%
  \BibitemOpen
  \bibfield  {author} {\bibinfo {author} {\bibfnamefont {A.~E.}\ \bibnamefont
  {Allahverdyan}}, \bibinfo {author} {\bibfnamefont {K.~V.}\ \bibnamefont
  {Hovhannisyan}}, \bibinfo {author} {\bibfnamefont {A.~V.}\ \bibnamefont
  {Melkikh}},\ and\ \bibinfo {author} {\bibfnamefont {S.~G.}\ \bibnamefont
  {Gevorkian}},\ }\bibfield  {title} {\bibinfo {title} {Carnot cycle at finite
  power: Attainability of maximal efficiency},\ }\href
  {https://doi.org/10.1103/PhysRevLett.111.050601} {\bibfield  {journal}
  {\bibinfo  {journal} {Phys. Rev. Lett.}\ }\textbf {\bibinfo {volume} {111}},\
  \bibinfo {pages} {050601} (\bibinfo {year} {2013})}\BibitemShut {NoStop}%
\bibitem [{\citenamefont {Benenti}\ \emph {et~al.}(2011)\citenamefont
  {Benenti}, \citenamefont {Saito},\ and\ \citenamefont {Casati}}]{ref6}%
  \BibitemOpen
  \bibfield  {author} {\bibinfo {author} {\bibfnamefont {G.}~\bibnamefont
  {Benenti}}, \bibinfo {author} {\bibfnamefont {K.}~\bibnamefont {Saito}},\
  and\ \bibinfo {author} {\bibfnamefont {G.}~\bibnamefont {Casati}},\
  }\bibfield  {title} {\bibinfo {title} {Thermodynamic bounds on efficiency for
  systems with broken time-reversal symmetry},\ }\href@noop {} {\bibfield
  {journal} {\bibinfo  {journal} {Phys. Rev. Lett.}\ }\textbf {\bibinfo
  {volume} {106}},\ \bibinfo {pages} {230602} (\bibinfo {year}
  {2011})}\BibitemShut {NoStop}%
\bibitem [{\citenamefont {Brandner}\ \emph {et~al.}(2013)\citenamefont
  {Brandner}, \citenamefont {Saito},\ and\ \citenamefont {Seifert}}]{ref18}%
  \BibitemOpen
  \bibfield  {author} {\bibinfo {author} {\bibfnamefont {K.}~\bibnamefont
  {Brandner}}, \bibinfo {author} {\bibfnamefont {K.}~\bibnamefont {Saito}},\
  and\ \bibinfo {author} {\bibfnamefont {U.}~\bibnamefont {Seifert}},\
  }\bibfield  {title} {\bibinfo {title} {Strong bounds on onsager coefficients
  and efficiency for three-terminal thermoelectric transport in a magnetic
  field},\ }\href@noop {} {\bibfield  {journal} {\bibinfo  {journal} {Phys.
  Rev. Lett.}\ }\textbf {\bibinfo {volume} {110}},\ \bibinfo {pages} {070603}
  (\bibinfo {year} {2013})}\BibitemShut {NoStop}%
\bibitem [{\citenamefont {Brandner}\ and\ \citenamefont
  {Seifert}(2015)}]{engine13}%
  \BibitemOpen
  \bibfield  {author} {\bibinfo {author} {\bibfnamefont {K.}~\bibnamefont
  {Brandner}}\ and\ \bibinfo {author} {\bibfnamefont {U.}~\bibnamefont
  {Seifert}},\ }\bibfield  {title} {\bibinfo {title} {Bound on thermoelectric
  power in a magnetic field within linear response},\ }\href
  {https://doi.org/10.1103/PhysRevE.91.012121} {\bibfield  {journal} {\bibinfo
  {journal} {Phys. Rev. E}\ }\textbf {\bibinfo {volume} {91}},\ \bibinfo
  {pages} {012121} (\bibinfo {year} {2015})}\BibitemShut {NoStop}%
\bibitem [{\citenamefont {Brandner}\ and\ \citenamefont
  {Seifert}(2013)}]{ref19}%
  \BibitemOpen
  \bibfield  {author} {\bibinfo {author} {\bibfnamefont {K.}~\bibnamefont
  {Brandner}}\ and\ \bibinfo {author} {\bibfnamefont {U.}~\bibnamefont
  {Seifert}},\ }\bibfield  {title} {\bibinfo {title} {Multi-terminal
  thermoelectric transport in a magnetic field: Bounds on onsager coefficients
  and efficiency},\ }\href@noop {} {\bibfield  {journal} {\bibinfo  {journal}
  {New J. Phys.}\ }\textbf {\bibinfo {volume} {15}},\ \bibinfo {pages} {105003}
  (\bibinfo {year} {2013})}\BibitemShut {NoStop}%
\bibitem [{\citenamefont {Xiong}\ and\ \citenamefont
  {Liang}(2005)}]{XIONG2005216}%
  \BibitemOpen
  \bibfield  {author} {\bibinfo {author} {\bibfnamefont {Y.-J.}\ \bibnamefont
  {Xiong}}\ and\ \bibinfo {author} {\bibfnamefont {X.-T.}\ \bibnamefont
  {Liang}},\ }\bibfield  {title} {\bibinfo {title} {The phase of fano resonance
  in aharonov–bohm interferometer with a quantum dot embedded},\ }\href
  {https://doi.org/https://doi.org/10.1016/j.physb.2004.10.094} {\bibfield
  {journal} {\bibinfo  {journal} {Physica B: Condensed Matter}\ }\textbf
  {\bibinfo {volume} {355}},\ \bibinfo {pages} {216} (\bibinfo {year}
  {2005})}\BibitemShut {NoStop}%
\bibitem [{\citenamefont {Saito}\ \emph
  {et~al.}(2011{\natexlab{a}})\citenamefont {Saito}, \citenamefont {Benenti},
  \citenamefont {Casati},\ and\ \citenamefont {Prosen}}]{saito2011}%
  \BibitemOpen
  \bibfield  {author} {\bibinfo {author} {\bibfnamefont {K.}~\bibnamefont
  {Saito}}, \bibinfo {author} {\bibfnamefont {G.}~\bibnamefont {Benenti}},
  \bibinfo {author} {\bibfnamefont {G.}~\bibnamefont {Casati}},\ and\ \bibinfo
  {author} {\bibfnamefont {T.}~\bibnamefont {Prosen}},\ }\bibfield  {title}
  {\bibinfo {title} {Thermopower with broken time-reversal symmetry},\ }\href
  {https://doi.org/10.1103/PhysRevB.84.201306} {\bibfield  {journal} {\bibinfo
  {journal} {Phys. Rev. B}\ }\textbf {\bibinfo {volume} {84}},\ \bibinfo
  {pages} {201306} (\bibinfo {year} {2011}{\natexlab{a}})}\BibitemShut
  {NoStop}%
\bibitem [{\citenamefont {Balachandran}\ \emph {et~al.}(2013)\citenamefont
  {Balachandran}, \citenamefont {Benenti},\ and\ \citenamefont
  {Casati}}]{balachandran2013}%
  \BibitemOpen
  \bibfield  {author} {\bibinfo {author} {\bibfnamefont {V.}~\bibnamefont
  {Balachandran}}, \bibinfo {author} {\bibfnamefont {G.}~\bibnamefont
  {Benenti}},\ and\ \bibinfo {author} {\bibfnamefont {G.}~\bibnamefont
  {Casati}},\ }\bibfield  {title} {\bibinfo {title} {Efficiency of
  three-terminal thermoelectric transport under broken time-reversal
  symmetry},\ }\href@noop {} {\bibfield  {journal} {\bibinfo  {journal} {Phys.
  Rev. B}\ }\textbf {\bibinfo {volume} {87}},\ \bibinfo {pages} {165419}
  (\bibinfo {year} {2013})}\BibitemShut {NoStop}%
\bibitem [{\citenamefont {Onsager}(1931{\natexlab{a}})}]{onsager1931a}%
  \BibitemOpen
  \bibfield  {author} {\bibinfo {author} {\bibfnamefont {L.}~\bibnamefont
  {Onsager}},\ }\bibfield  {title} {\bibinfo {title} {Phys. rev. 37, 405
  (1931)},\ }\href@noop {} {\bibfield  {journal} {\bibinfo  {journal} {Physical
  Review}\ }\textbf {\bibinfo {volume} {37}},\ \bibinfo {pages} {405} (\bibinfo
  {year} {1931}{\natexlab{a}})}\BibitemShut {NoStop}%
\bibitem [{\citenamefont {Onsager}(1931{\natexlab{b}})}]{onsager1931b}%
  \BibitemOpen
  \bibfield  {author} {\bibinfo {author} {\bibfnamefont {L.}~\bibnamefont
  {Onsager}},\ }\bibfield  {title} {\bibinfo {title} {Phys. rev. 38, 2265
  (1931)},\ }\href@noop {} {\bibfield  {journal} {\bibinfo  {journal} {Physical
  Review}\ }\textbf {\bibinfo {volume} {38}},\ \bibinfo {pages} {2265}
  (\bibinfo {year} {1931}{\natexlab{b}})}\BibitemShut {NoStop}%
\bibitem [{\citenamefont {Casimir}(1945)}]{casimir1945}%
  \BibitemOpen
  \bibfield  {author} {\bibinfo {author} {\bibfnamefont {H.~B.~G.}\
  \bibnamefont {Casimir}},\ }\bibfield  {title} {\bibinfo {title} {Rev. mod.
  phys. 17, 343 (1945)},\ }\href@noop {} {\bibfield  {journal} {\bibinfo
  {journal} {Reviews of Modern Physics}\ }\textbf {\bibinfo {volume} {17}},\
  \bibinfo {pages} {343} (\bibinfo {year} {1945})}\BibitemShut {NoStop}%
\bibitem [{\citenamefont {Keldysh}(1965)}]{Keldysh1965}%
  \BibitemOpen
  \bibfield  {author} {\bibinfo {author} {\bibfnamefont {L.}~\bibnamefont
  {Keldysh}},\ }\bibfield  {title} {\bibinfo {title} {Diagram technique for
  nonequilibrium processes},\ }\href@noop {} {\bibfield  {journal} {\bibinfo
  {journal} {Soviet Physics JETP}\ }\textbf {\bibinfo {volume} {20}},\ \bibinfo
  {pages} {1018} (\bibinfo {year} {1965})}\BibitemShut {NoStop}%
\bibitem [{\citenamefont {Schwinger}(1961)}]{Schwinger1961}%
  \BibitemOpen
  \bibfield  {author} {\bibinfo {author} {\bibfnamefont {J.}~\bibnamefont
  {Schwinger}},\ }\href@noop {} {\bibfield  {journal} {\bibinfo  {journal}
  {Journal of Mathematical Physics}\ }\textbf {\bibinfo {volume} {2}},\
  \bibinfo {pages} {407} (\bibinfo {year} {1961})}\BibitemShut {NoStop}%
\bibitem [{\citenamefont {Kadanoff}\ and\ \citenamefont
  {Baym}(1962)}]{Kadanoff1962}%
  \BibitemOpen
  \bibfield  {author} {\bibinfo {author} {\bibfnamefont {L.}~\bibnamefont
  {Kadanoff}}\ and\ \bibinfo {author} {\bibfnamefont {G.}~\bibnamefont
  {Baym}},\ }\href@noop {} {\emph {\bibinfo {title} {Quantum Statistical
  Mechanics}}}\ (\bibinfo  {publisher} {Benjamin/Cummings},\ \bibinfo {year}
  {1962})\BibitemShut {NoStop}%
\bibitem [{\citenamefont {Bedkihal}\ and\ \citenamefont
  {Segal}(2012)}]{SBPath}%
  \BibitemOpen
  \bibfield  {author} {\bibinfo {author} {\bibfnamefont {S.}~\bibnamefont
  {Bedkihal}}\ and\ \bibinfo {author} {\bibfnamefont {D.}~\bibnamefont
  {Segal}},\ }\bibfield  {title} {\bibinfo {title} {Dynamics of coherences in
  the interacting double-dot aharonov-bohm interferometer: Exact numerical
  simulations},\ }\href {https://doi.org/10.1103/PhysRevB.85.155324} {\bibfield
   {journal} {\bibinfo  {journal} {Phys. Rev. B}\ }\textbf {\bibinfo {volume}
  {85}},\ \bibinfo {pages} {155324} (\bibinfo {year} {2012})}\BibitemShut
  {NoStop}%
\bibitem [{\citenamefont {Caroli}\ \emph {et~al.}(1971)\citenamefont {Caroli},
  \citenamefont {Combescot}, \citenamefont {Nozieres},\ and\ \citenamefont
  {Saint-James}}]{Caroli1971}%
  \BibitemOpen
  \bibfield  {author} {\bibinfo {author} {\bibfnamefont {C.}~\bibnamefont
  {Caroli}}, \bibinfo {author} {\bibfnamefont {R.}~\bibnamefont {Combescot}},
  \bibinfo {author} {\bibfnamefont {P.}~\bibnamefont {Nozieres}},\ and\
  \bibinfo {author} {\bibfnamefont {D.}~\bibnamefont {Saint-James}},\
  }\href@noop {} {\bibfield  {journal} {\bibinfo  {journal} {Journal of Physics
  C: Solid State Physics}\ }\textbf {\bibinfo {volume} {4}},\ \bibinfo {pages}
  {916} (\bibinfo {year} {1971})}\BibitemShut {NoStop}%
\bibitem [{\citenamefont {Landauer}(1957{\natexlab{a}})}]{Landauer1957}%
  \BibitemOpen
  \bibfield  {author} {\bibinfo {author} {\bibfnamefont {R.}~\bibnamefont
  {Landauer}},\ }\href@noop {} {\bibfield  {journal} {\bibinfo  {journal} {IBM
  Journal of Research and Development}\ }\textbf {\bibinfo {volume} {1}},\
  \bibinfo {pages} {223} (\bibinfo {year} {1957}{\natexlab{a}})}\BibitemShut
  {NoStop}%
\bibitem [{\citenamefont {Büttiker}\ \emph {et~al.}(1983)\citenamefont
  {Büttiker}, \citenamefont {Imry}, ,\ and\ \citenamefont
  {Landauer}}]{buttiker1984}%
  \BibitemOpen
  \bibfield  {author} {\bibinfo {author} {\bibfnamefont {M.}~\bibnamefont
  {Büttiker}}, \bibinfo {author} {\bibfnamefont {Y.}~\bibnamefont {Imry}}, ,\
  and\ \bibinfo {author} {\bibfnamefont {R.}~\bibnamefont {Landauer}},\
  }\bibfield  {title} {\bibinfo {title} {Josephson behavior in small normal
  one-dimensional rings},\ }\href@noop {} {\bibfield  {journal} {\bibinfo
  {journal} {Phys. Lett. A}\ }\textbf {\bibinfo {volume} {96}},\ \bibinfo
  {pages} {365} (\bibinfo {year} {1983})}\BibitemShut {NoStop}%
\bibitem [{\citenamefont {B\"uttiker}(1986{\natexlab{a}})}]{Buttiker1986}%
  \BibitemOpen
  \bibfield  {author} {\bibinfo {author} {\bibfnamefont {M.}~\bibnamefont
  {B\"uttiker}},\ }\href@noop {} {\bibfield  {journal} {\bibinfo  {journal}
  {Physical Review Letters}\ }\textbf {\bibinfo {volume} {57}},\ \bibinfo
  {pages} {1761} (\bibinfo {year} {1986}{\natexlab{a}})}\BibitemShut {NoStop}%
\bibitem [{\citenamefont {B\"uttiker}(1986{\natexlab{b}})}]{Buttiker1986b}%
  \BibitemOpen
  \bibfield  {author} {\bibinfo {author} {\bibfnamefont {M.}~\bibnamefont
  {B\"uttiker}},\ }\href@noop {} {\bibfield  {journal} {\bibinfo  {journal}
  {Physical Review B}\ }\textbf {\bibinfo {volume} {33}},\ \bibinfo {pages}
  {3020} (\bibinfo {year} {1986}{\natexlab{b}})}\BibitemShut {NoStop}%
\bibitem [{\citenamefont {Ambrosi}\ \emph {et~al.}(2019)\citenamefont
  {Ambrosi}, \citenamefont {Williams}, \citenamefont {Watkinson} \emph
  {et~al.}}]{ambrosi2019rtgs}%
  \BibitemOpen
  \bibfield  {author} {\bibinfo {author} {\bibfnamefont {R.}~\bibnamefont
  {Ambrosi}}, \bibinfo {author} {\bibfnamefont {H.}~\bibnamefont {Williams}},
  \bibinfo {author} {\bibfnamefont {E.}~\bibnamefont {Watkinson}}, \emph
  {et~al.},\ }\bibfield  {title} {\bibinfo {title} {European radioisotope
  thermoelectric generators (rtgs) and radioisotope heater units (rhus) for
  space science and exploration},\ }\href
  {https://doi.org/10.1007/s11214-019-0623-9} {\bibfield  {journal} {\bibinfo
  {journal} {Space Science Reviews}\ }\textbf {\bibinfo {volume} {215}},\
  \bibinfo {pages} {55} (\bibinfo {year} {2019})}\BibitemShut {NoStop}%
\bibitem [{\citenamefont {Kim}\ \emph {et~al.}(2009)\citenamefont {Kim},
  \citenamefont {Datta},\ and\ \citenamefont {Lundstrom}}]{kim2009influence}%
  \BibitemOpen
  \bibfield  {author} {\bibinfo {author} {\bibfnamefont {R.}~\bibnamefont
  {Kim}}, \bibinfo {author} {\bibfnamefont {S.}~\bibnamefont {Datta}},\ and\
  \bibinfo {author} {\bibfnamefont {M.~S.}\ \bibnamefont {Lundstrom}},\
  }\bibfield  {title} {\bibinfo {title} {Influence of dimensionality on
  thermoelectric device performance},\ }\href
  {https://doi.org/10.1063/1.3064184} {\bibfield  {journal} {\bibinfo
  {journal} {Journal of Applied Physics}\ }\textbf {\bibinfo {volume} {105}},\
  \bibinfo {pages} {034506} (\bibinfo {year} {2009})}\BibitemShut {NoStop}%
\bibitem [{\citenamefont {Park}\ \emph {et~al.}(2023)\citenamefont {Park},
  \citenamefont {Park}, \citenamefont {Kim}, \citenamefont {Lee}, \citenamefont
  {Heo},\ and\ \citenamefont {il~Kim}}]{PARK2023107723}%
  \BibitemOpen
  \bibfield  {author} {\bibinfo {author} {\bibfnamefont {O.}~\bibnamefont
  {Park}}, \bibinfo {author} {\bibfnamefont {S.~J.}\ \bibnamefont {Park}},
  \bibinfo {author} {\bibfnamefont {H.-S.}\ \bibnamefont {Kim}}, \bibinfo
  {author} {\bibfnamefont {S.~W.}\ \bibnamefont {Lee}}, \bibinfo {author}
  {\bibfnamefont {M.}~\bibnamefont {Heo}},\ and\ \bibinfo {author}
  {\bibfnamefont {S.}~\bibnamefont {il~Kim}},\ }\bibfield  {title} {\bibinfo
  {title} {Enhanced thermoelectric transport properties of bi2te3
  polycrystalline alloys via carrier type change arising from slight pb
  doping},\ }\href {https://doi.org/https://doi.org/10.1016/j.mssp.2023.107723}
  {\bibfield  {journal} {\bibinfo  {journal} {Materials Science in
  Semiconductor Processing}\ }\textbf {\bibinfo {volume} {166}},\ \bibinfo
  {pages} {107723} (\bibinfo {year} {2023})}\BibitemShut {NoStop}%
\bibitem [{\citenamefont {Kwon}\ \emph {et~al.}(2023)\citenamefont {Kwon},
  \citenamefont {Kim}, \citenamefont {Doh}, \citenamefont {Yu}, \citenamefont
  {Song},\ and\ \citenamefont {Bae}}]{KWON2023105691}%
  \BibitemOpen
  \bibfield  {author} {\bibinfo {author} {\bibfnamefont {D.}~\bibnamefont
  {Kwon}}, \bibinfo {author} {\bibfnamefont {B.-K.}\ \bibnamefont {Kim}},
  \bibinfo {author} {\bibfnamefont {Y.-J.}\ \bibnamefont {Doh}}, \bibinfo
  {author} {\bibfnamefont {D.}~\bibnamefont {Yu}}, \bibinfo {author}
  {\bibfnamefont {J.}~\bibnamefont {Song}},\ and\ \bibinfo {author}
  {\bibfnamefont {M.-H.}\ \bibnamefont {Bae}},\ }\bibfield  {title} {\bibinfo
  {title} {Quantum interference probed by the thermo-voltage in sb-doped bi2se3
  nanowires},\ }\href
  {https://doi.org/https://doi.org/10.1016/j.isci.2022.105691} {\bibfield
  {journal} {\bibinfo  {journal} {iScience}\ }\textbf {\bibinfo {volume}
  {26}},\ \bibinfo {pages} {105691} (\bibinfo {year} {2023})}\BibitemShut
  {NoStop}%
\bibitem [{\citenamefont {Lambert}\ \emph {et~al.}(2016)\citenamefont
  {Lambert}, \citenamefont {Sadeghi},\ and\ \citenamefont
  {Al-Galiby}}]{lambert2016mesoscopic}%
  \BibitemOpen
  \bibfield  {author} {\bibinfo {author} {\bibfnamefont {C.~J.}\ \bibnamefont
  {Lambert}}, \bibinfo {author} {\bibfnamefont {H.}~\bibnamefont {Sadeghi}},\
  and\ \bibinfo {author} {\bibfnamefont {Q.~H.}\ \bibnamefont {Al-Galiby}},\
  }\bibfield  {title} {\bibinfo {title} {Mesoscopic thermoelectric phenomena /
  phénomènes thermoélectriques mésoscopiques: Quantum-interference-enhanced
  thermoelectricity in single molecules and molecular films / effets
  thermoélectriques amplifiés par interférences quantiques dans les
  molécules et les films moléculaires},\ }\href
  {https://doi.org/10.1063/1.4946892} {\bibfield  {journal} {\bibinfo
  {journal} {Journal of Applied Physics}\ }\textbf {\bibinfo {volume} {119}},\
  \bibinfo {pages} {154304} (\bibinfo {year} {2016})}\BibitemShut {NoStop}%
\bibitem [{\citenamefont {Behera}\ \emph {et~al.}(2023)\citenamefont {Behera},
  \citenamefont {Bedkihal}, \citenamefont {Agarwalla},\ and\ \citenamefont
  {Bandyopadhyay}}]{PhysRevB.108.165419}%
  \BibitemOpen
  \bibfield  {author} {\bibinfo {author} {\bibfnamefont {J.}~\bibnamefont
  {Behera}}, \bibinfo {author} {\bibfnamefont {S.}~\bibnamefont {Bedkihal}},
  \bibinfo {author} {\bibfnamefont {B.~K.}\ \bibnamefont {Agarwalla}},\ and\
  \bibinfo {author} {\bibfnamefont {M.}~\bibnamefont {Bandyopadhyay}},\
  }\bibfield  {title} {\bibinfo {title} {Quantum coherent control of nonlinear
  thermoelectric transport in a triple-dot aharonov-bohm heat engine},\ }\href
  {https://doi.org/10.1103/PhysRevB.108.165419} {\bibfield  {journal} {\bibinfo
   {journal} {Phys. Rev. B}\ }\textbf {\bibinfo {volume} {108}},\ \bibinfo
  {pages} {165419} (\bibinfo {year} {2023})}\BibitemShut {NoStop}%
\bibitem [{\citenamefont {Menichetti}\ \emph {et~al.}(2018)\citenamefont
  {Menichetti}, \citenamefont {Grosso},\ and\ \citenamefont
  {Parravicini}}]{Menichetti_2018}%
  \BibitemOpen
  \bibfield  {author} {\bibinfo {author} {\bibfnamefont {G.}~\bibnamefont
  {Menichetti}}, \bibinfo {author} {\bibfnamefont {G.}~\bibnamefont {Grosso}},\
  and\ \bibinfo {author} {\bibfnamefont {G.~P.}\ \bibnamefont {Parravicini}},\
  }\bibfield  {title} {\bibinfo {title} {Analytic treatment of the
  thermoelectric properties for two coupled quantum dots threaded by magnetic
  fields},\ }\href {https://doi.org/10.1088/2399-6528/aac423} {\bibfield
  {journal} {\bibinfo  {journal} {Journal of Physics Communications}\ }\textbf
  {\bibinfo {volume} {2}},\ \bibinfo {pages} {055026} (\bibinfo {year}
  {2018})}\BibitemShut {NoStop}%
\bibitem [{\citenamefont {Kim}\ and\ \citenamefont
  {Hershfield}(2003)}]{PhysRevB.67.165313}%
  \BibitemOpen
  \bibfield  {author} {\bibinfo {author} {\bibfnamefont {T.-S.}\ \bibnamefont
  {Kim}}\ and\ \bibinfo {author} {\bibfnamefont {S.}~\bibnamefont
  {Hershfield}},\ }\bibfield  {title} {\bibinfo {title} {Thermoelectric effects
  of an aharonov-bohm interferometer with an embedded quantum dot in the kondo
  regime},\ }\href {https://doi.org/10.1103/PhysRevB.67.165313} {\bibfield
  {journal} {\bibinfo  {journal} {Phys. Rev. B}\ }\textbf {\bibinfo {volume}
  {67}},\ \bibinfo {pages} {165313} (\bibinfo {year} {2003})}\BibitemShut
  {NoStop}%
\bibitem [{\citenamefont {Liu}\ \emph {et~al.}(2011)\citenamefont {Liu},
  \citenamefont {Zhang}, \citenamefont {Yang},\ and\ \citenamefont
  {Feng}}]{liu2011role}%
  \BibitemOpen
  \bibfield  {author} {\bibinfo {author} {\bibfnamefont {Y.-S.}\ \bibnamefont
  {Liu}}, \bibinfo {author} {\bibfnamefont {D.-B.}\ \bibnamefont {Zhang}},
  \bibinfo {author} {\bibfnamefont {X.-F.}\ \bibnamefont {Yang}},\ and\
  \bibinfo {author} {\bibfnamefont {J.-F.}\ \bibnamefont {Feng}},\ }\bibfield
  {title} {\bibinfo {title} {The role of coulomb interaction in thermoelectric
  effects of an aharonov–bohm interferometer},\ }\href
  {https://doi.org/10.1088/0957-4484/22/22/225201} {\bibfield  {journal}
  {\bibinfo  {journal} {Nanotechnology}\ }\textbf {\bibinfo {volume} {22}},\
  \bibinfo {pages} {225201} (\bibinfo {year} {2011})}\BibitemShut {NoStop}%
\bibitem [{\citenamefont {Pye}\ \emph {et~al.}(2016)\citenamefont {Pye},
  \citenamefont {Faux},\ and\ \citenamefont {Kearney}}]{Pye2016}%
  \BibitemOpen
  \bibfield  {author} {\bibinfo {author} {\bibfnamefont {A.~J.}\ \bibnamefont
  {Pye}}, \bibinfo {author} {\bibfnamefont {D.~A.}\ \bibnamefont {Faux}},\ and\
  \bibinfo {author} {\bibfnamefont {M.~J.}\ \bibnamefont {Kearney}},\
  }\bibfield  {title} {\bibinfo {title} {Thermoelectric effects in a
  rectangular aharonov-bohm geometry},\ }\href
  {https://doi.org/10.1063/1.4946892} {\bibfield  {journal} {\bibinfo
  {journal} {Journal of Applied Physics}\ }\textbf {\bibinfo {volume} {119}},\
  \bibinfo {pages} {154304} (\bibinfo {year} {2016})}\BibitemShut {NoStop}%
\bibitem [{\citenamefont {Mishra}\ \emph
  {et~al.}(2023{\natexlab{a}})\citenamefont {Mishra}, \citenamefont {Das},\
  and\ \citenamefont {Benjamin}}]{mishra2023majorana}%
  \BibitemOpen
  \bibfield  {author} {\bibinfo {author} {\bibfnamefont {S.}~\bibnamefont
  {Mishra}}, \bibinfo {author} {\bibfnamefont {R.}~\bibnamefont {Das}},\ and\
  \bibinfo {author} {\bibfnamefont {C.}~\bibnamefont {Benjamin}},\ }\bibfield
  {title} {\bibinfo {title} {Majorana thermoelectrics and refrigeration},\
  }\href {https://arxiv.org/abs/2305.12462} {\bibfield  {journal} {\bibinfo
  {journal} {e-Print}\ } (\bibinfo {year} {2023}{\natexlab{a}})},\ \bibinfo
  {note} {arXiv:2305.12462 [cond-mat.mes-hall]}\BibitemShut {NoStop}%
\bibitem [{\citenamefont {Mishra}\ \emph
  {et~al.}(2023{\natexlab{b}})\citenamefont {Mishra}, \citenamefont {Das},\
  and\ \citenamefont {Benjamin}}]{Mishra2023}%
  \BibitemOpen
  \bibfield  {author} {\bibinfo {author} {\bibfnamefont {S.}~\bibnamefont
  {Mishra}}, \bibinfo {author} {\bibfnamefont {R.}~\bibnamefont {Das}},\ and\
  \bibinfo {author} {\bibfnamefont {C.}~\bibnamefont {Benjamin}},\ }\bibfield
  {title} {\bibinfo {title} {Majorana thermoelectrics and refrigeration},\
  }\href@noop {} {\bibfield  {journal} {\bibinfo  {journal} {arXiv preprint}\ }
  (\bibinfo {year} {2023}{\natexlab{b}})},\ \Eprint
  {https://arxiv.org/abs/2305.12462} {arXiv:2305.12462 [cond-mat.mes-hall]}
  \BibitemShut {NoStop}%
\bibitem [{\citenamefont {Hwang}\ \emph {et~al.}(2024)\citenamefont {Hwang},
  \citenamefont {Sothmann},\ and\ \citenamefont
  {L\'opez}}]{PhysRevResearch.6.013215}%
  \BibitemOpen
  \bibfield  {author} {\bibinfo {author} {\bibfnamefont {S.-Y.}\ \bibnamefont
  {Hwang}}, \bibinfo {author} {\bibfnamefont {B.}~\bibnamefont {Sothmann}},\
  and\ \bibinfo {author} {\bibfnamefont {R.}~\bibnamefont {L\'opez}},\
  }\bibfield  {title} {\bibinfo {title} {Phase-controlled heat modulation with
  aharonov-bohm interferometers},\ }\href
  {https://doi.org/10.1103/PhysRevResearch.6.013215} {\bibfield  {journal}
  {\bibinfo  {journal} {Phys. Rev. Res.}\ }\textbf {\bibinfo {volume} {6}},\
  \bibinfo {pages} {013215} (\bibinfo {year} {2024})}\BibitemShut {NoStop}%
\bibitem [{\citenamefont {Briones-Torres}\ \emph {et~al.}(2021)\citenamefont
  {Briones-Torres}, \citenamefont {Pérez-Álvarez}, \citenamefont
  {Molina-Valdovinos},\ and\ \citenamefont
  {Rodríguez-Vargas}}]{Briones-Torres2021}%
  \BibitemOpen
  \bibfield  {author} {\bibinfo {author} {\bibfnamefont {J.~A.}\ \bibnamefont
  {Briones-Torres}}, \bibinfo {author} {\bibfnamefont {R.}~\bibnamefont
  {Pérez-Álvarez}}, \bibinfo {author} {\bibfnamefont {S.}~\bibnamefont
  {Molina-Valdovinos}},\ and\ \bibinfo {author} {\bibfnamefont
  {I.}~\bibnamefont {Rodríguez-Vargas}},\ }\bibfield  {title} {\bibinfo
  {title} {Enhancement of the thermoelectric properties in bilayer graphene
  structures induced by fano resonances},\ }\href
  {https://doi.org/10.1038/s41598-021-93106-4} {\bibfield  {journal} {\bibinfo
  {journal} {Scientific Reports}\ }\textbf {\bibinfo {volume} {11}},\ \bibinfo
  {pages} {13872} (\bibinfo {year} {2021})}\BibitemShut {NoStop}%
\bibitem [{\citenamefont {Klein}\ and\ \citenamefont
  {Michaeli}(2023)}]{PhysRevB.108.075407}%
  \BibitemOpen
  \bibfield  {author} {\bibinfo {author} {\bibfnamefont {D.}~\bibnamefont
  {Klein}}\ and\ \bibinfo {author} {\bibfnamefont {K.}~\bibnamefont
  {Michaeli}},\ }\bibfield  {title} {\bibinfo {title} {Chiral molecules and
  magnets as efficient thermoelectric converters},\ }\href
  {https://doi.org/10.1103/PhysRevB.108.075407} {\bibfield  {journal} {\bibinfo
   {journal} {Phys. Rev. B}\ }\textbf {\bibinfo {volume} {108}},\ \bibinfo
  {pages} {075407} (\bibinfo {year} {2023})}\BibitemShut {NoStop}%
\bibitem [{\citenamefont {Tsuji}\ \emph {et~al.}(2018)\citenamefont {Tsuji},
  \citenamefont {Estrada}, \citenamefont {Movassagh},\ and\ \citenamefont
  {Hoffmann}}]{tsuji2018quantum}%
  \BibitemOpen
  \bibfield  {author} {\bibinfo {author} {\bibfnamefont {Y.}~\bibnamefont
  {Tsuji}}, \bibinfo {author} {\bibfnamefont {E.}~\bibnamefont {Estrada}},
  \bibinfo {author} {\bibfnamefont {R.}~\bibnamefont {Movassagh}},\ and\
  \bibinfo {author} {\bibfnamefont {R.}~\bibnamefont {Hoffmann}},\ }\bibfield
  {title} {\bibinfo {title} {Quantum interference, graphs, walks, and
  polynomials},\ }\href {https://doi.org/10.1021/acs.chemrev.7b00733}
  {\bibfield  {journal} {\bibinfo  {journal} {Chemical Reviews}\ }\textbf
  {\bibinfo {volume} {118}},\ \bibinfo {pages} {4887} (\bibinfo {year}
  {2018})}\BibitemShut {NoStop}%
\bibitem [{\citenamefont {Bedkihal}\ \emph
  {et~al.}(2013{\natexlab{b}})\citenamefont {Bedkihal}, \citenamefont
  {Bandyopadhyay},\ and\ \citenamefont {Segal}}]{PhysRevB.88.155407}%
  \BibitemOpen
  \bibfield  {author} {\bibinfo {author} {\bibfnamefont {S.}~\bibnamefont
  {Bedkihal}}, \bibinfo {author} {\bibfnamefont {M.}~\bibnamefont
  {Bandyopadhyay}},\ and\ \bibinfo {author} {\bibfnamefont {D.}~\bibnamefont
  {Segal}},\ }\bibfield  {title} {\bibinfo {title} {Magnetic field symmetries
  of nonlinear transport with elastic and inelastic scattering},\ }\href
  {https://doi.org/10.1103/PhysRevB.88.155407} {\bibfield  {journal} {\bibinfo
  {journal} {Phys. Rev. B}\ }\textbf {\bibinfo {volume} {88}},\ \bibinfo
  {pages} {155407} (\bibinfo {year} {2013}{\natexlab{b}})}\BibitemShut
  {NoStop}%
\bibitem [{\citenamefont {Wei}\ \emph {et~al.}(2005)\citenamefont {Wei},
  \citenamefont {Shimogawa}, \citenamefont {Wang}, \citenamefont {Radu},
  \citenamefont {Dormaier},\ and\ \citenamefont {Cobden}}]{wei2005phys}%
  \BibitemOpen
  \bibfield  {author} {\bibinfo {author} {\bibfnamefont {J.}~\bibnamefont
  {Wei}}, \bibinfo {author} {\bibfnamefont {M.}~\bibnamefont {Shimogawa}},
  \bibinfo {author} {\bibfnamefont {Z.}~\bibnamefont {Wang}}, \bibinfo {author}
  {\bibfnamefont {I.}~\bibnamefont {Radu}}, \bibinfo {author} {\bibfnamefont
  {R.}~\bibnamefont {Dormaier}},\ and\ \bibinfo {author} {\bibfnamefont
  {D.~H.}\ \bibnamefont {Cobden}},\ }\href@noop {} {\bibfield  {journal}
  {\bibinfo  {journal} {Phys. Rev. Lett.}\ }\textbf {\bibinfo {volume} {95}},\
  \bibinfo {pages} {256601} (\bibinfo {year} {2005})}\BibitemShut {NoStop}%
\bibitem [{\citenamefont {Angers}\ \emph {et~al.}(2007)\citenamefont {Angers},
  \citenamefont {Zakka-Bajjani}, \citenamefont {Deblock}, \citenamefont
  {Gueron}, \citenamefont {Bouchiat}, \citenamefont {Cavanna}, \citenamefont
  {Gennser},\ and\ \citenamefont {Polianski}}]{angers2007phys}%
  \BibitemOpen
  \bibfield  {author} {\bibinfo {author} {\bibfnamefont {L.}~\bibnamefont
  {Angers}}, \bibinfo {author} {\bibfnamefont {E.}~\bibnamefont
  {Zakka-Bajjani}}, \bibinfo {author} {\bibfnamefont {R.}~\bibnamefont
  {Deblock}}, \bibinfo {author} {\bibfnamefont {S.}~\bibnamefont {Gueron}},
  \bibinfo {author} {\bibfnamefont {H.}~\bibnamefont {Bouchiat}}, \bibinfo
  {author} {\bibfnamefont {A.}~\bibnamefont {Cavanna}}, \bibinfo {author}
  {\bibfnamefont {U.}~\bibnamefont {Gennser}},\ and\ \bibinfo {author}
  {\bibfnamefont {M.}~\bibnamefont {Polianski}},\ }\href@noop {} {\bibfield
  {journal} {\bibinfo  {journal} {Phys. Rev. B}\ }\textbf {\bibinfo {volume}
  {75}},\ \bibinfo {pages} {115309} (\bibinfo {year} {2007})}\BibitemShut
  {NoStop}%
\bibitem [{\citenamefont {Sanchez}\ and\ \citenamefont
  {Lopez}(2013)}]{sanchez2013phys}%
  \BibitemOpen
  \bibfield  {author} {\bibinfo {author} {\bibfnamefont {D.}~\bibnamefont
  {Sanchez}}\ and\ \bibinfo {author} {\bibfnamefont {R.}~\bibnamefont
  {Lopez}},\ }\href@noop {} {\bibfield  {journal} {\bibinfo  {journal} {Phys.
  Rev. Lett.}\ }\textbf {\bibinfo {volume} {110}},\ \bibinfo {pages} {026804}
  (\bibinfo {year} {2013})}\BibitemShut {NoStop}%
\bibitem [{\citenamefont {Hwang}\ \emph {et~al.}(2013)\citenamefont {Hwang},
  \citenamefont {Sanchez}, \citenamefont {Lee},\ and\ \citenamefont
  {Lopez}}]{hwang2013njp}%
  \BibitemOpen
  \bibfield  {author} {\bibinfo {author} {\bibfnamefont {S.}~\bibnamefont
  {Hwang}}, \bibinfo {author} {\bibfnamefont {D.}~\bibnamefont {Sanchez}},
  \bibinfo {author} {\bibfnamefont {M.}~\bibnamefont {Lee}},\ and\ \bibinfo
  {author} {\bibfnamefont {R.}~\bibnamefont {Lopez}},\ }\href@noop {}
  {\bibfield  {journal} {\bibinfo  {journal} {New Journal of Physics}\ }\textbf
  {\bibinfo {volume} {15}},\ \bibinfo {pages} {105012} (\bibinfo {year}
  {2013})}\BibitemShut {NoStop}%
\bibitem [{\citenamefont {Jacquet}(2009)}]{jacquet2009stat}%
  \BibitemOpen
  \bibfield  {author} {\bibinfo {author} {\bibfnamefont {P.~A.}\ \bibnamefont
  {Jacquet}},\ }\href@noop {} {\bibfield  {journal} {\bibinfo  {journal} {J.
  Stat. Phys.}\ }\textbf {\bibinfo {volume} {134}},\ \bibinfo {pages} {709}
  (\bibinfo {year} {2009})}\BibitemShut {NoStop}%
\bibitem [{\citenamefont {Jacquet}\ and\ \citenamefont
  {Pillet}(2012)}]{jacquet2012phys}%
  \BibitemOpen
  \bibfield  {author} {\bibinfo {author} {\bibfnamefont {P.}~\bibnamefont
  {Jacquet}}\ and\ \bibinfo {author} {\bibfnamefont {C.}~\bibnamefont
  {Pillet}},\ }\href@noop {} {\bibfield  {journal} {\bibinfo  {journal} {Phys.
  Rev. B}\ }\textbf {\bibinfo {volume} {85}},\ \bibinfo {pages} {125120}
  (\bibinfo {year} {2012})}\BibitemShut {NoStop}%
\bibitem [{\citenamefont {D’Amato}\ and\ \citenamefont
  {Pastawski}(1990)}]{damato1990phys}%
  \BibitemOpen
  \bibfield  {author} {\bibinfo {author} {\bibfnamefont {J.~L.}\ \bibnamefont
  {D’Amato}}\ and\ \bibinfo {author} {\bibfnamefont {H.~M.}\ \bibnamefont
  {Pastawski}},\ }\href@noop {} {\bibfield  {journal} {\bibinfo  {journal}
  {Phys. Rev. B}\ }\textbf {\bibinfo {volume} {41}},\ \bibinfo {pages} {7411}
  (\bibinfo {year} {1990})}\BibitemShut {NoStop}%
\bibitem [{\citenamefont {Bergfield}\ \emph {et~al.}(2013)\citenamefont
  {Bergfield}, \citenamefont {Story}, \citenamefont {Stafford},\ and\
  \citenamefont {Stafford}}]{bergfield2013acs}%
  \BibitemOpen
  \bibfield  {author} {\bibinfo {author} {\bibfnamefont {J.~P.}\ \bibnamefont
  {Bergfield}}, \bibinfo {author} {\bibfnamefont {S.~M.}\ \bibnamefont
  {Story}}, \bibinfo {author} {\bibfnamefont {R.~C.}\ \bibnamefont
  {Stafford}},\ and\ \bibinfo {author} {\bibfnamefont {C.~A.}\ \bibnamefont
  {Stafford}},\ }\href@noop {} {\bibfield  {journal} {\bibinfo  {journal} {ACS
  Nano}\ }\textbf {\bibinfo {volume} {7}},\ \bibinfo {pages} {4429} (\bibinfo
  {year} {2013})}\BibitemShut {NoStop}%
\bibitem [{\citenamefont {F\"orster}\ \emph {et~al.}(2007)\citenamefont
  {F\"orster}, \citenamefont {Samuelsson},\ and\ \citenamefont
  {B\"uttiker}}]{forster2007new}%
  \BibitemOpen
  \bibfield  {author} {\bibinfo {author} {\bibfnamefont {H.}~\bibnamefont
  {F\"orster}}, \bibinfo {author} {\bibfnamefont {P.}~\bibnamefont
  {Samuelsson}},\ and\ \bibinfo {author} {\bibfnamefont {M.}~\bibnamefont
  {B\"uttiker}},\ }\href@noop {} {\bibfield  {journal} {\bibinfo  {journal}
  {New J. Phys.}\ }\textbf {\bibinfo {volume} {9}},\ \bibinfo {pages} {117}
  (\bibinfo {year} {2007})}\BibitemShut {NoStop}%
\bibitem [{\citenamefont {Roy}\ and\ \citenamefont
  {Dhar}(2007)}]{PhysRevB.75.195110}%
  \BibitemOpen
  \bibfield  {author} {\bibinfo {author} {\bibfnamefont {D.}~\bibnamefont
  {Roy}}\ and\ \bibinfo {author} {\bibfnamefont {A.}~\bibnamefont {Dhar}},\
  }\bibfield  {title} {\bibinfo {title} {Electron transport in a one
  dimensional conductor with inelastic scattering by self-consistent
  reservoirs},\ }\href {https://doi.org/10.1103/PhysRevB.75.195110} {\bibfield
  {journal} {\bibinfo  {journal} {Phys. Rev. B}\ }\textbf {\bibinfo {volume}
  {75}},\ \bibinfo {pages} {195110} (\bibinfo {year} {2007})}\BibitemShut
  {NoStop}%
\bibitem [{\citenamefont {Bredol}\ \emph {et~al.}(2021)\citenamefont {Bredol},
  \citenamefont {Boschker}, \citenamefont {Braak},\ and\ \citenamefont
  {Mannhart}}]{PhysRevB.104.115413}%
  \BibitemOpen
  \bibfield  {author} {\bibinfo {author} {\bibfnamefont {P.}~\bibnamefont
  {Bredol}}, \bibinfo {author} {\bibfnamefont {H.}~\bibnamefont {Boschker}},
  \bibinfo {author} {\bibfnamefont {D.}~\bibnamefont {Braak}},\ and\ \bibinfo
  {author} {\bibfnamefont {J.}~\bibnamefont {Mannhart}},\ }\bibfield  {title}
  {\bibinfo {title} {Decoherence effects break reciprocity in matter
  transport},\ }\href {https://doi.org/10.1103/PhysRevB.104.115413} {\bibfield
  {journal} {\bibinfo  {journal} {Phys. Rev. B}\ }\textbf {\bibinfo {volume}
  {104}},\ \bibinfo {pages} {115413} (\bibinfo {year} {2021})}\BibitemShut
  {NoStop}%
\bibitem [{\citenamefont {Blasi}\ \emph {et~al.}(2023)\citenamefont {Blasi}
  \emph {et~al.}}]{Blasi2023}%
  \BibitemOpen
  \bibfield  {author} {\bibinfo {author} {\bibfnamefont {G.}~\bibnamefont
  {Blasi}} \emph {et~al.},\ }\bibfield  {title} {\bibinfo {title} {Citation
  gianmichele blasi et al 2023 quantum sci. technol. 8 015023},\ }\href
  {https://doi.org/10.1088/2058-9565/acacbf} {\bibfield  {journal} {\bibinfo
  {journal} {Quantum Sci. Technol.}\ }\textbf {\bibinfo {volume} {8}},\
  \bibinfo {pages} {015023} (\bibinfo {year} {2023})}\BibitemShut {NoStop}%
\bibitem [{\citenamefont {Sela}\ and\ \citenamefont
  {Affleck}(2009)}]{PhysRevB.79.125110}%
  \BibitemOpen
  \bibfield  {author} {\bibinfo {author} {\bibfnamefont {E.}~\bibnamefont
  {Sela}}\ and\ \bibinfo {author} {\bibfnamefont {I.}~\bibnamefont {Affleck}},\
  }\bibfield  {title} {\bibinfo {title} {Nonequilibrium critical behavior for
  electron tunneling through quantum dots in an aharonov-bohm circuit},\ }\href
  {https://doi.org/10.1103/PhysRevB.79.125110} {\bibfield  {journal} {\bibinfo
  {journal} {Phys. Rev. B}\ }\textbf {\bibinfo {volume} {79}},\ \bibinfo
  {pages} {125110} (\bibinfo {year} {2009})}\BibitemShut {NoStop}%
\bibitem [{\citenamefont {Li}\ \emph {et~al.}(2014)\citenamefont {Li},
  \citenamefont {Zhou}, \citenamefont {Marchesoni},\ and\ \citenamefont
  {Li}}]{li2014scientific}%
  \BibitemOpen
  \bibfield  {author} {\bibinfo {author} {\bibfnamefont {Y.}~\bibnamefont
  {Li}}, \bibinfo {author} {\bibfnamefont {J.}~\bibnamefont {Zhou}}, \bibinfo
  {author} {\bibfnamefont {F.}~\bibnamefont {Marchesoni}},\ and\ \bibinfo
  {author} {\bibfnamefont {B.}~\bibnamefont {Li}},\ }\href@noop {} {\bibfield
  {journal} {\bibinfo  {journal} {Scientific Reports}\ }\textbf {\bibinfo
  {volume} {4}},\ \bibinfo {pages} {4566} (\bibinfo {year} {2014})}\BibitemShut
  {NoStop}%
\bibitem [{\citenamefont {Li}\ and\ \citenamefont
  {Leijnse}(2019)}]{PhysRevB.99.125406}%
  \BibitemOpen
  \bibfield  {author} {\bibinfo {author} {\bibfnamefont {Z.-Z.}\ \bibnamefont
  {Li}}\ and\ \bibinfo {author} {\bibfnamefont {M.}~\bibnamefont {Leijnse}},\
  }\bibfield  {title} {\bibinfo {title} {Quantum interference in transport
  through almost symmetric double quantum dots},\ }\href
  {https://doi.org/10.1103/PhysRevB.99.125406} {\bibfield  {journal} {\bibinfo
  {journal} {Phys. Rev. B}\ }\textbf {\bibinfo {volume} {99}},\ \bibinfo
  {pages} {125406} (\bibinfo {year} {2019})}\BibitemShut {NoStop}%
\bibitem [{\citenamefont {Balduque}\ \emph {et~al.}(2023)\citenamefont
  {Balduque}, \citenamefont {Mecha},\ and\ \citenamefont
  {Sánchez}}]{balduque2023arxiv}%
  \BibitemOpen
  \bibfield  {author} {\bibinfo {author} {\bibfnamefont {J.}~\bibnamefont
  {Balduque}}, \bibinfo {author} {\bibfnamefont {A.}~\bibnamefont {Mecha}},\
  and\ \bibinfo {author} {\bibfnamefont {R.}~\bibnamefont {Sánchez}},\ }\href
  {https://arxiv.org/abs/2405.05637} {\bibfield  {journal} {\bibinfo  {journal}
  {arXiv preprint arXiv:2405.05637}\ } (\bibinfo {year} {2023})},\ \bibinfo
  {note} {[cond-mat.mes-hall]}\BibitemShut {NoStop}%
\bibitem [{\citenamefont {Dresselhaus}\ \emph {et~al.}(2007)\citenamefont
  {Dresselhaus}, \citenamefont {Chen}, \citenamefont {Tang}, \citenamefont
  {Yang}, \citenamefont {Lee}, \citenamefont {Wang}, \citenamefont {Ren},
  \citenamefont {Fleurial},\ and\ \citenamefont {Gogna}}]{ref1}%
  \BibitemOpen
  \bibfield  {author} {\bibinfo {author} {\bibfnamefont {M.~S.}\ \bibnamefont
  {Dresselhaus}}, \bibinfo {author} {\bibfnamefont {G.}~\bibnamefont {Chen}},
  \bibinfo {author} {\bibfnamefont {M.~Y.}\ \bibnamefont {Tang}}, \bibinfo
  {author} {\bibfnamefont {R.}~\bibnamefont {Yang}}, \bibinfo {author}
  {\bibfnamefont {H.}~\bibnamefont {Lee}}, \bibinfo {author} {\bibfnamefont
  {D.}~\bibnamefont {Wang}}, \bibinfo {author} {\bibfnamefont {Z.}~\bibnamefont
  {Ren}}, \bibinfo {author} {\bibfnamefont {J.-P.}\ \bibnamefont {Fleurial}},\
  and\ \bibinfo {author} {\bibfnamefont {P.}~\bibnamefont {Gogna}},\ }\bibfield
   {title} {\bibinfo {title} {New directions for low-dimensional thermoelectric
  materials},\ }\href@noop {} {\bibfield  {journal} {\bibinfo  {journal} {Adv.
  Mater.}\ }\textbf {\bibinfo {volume} {19}},\ \bibinfo {pages} {1043}
  (\bibinfo {year} {2007})}\BibitemShut {NoStop}%
\bibitem [{\citenamefont {Snyder}\ and\ \citenamefont {Toberer}(2008)}]{ref2}%
  \BibitemOpen
  \bibfield  {author} {\bibinfo {author} {\bibfnamefont {G.~J.}\ \bibnamefont
  {Snyder}}\ and\ \bibinfo {author} {\bibfnamefont {S.}~\bibnamefont
  {Toberer}},\ }\bibfield  {title} {\bibinfo {title} {Complex thermoelectric
  materials},\ }\href@noop {} {\bibfield  {journal} {\bibinfo  {journal}
  {Nature Mater.}\ }\textbf {\bibinfo {volume} {7}},\ \bibinfo {pages} {105}
  (\bibinfo {year} {2008})}\BibitemShut {NoStop}%
\bibitem [{\citenamefont {Buttiker}(1986{\natexlab{a}})}]{ref9}%
  \BibitemOpen
  \bibfield  {author} {\bibinfo {author} {\bibfnamefont {M.}~\bibnamefont
  {Buttiker}},\ }\bibfield  {title} {\bibinfo {title} {Role of quantum
  coherence in series resistors},\ }\href@noop {} {\bibfield  {journal}
  {\bibinfo  {journal} {Phys. Rev. B}\ }\textbf {\bibinfo {volume} {33}},\
  \bibinfo {pages} {3020} (\bibinfo {year} {1986}{\natexlab{a}})}\BibitemShut
  {NoStop}%
\bibitem [{\citenamefont {Landauer}(1957{\natexlab{b}})}]{ref7}%
  \BibitemOpen
  \bibfield  {author} {\bibinfo {author} {\bibfnamefont {R.}~\bibnamefont
  {Landauer}},\ }\bibfield  {title} {\bibinfo {title} {Spatial variations of
  currents and fields due to localized scatterers in metallic conduction},\
  }\href@noop {} {\bibfield  {journal} {\bibinfo  {journal} {IBM J. Res. Dev.}\
  }\textbf {\bibinfo {volume} {1}},\ \bibinfo {pages} {223} (\bibinfo {year}
  {1957}{\natexlab{b}})}\BibitemShut {NoStop}%
\bibitem [{\citenamefont {Buttiker}\ \emph {et~al.}(1985)\citenamefont
  {Buttiker}, \citenamefont {Imry}, \citenamefont {Landauer},\ and\
  \citenamefont {Pinhas}}]{ref8}%
  \BibitemOpen
  \bibfield  {author} {\bibinfo {author} {\bibfnamefont {M.}~\bibnamefont
  {Buttiker}}, \bibinfo {author} {\bibfnamefont {Y.}~\bibnamefont {Imry}},
  \bibinfo {author} {\bibfnamefont {R.}~\bibnamefont {Landauer}},\ and\
  \bibinfo {author} {\bibfnamefont {S.}~\bibnamefont {Pinhas}},\ }\bibfield
  {title} {\bibinfo {title} {Generalized many-channel conductance formula with
  application to small rings},\ }\href@noop {} {\bibfield  {journal} {\bibinfo
  {journal} {Phys. Rev. B}\ }\textbf {\bibinfo {volume} {31}},\ \bibinfo
  {pages} {6207} (\bibinfo {year} {1985})}\BibitemShut {NoStop}%
\bibitem [{\citenamefont {Brandner}(2015)}]{ref20}%
  \BibitemOpen
  \bibfield  {author} {\bibinfo {author} {\bibfnamefont {K.}~\bibnamefont
  {Brandner}},\ }\bibfield  {title} {\bibinfo {title} {Universal bounds on
  efficiency and power of heat engines with broken time-reversal symmetry},\
  }\href@noop {} {\bibfield  {journal} {\bibinfo  {journal} {PhD Thesis}\ ,\
  \bibinfo {pages} {125}} (\bibinfo {year} {2015})}\BibitemShut {NoStop}%
\bibitem [{\citenamefont {Buttiker}(1986{\natexlab{b}})}]{ref10}%
  \BibitemOpen
  \bibfield  {author} {\bibinfo {author} {\bibfnamefont {M.}~\bibnamefont
  {Buttiker}},\ }\bibfield  {title} {\bibinfo {title} {Four-terminal
  phase-coherent conductance},\ }\href@noop {} {\bibfield  {journal} {\bibinfo
  {journal} {Phys. Rev. Lett.}\ }\textbf {\bibinfo {volume} {57}},\ \bibinfo
  {pages} {1761} (\bibinfo {year} {1986}{\natexlab{b}})}\BibitemShut {NoStop}%
\bibitem [{\citenamefont {Buttiker}(1988)}]{ref11}%
  \BibitemOpen
  \bibfield  {author} {\bibinfo {author} {\bibfnamefont {M.}~\bibnamefont
  {Buttiker}},\ }\bibfield  {title} {\bibinfo {title} {Symmetry of electrical
  conduction},\ }\href@noop {} {\bibfield  {journal} {\bibinfo  {journal} {IBM
  J. Res. Dev.}\ }\textbf {\bibinfo {volume} {32}},\ \bibinfo {pages} {317}
  (\bibinfo {year} {1988})}\BibitemShut {NoStop}%
\bibitem [{\citenamefont {Callen}(1985)}]{ref12}%
  \BibitemOpen
  \bibfield  {author} {\bibinfo {author} {\bibfnamefont {H.~B.}\ \bibnamefont
  {Callen}},\ }\href@noop {} {\emph {\bibinfo {title} {Thermodynamics and an
  Introduction to Thermostatics}}},\ \bibinfo {edition} {2nd}\ ed.\ (\bibinfo
  {publisher} {John Wiley \& Sons},\ \bibinfo {address} {New York},\ \bibinfo
  {year} {1985})\BibitemShut {NoStop}%
\bibitem [{\citenamefont {Goupil}\ \emph {et~al.}(2011)\citenamefont {Goupil},
  \citenamefont {Seifert}, \citenamefont {Zabrocki}, \citenamefont {Muller},\
  and\ \citenamefont {Snyder}}]{ref13}%
  \BibitemOpen
  \bibfield  {author} {\bibinfo {author} {\bibfnamefont {C.}~\bibnamefont
  {Goupil}}, \bibinfo {author} {\bibfnamefont {W.}~\bibnamefont {Seifert}},
  \bibinfo {author} {\bibfnamefont {K.}~\bibnamefont {Zabrocki}}, \bibinfo
  {author} {\bibfnamefont {E.}~\bibnamefont {Muller}},\ and\ \bibinfo {author}
  {\bibfnamefont {G.~J.}\ \bibnamefont {Snyder}},\ }\bibfield  {title}
  {\bibinfo {title} {Thermodynamics of thermoelectric phenomena and
  applications},\ }\href@noop {} {\bibfield  {journal} {\bibinfo  {journal}
  {Entropy}\ }\textbf {\bibinfo {volume} {13}},\ \bibinfo {pages} {1481}
  (\bibinfo {year} {2011})}\BibitemShut {NoStop}%
\bibitem [{\citenamefont {Bell}(2008)}]{Bell}%
  \BibitemOpen
  \bibfield  {author} {\bibinfo {author} {\bibfnamefont {L.~E.}\ \bibnamefont
  {Bell}},\ }\bibfield  {title} {\bibinfo {title} {Cooling, heating, generating
  power, and recovering waste heat with thermoelectric systems},\ }\href
  {https://doi.org/10.1126/science.1158899} {\bibfield  {journal} {\bibinfo
  {journal} {Science}\ }\textbf {\bibinfo {volume} {321}},\ \bibinfo {pages}
  {1457} (\bibinfo {year} {2008})}\BibitemShut {NoStop}%
\bibitem [{\citenamefont {Godijn}\ \emph {et~al.}(1999)\citenamefont {Godijn},
  \citenamefont {Möller}, \citenamefont {Buhmann}, \citenamefont {Molenkamp},
  ,\ and\ \citenamefont {van Langen}}]{ref21}%
  \BibitemOpen
  \bibfield  {author} {\bibinfo {author} {\bibfnamefont {S.~F.}\ \bibnamefont
  {Godijn}}, \bibinfo {author} {\bibfnamefont {S.}~\bibnamefont {Möller}},
  \bibinfo {author} {\bibfnamefont {H.}~\bibnamefont {Buhmann}}, \bibinfo
  {author} {\bibfnamefont {L.~W.}\ \bibnamefont {Molenkamp}}, ,\ and\ \bibinfo
  {author} {\bibfnamefont {S.~A.}\ \bibnamefont {van Langen}},\ }\bibfield
  {title} {\bibinfo {title} {Thermopower of a chaotic quantum dot},\
  }\href@noop {} {\bibfield  {journal} {\bibinfo  {journal} {Phys. Rev. Lett.}\
  }\textbf {\bibinfo {volume} {82}},\ \bibinfo {pages} {2927} (\bibinfo {year}
  {1999})}\BibitemShut {NoStop}%
\bibitem [{\citenamefont {Eom}\ \emph {et~al.}(1998)\citenamefont {Eom},
  \citenamefont {Chien},\ and\ \citenamefont {Chandrasekhar}}]{ref22}%
  \BibitemOpen
  \bibfield  {author} {\bibinfo {author} {\bibfnamefont {J.}~\bibnamefont
  {Eom}}, \bibinfo {author} {\bibfnamefont {C.-J.}\ \bibnamefont {Chien}},\
  and\ \bibinfo {author} {\bibfnamefont {V.}~\bibnamefont {Chandrasekhar}},\
  }\bibfield  {title} {\bibinfo {title} {Phase dependent thermopower in andreev
  interferometers},\ }\href@noop {} {\bibfield  {journal} {\bibinfo  {journal}
  {Phys. Rev. Lett.}\ }\textbf {\bibinfo {volume} {81}},\ \bibinfo {pages}
  {437} (\bibinfo {year} {1998})}\BibitemShut {NoStop}%
\bibitem [{\citenamefont {Brandner}\ and\ \citenamefont
  {Seifert}(2010)}]{ref23}%
  \BibitemOpen
  \bibfield  {author} {\bibinfo {author} {\bibfnamefont {K.}~\bibnamefont
  {Brandner}}\ and\ \bibinfo {author} {\bibfnamefont {U.}~\bibnamefont
  {Seifert}},\ }\bibfield  {title} {\bibinfo {title} {Multi-terminal
  thermoelectric transport in a magnetic field: Bounds on onsager coefficients
  and efficiency},\ }\href@noop {} {\bibfield  {journal} {\bibinfo  {journal}
  {Eur. Phys. lett.}\ }\textbf {\bibinfo {volume} {91}},\ \bibinfo {pages}
  {67009} (\bibinfo {year} {2010})}\BibitemShut {NoStop}%
\bibitem [{\citenamefont {Izumida}\ and\ \citenamefont
  {Okuda}(2012)}]{Izumida1}%
  \BibitemOpen
  \bibfield  {author} {\bibinfo {author} {\bibfnamefont {Y.}~\bibnamefont
  {Izumida}}\ and\ \bibinfo {author} {\bibfnamefont {K.}~\bibnamefont
  {Okuda}},\ }\bibfield  {title} {\bibinfo {title} {Efficient performance of
  minimally nonlinear irreversible heat engines},\ }\href@noop {} {\bibfield
  {journal} {\bibinfo  {journal} {EPL}\ }\textbf {\bibinfo {volume} {97}},\
  \bibinfo {pages} {10004} (\bibinfo {year} {2012})}\BibitemShut {NoStop}%
\bibitem [{\citenamefont {Izumida}\ \emph {et~al.}(2015)\citenamefont
  {Izumida}, \citenamefont {Okuda}, \citenamefont {Roco},\ and\ \citenamefont
  {Hernandez}}]{Izumida2}%
  \BibitemOpen
  \bibfield  {author} {\bibinfo {author} {\bibfnamefont {Y.}~\bibnamefont
  {Izumida}}, \bibinfo {author} {\bibfnamefont {K.}~\bibnamefont {Okuda}},
  \bibinfo {author} {\bibfnamefont {J.~M.~M.}\ \bibnamefont {Roco}},\ and\
  \bibinfo {author} {\bibfnamefont {A.~C.}\ \bibnamefont {Hernandez}},\
  }\bibfield  {title} {\bibinfo {title} {Linear irreversible heat engines based
  on strongly coupled thermodynamic systems},\ }\href@noop {} {\bibfield
  {journal} {\bibinfo  {journal} {Phys. Rev. E}\ }\textbf {\bibinfo {volume}
  {91}},\ \bibinfo {pages} {052140} (\bibinfo {year} {2015})}\BibitemShut
  {NoStop}%
\bibitem [{\citenamefont {Liu}\ \emph {et~al.}(2019)\citenamefont {Liu},
  \citenamefont {Li}, \citenamefont {Zhang}, \citenamefont {He},\ and\
  \citenamefont {Wang}}]{entropy_Liu}%
  \BibitemOpen
  \bibfield  {author} {\bibinfo {author} {\bibfnamefont {Q.}~\bibnamefont
  {Liu}}, \bibinfo {author} {\bibfnamefont {W.}~\bibnamefont {Li}}, \bibinfo
  {author} {\bibfnamefont {M.}~\bibnamefont {Zhang}}, \bibinfo {author}
  {\bibfnamefont {J.}~\bibnamefont {He}},\ and\ \bibinfo {author}
  {\bibfnamefont {J.}~\bibnamefont {Wang}},\ }\bibfield  {title} {\bibinfo
  {title} {Entropy production and efficiency in quantum heat engines},\
  }\href@noop {} {\bibfield  {journal} {\bibinfo  {journal} {Entropy}\ }\textbf
  {\bibinfo {volume} {21}},\ \bibinfo {pages} {717} (\bibinfo {year}
  {2019})}\BibitemShut {NoStop}%
\bibitem [{\citenamefont {Long}\ \emph {et~al.}(2014)\citenamefont {Long},
  \citenamefont {Liu},\ and\ \citenamefont {Liu}}]{Long1}%
  \BibitemOpen
  \bibfield  {author} {\bibinfo {author} {\bibfnamefont {R.}~\bibnamefont
  {Long}}, \bibinfo {author} {\bibfnamefont {Z.}~\bibnamefont {Liu}},\ and\
  \bibinfo {author} {\bibfnamefont {W.}~\bibnamefont {Liu}},\ }\bibfield
  {title} {\bibinfo {title} {Performance analysis of heat engines based on
  strongly coupled thermoelectric systems},\ }\href@noop {} {\bibfield
  {journal} {\bibinfo  {journal} {Phys. Rev. E}\ }\textbf {\bibinfo {volume}
  {89}},\ \bibinfo {pages} {062119} (\bibinfo {year} {2014})}\BibitemShut
  {NoStop}%
\bibitem [{\citenamefont {Long}\ and\ \citenamefont {Liu}(2016)}]{Long2}%
  \BibitemOpen
  \bibfield  {author} {\bibinfo {author} {\bibfnamefont {R.}~\bibnamefont
  {Long}}\ and\ \bibinfo {author} {\bibfnamefont {W.}~\bibnamefont {Liu}},\
  }\bibfield  {title} {\bibinfo {title} {Thermodynamic analysis of quantum heat
  engines using feynman ratchets},\ }\href@noop {} {\bibfield  {journal}
  {\bibinfo  {journal} {Phys. Rev. E}\ }\textbf {\bibinfo {volume} {94}},\
  \bibinfo {pages} {052114} (\bibinfo {year} {2016})}\BibitemShut {NoStop}%
\bibitem [{\citenamefont {Iyyappan}\ and\ \citenamefont
  {Ponmurugan}(2018)}]{Iyyappan1}%
  \BibitemOpen
  \bibfield  {author} {\bibinfo {author} {\bibfnamefont {I.}~\bibnamefont
  {Iyyappan}}\ and\ \bibinfo {author} {\bibfnamefont {M.}~\bibnamefont
  {Ponmurugan}},\ }\bibfield  {title} {\bibinfo {title} {Thermodynamics of
  quantum heat engines: A symmetry analysis},\ }\href@noop {} {\bibfield
  {journal} {\bibinfo  {journal} {Phys. Rev. E}\ }\textbf {\bibinfo {volume}
  {97}},\ \bibinfo {pages} {012141} (\bibinfo {year} {2018})}\BibitemShut
  {NoStop}%
\bibitem [{\citenamefont {Ponmurugan}(2019)}]{Ponmurugan}%
  \BibitemOpen
  \bibfield  {author} {\bibinfo {author} {\bibfnamefont {M.}~\bibnamefont
  {Ponmurugan}},\ }\bibfield  {title} {\bibinfo {title} {Quantum heat engines
  and thermal machines},\ }\href@noop {} {\bibfield  {journal} {\bibinfo
  {journal} {J. Non-Equilib. Thermodyn.}\ }\textbf {\bibinfo {volume} {44}},\
  \bibinfo {pages} {143} (\bibinfo {year} {2019})}\BibitemShut {NoStop}%
\bibitem [{\citenamefont {Bai}\ \emph {et~al.}(2023)\citenamefont {Bai},
  \citenamefont {Li},\ and\ \citenamefont {Zhang}}]{Bai}%
  \BibitemOpen
  \bibfield  {author} {\bibinfo {author} {\bibfnamefont {L.}~\bibnamefont
  {Bai}}, \bibinfo {author} {\bibfnamefont {L.}~\bibnamefont {Li}},\ and\
  \bibinfo {author} {\bibfnamefont {R.}~\bibnamefont {Zhang}},\ }\bibfield
  {title} {\bibinfo {title} {Thermodynamic optimization of quantum heat
  engines},\ }\href@noop {} {\bibfield  {journal} {\bibinfo  {journal} {Chinese
  Journal of Physics}\ }\textbf {\bibinfo {volume} {83}},\ \bibinfo {pages}
  {418} (\bibinfo {year} {2023})}\BibitemShut {NoStop}%
\bibitem [{\citenamefont {Zhang}\ \emph {et~al.}(2018)\citenamefont {Zhang},
  \citenamefont {Liu}, \citenamefont {Li}, \citenamefont {Zhang},\ and\
  \citenamefont {Bai}}]{Zhang2}%
  \BibitemOpen
  \bibfield  {author} {\bibinfo {author} {\bibfnamefont {R.}~\bibnamefont
  {Zhang}}, \bibinfo {author} {\bibfnamefont {W.}~\bibnamefont {Liu}}, \bibinfo
  {author} {\bibfnamefont {Q.}~\bibnamefont {Li}}, \bibinfo {author}
  {\bibfnamefont {L.}~\bibnamefont {Zhang}},\ and\ \bibinfo {author}
  {\bibfnamefont {L.}~\bibnamefont {Bai}},\ }\bibfield  {title} {\bibinfo
  {title} {Analysis of the performance of quantum heat engines using different
  working substances},\ }\href@noop {} {\bibfield  {journal} {\bibinfo
  {journal} {Phys. Lett. A}\ }\textbf {\bibinfo {volume} {382}},\ \bibinfo
  {pages} {20} (\bibinfo {year} {2018})}\BibitemShut {NoStop}%
\bibitem [{\citenamefont {den Broeck}(2005)}]{Broeck}%
  \BibitemOpen
  \bibfield  {author} {\bibinfo {author} {\bibfnamefont {C.~V.}\ \bibnamefont
  {den Broeck}},\ }\bibfield  {title} {\bibinfo {title} {Thermodynamic
  efficiency at maximum power},\ }\href@noop {} {\bibfield  {journal} {\bibinfo
   {journal} {Phys. Rev. Lett.}\ }\textbf {\bibinfo {volume} {95}},\ \bibinfo
  {pages} {190602} (\bibinfo {year} {2005})}\BibitemShut {NoStop}%
\bibitem [{\citenamefont {Esposito}\ \emph {et~al.}(2010)\citenamefont
  {Esposito}, \citenamefont {Kawai}, \citenamefont {Lindenberg},\ and\
  \citenamefont {den Broeck}}]{lowCarnot}%
  \BibitemOpen
  \bibfield  {author} {\bibinfo {author} {\bibfnamefont {M.}~\bibnamefont
  {Esposito}}, \bibinfo {author} {\bibfnamefont {R.}~\bibnamefont {Kawai}},
  \bibinfo {author} {\bibfnamefont {K.}~\bibnamefont {Lindenberg}},\ and\
  \bibinfo {author} {\bibfnamefont {C.~V.}\ \bibnamefont {den Broeck}},\
  }\bibfield  {title} {\bibinfo {title} {Efficiency at maximum power of
  low-dissipation carnot engines},\ }\href@noop {} {\bibfield  {journal}
  {\bibinfo  {journal} {Phys. Rev. Lett.}\ }\textbf {\bibinfo {volume} {105}},\
  \bibinfo {pages} {150603} (\bibinfo {year} {2010})}\BibitemShut {NoStop}%
\bibitem [{\citenamefont {Saito}\ \emph
  {et~al.}(2011{\natexlab{b}})\citenamefont {Saito}, \citenamefont {Benenti},
  \citenamefont {Casati},\ and\ \citenamefont {Prosen}}]{Saito}%
  \BibitemOpen
  \bibfield  {author} {\bibinfo {author} {\bibfnamefont {K.}~\bibnamefont
  {Saito}}, \bibinfo {author} {\bibfnamefont {G.}~\bibnamefont {Benenti}},
  \bibinfo {author} {\bibfnamefont {G.}~\bibnamefont {Casati}},\ and\ \bibinfo
  {author} {\bibfnamefont {T.}~\bibnamefont {Prosen}},\ }\bibfield  {title}
  {\bibinfo {title} {Symmetry of nonlinear response and efficiency of molecular
  machines},\ }\href@noop {} {\bibfield  {journal} {\bibinfo  {journal} {Phys.
  Rev. B}\ }\textbf {\bibinfo {volume} {84}},\ \bibinfo {pages} {201306(R)}
  (\bibinfo {year} {2011}{\natexlab{b}})}\BibitemShut {NoStop}%
\bibitem [{\citenamefont {Zhang}\ \emph {et~al.}(2017)\citenamefont {Zhang},
  \citenamefont {Li}, \citenamefont {Tang}, \citenamefont {Yang},\ and\
  \citenamefont {Bai}}]{RongZhang}%
  \BibitemOpen
  \bibfield  {author} {\bibinfo {author} {\bibfnamefont {R.}~\bibnamefont
  {Zhang}}, \bibinfo {author} {\bibfnamefont {Q.-W.}\ \bibnamefont {Li}},
  \bibinfo {author} {\bibfnamefont {F.~R.}\ \bibnamefont {Tang}}, \bibinfo
  {author} {\bibfnamefont {X.~Q.}\ \bibnamefont {Yang}},\ and\ \bibinfo
  {author} {\bibfnamefont {L.}~\bibnamefont {Bai}},\ }\bibfield  {title}
  {\bibinfo {title} {Route towards the optimization at given power of
  thermoelectric heat engines with broken time-reversal symmetry},\ }\href
  {https://doi.org/10.1103/PhysRevE.96.022133} {\bibfield  {journal} {\bibinfo
  {journal} {Phys. Rev. E}\ }\textbf {\bibinfo {volume} {96}},\ \bibinfo
  {pages} {022133} (\bibinfo {year} {2017})}\BibitemShut {NoStop}%
\bibitem [{\citenamefont {Sartipi}\ and\ \citenamefont {Vahedi}(2023)}]{Zahra}%
  \BibitemOpen
  \bibfield  {author} {\bibinfo {author} {\bibfnamefont {Z.}~\bibnamefont
  {Sartipi}}\ and\ \bibinfo {author} {\bibfnamefont {J.}~\bibnamefont
  {Vahedi}},\ }\bibfield  {title} {\bibinfo {title} {Effect of quantum
  coherence on the performance of a quantum thermoelectric heat engine},\
  }\href@noop {} {\bibfield  {journal} {\bibinfo  {journal} {Phys. Rev. B}\
  }\textbf {\bibinfo {volume} {108}},\ \bibinfo {pages} {195435} (\bibinfo
  {year} {2023})}\BibitemShut {NoStop}%
\bibitem [{\citenamefont {Aharonov}\ and\ \citenamefont
  {Bohm}(1959)}]{aharonov1959}%
  \BibitemOpen
  \bibfield  {author} {\bibinfo {author} {\bibfnamefont {Y.}~\bibnamefont
  {Aharonov}}\ and\ \bibinfo {author} {\bibfnamefont {D.}~\bibnamefont
  {Bohm}},\ }\bibfield  {title} {\bibinfo {title} {Significance of
  electromagnetic potentials in the quantum theory},\ }\href@noop {} {\bibfield
   {journal} {\bibinfo  {journal} {Phys. Rev.}\ }\textbf {\bibinfo {volume}
  {115}},\ \bibinfo {pages} {481} (\bibinfo {year} {1959})}\BibitemShut
  {NoStop}%
\end{thebibliography}%
\end{document}